\documentclass[a4paper,11pt]{article}
\usepackage[margin=1in]{geometry}
\usepackage{epsfig}
\usepackage{cite}
\usepackage{graphicx}
\DeclareGraphicsExtensions{%
    .png,.PNG,%
    .pdf,.PDF,%
    .jpg,.mps,.jpeg,.jbig2,.jb2,.JPG,.JPEG,.JBIG2,.JB2}
\usepackage{epstopdf}
\usepackage{color}
\UseRawInputEncoding
\usepackage[nottoc]{tocbibind}
\usepackage[english]{babel}
\usepackage{tabularx}
\usepackage{xparse}
\usepackage{braket}
\usepackage{amsmath}
\usepackage{hyperref}
\usepackage{grfext}
\usepackage{amssymb}
\usepackage{caption}
\usepackage{mathtools}
\usepackage{subcaption}
\usepackage{algorithm2e}
\usepackage{float}
\usepackage{subcaption}
\usepackage{multirow}
\restylefloat{table}
\usepackage{booktabs}
\usepackage[dvipsnames]{xcolor}
\usepackage{tikz}
\usepackage{pdflscape}
\usepackage{lscape}
\date{\today}
 \normalsize

\newcommand{\bmat}{\left(\begin{array}}
\newcommand{\emat}{\end{array}\right)}
\newcommand{\be}{\begin{equation}}
\newcommand{\ee}{\end{equation}}
\newcommand{\bea}{\begin{eqnarray}}
\newcommand{\eea}{\end{eqnarray}}

\def\gtwid{\mathrel{\raise.3ex\hbox{$>$\kern-.75em\lower1ex\hbox{$\sim$}}}}
\def\ltwid{\mathrel{\raise.3ex\hbox{$<$\kern-.75em\lower1ex\hbox{$\sim$}}}}
\def\gev{{\rm \, Ge\kern-0.125em V}}
\def\tev{{\rm \, Te\kern-0.125em V}}

\def    \be            {\begin{equation}}
\def    \ee            {\end{equation}}
\def    \bea           {\begin{eqnarray}}
\def    \eea           {\end{eqnarray}}
\def\eps{\epsilon}
\def\a{\alpha}
\def\b{\beta}

\def\d{\delta}
\def\n{\nu}

\def\m{\mu}

\def\d{\delta}

\def\s{\sigma}
\def\r{\rho}
\def\t{\theta}

 \normalsize

\begin{document}
\renewcommand{\thefootnote}{\fnsymbol{footnote}}
\vspace{.3cm}
\title{\Large\bf Texture of One Equality in Neutrino Mass Matrix}
\author
{ \hspace{-2.cm} \it \bf A. Ismael$^{1,2}$\thanks{ahmedEhusien@sci.asu.edu.eg},  E. I. Lashin$^{1,2,3}$\thanks{slashin@zewailcity.edu.eg, elashin@ictp.it}, M. AlKhateeb$^{4,5}$\thanks{mohkha88@gmail.com} and  N. Chamoun$^{6}$\thanks{nidal.chamoun@hiast.edu.sy},
 \\\hspace{-2.cm}
\footnotesize$^1$  Department of Physics, Faculty of Science, Ain Shams University, Cairo 11566,  Egypt.  \\\hspace{-2.cm}
\footnotesize$^2$ Centre for Fundamental Physics, Zewail City of Science and
Technology, %\\\hspace{-3.cm}
 \footnotesize  6 October City, Giza 12578, Egypt.  \\\hspace{-2.cm}
\footnotesize$^3$ The Abdus Salam, ICTP, P. O. Box 586, 34100 Trieste,  Italy.\\\hspace{-2.cm}
\footnotesize$^4$ Department of Physics, Faculty of Science, Damascus University,  Syria.\\\hspace{-2.cm}
\footnotesize$^5$ D$\acute{e}$partement de Physique, University$\acute{e}$ de Cergy-Pontoise, F-95302 Cergy-Pontoise, France.\\\hspace{-2.cm}
\footnotesize$^6$ Physics Department, HIAST, P.O.Box 31983, Damascus, Syria.
}
%\date{\formatdate{2}{1}{2021}}
\date{\today}
\maketitle
%\begin{center}
%\small{\bf Abstract}\\[3mm]
%\end{center}
\begin{abstract}
We carry out a phenomenological and analytical study of the texture structures for the Majorana neutrino mass matrix characterized by one single equality between two independent matrix elements, with vanishing non-physical phases whose role in the definition we clarify. We find that fourteen textures are viable in both types of hierarchy, whereas the remaining fifteenth one is viable only in normal hierarchy. We also present symmetry realizations for some patterns by using Abelian flavor symmetries within type-I, type-II and mixed type-(I+II) seesaw scenarios.
\end{abstract}
\maketitle
\vskip 0.3cm \hrule \vskip 0.5cm'
%**************************************************************
%**************************************************************
\section{Introduction}
The neutrino oscillation observations present conclusive evidence that neutrinos are massive and that their flavor states do oscillate. Moreover, they open the door for physics beyond the standard model \cite{PhysRevLett.81.1562,PhysRevLett.89.011301,PhysRevLett.90.021802,PhysRevLett.108.171803, PhysRevLett.90.041801}. If we work in the flavor basis, where charged lepton mass matrix $M_{l}$ is diagonal and assume the neutrinos are of Majorana-type, then the neutrino mass matrix $M_{\nu}$ becomes symmetric with $12$ free parameters. However, one can fix $3$ ``nonphysical phases'' without changing the physics leaving thus $9$ physical free parameters in $M_{\nu}$. This number is still greater than the number of parameters which can be measured by the experiments. There are experimental bounds only on the values of the mass-squared differences, the three mixing angles ($\theta_{12}$, $\theta_{23}$, $\theta_{13}$), and the Dirac phase $\delta$. However, no experimental constraints on the two Majorana-phases ($\rho$ and $\sigma$) up till now. In order to reduce the number of free parameters, some phenomenological models have been studied such as textures characterized by zero elements \cite{Fritzsch_2011,Frampton_2002,Xing_2002,Lashin_2012}, zero minors \cite{Lashin_2008,Lashin_2009,Dev_2011},
two vanishing subtraces \cite{Alhendi_2008}, exact or partial $\mu$-$\tau$ symmetry \cite{Mohapatra_1999,Grimus_2001, Harrison_2002, Gupta_2013}, and some corresponding variants \cite{Chamoun_2021}.

In light of the recent experimental data in \cite{de_Salas_2021}, we carry out in this paper a complete numerical investigation of all possible textures characterized by one equality between two independent matrix elements in $M_{\nu}$ at all $\sigma$-levels. We motivate our work as follows. First, studies about one zero textures are abundant in the literature (e.g. \cite{merlePrd2006,deepthiEPJC2012,liaoPrd2013,gengEpjc2015,barreirosJhep2019,KitabayashiPrd2020}). Whereas one-zero textures were generalized to textures characterized by one-zero minor in \cite{Lashin_2009} and by one-zero subtrace in \cite{ismael}, we further these studies in the present work by examining related textures. As we shall see, six one-equality textures can be related by unitary transformations to one-zero textures. This linkage between equality textures and zero textures helps in finding realization models for the former ones starting from `known' realization models for the latter ones, as was detailed in \cite{ismael}. In addition to this ``indirect" way of realization, we shall also content ourselves in this work by finding directly realization patterns, and it is the same six patterns related to zero textures where we could find ``direct'' realizations. Second, our work is furthermore motivated by existing studies in the literature involving equalities between the entries or cofactors \cite{Dev_2013} in $M_\nu$, or involving hybrid textures \cite{{Liu_2013}, {Han_2017}} where one imposes an equality constraint and another one-zero constraint. Actually, the work of \cite{Han_2017} can be considered as a special case of the present work by restricting the study to singular $M_{\nu}$. Third, it is well known that one needs to perturb the texture satisfying $\mu$-$\tau$-symmetry ($M_{e \mu}$= $M_{e \tau}$ and $M_{\mu \mu}$= $M_{\tau \tau}$), which leads to a vanishing $\theta_{13}$, in order to meet experimental constraints. One of the ways to do this is to break one of the two defining equalities, as was done in \cite{Lashin_2014}, so we can consider the one-equality texture as a generalization of the `perturbed' $\mu$-$\tau$-symmetry texture. Interestingly, we find that these `perturbed' matrices, when looked upon as one-equality textures, are the ones which can be related to one-zero textures. Fourth, although the studied texture is represented by one constraint, and so one might think naively that its predictive power is weakened compared to textures defined by more, say two, constraints, however, the way we define the texture by one constraint after fixing the nonphysical phases, increases the constraints, and the number of free parameters to be spanned is not more than in some other textures defined by two constraints. As a matter of fact, the two-equalities constraint in \cite{Dev_2013}, reduces the number of parameters in $M_{\nu}$ from 12 to 8, and the spanning was carried out over 7 physical parameters therein, exactly as in our present work where fixing the nonphysical phases makes $M_{\nu}$ of 9 parameters which are reduced to 7 upon imposing the one-equality constraint. It is not a trivial exercise that our one-equality texture, with $9-2=7$ degrees of freedom, can satisfy eight experimental constraints: four ranges of  ($\theta_{12}$, $\theta_{23}$, $\theta_{13}$, $\delta$) and two mass squares differences, added to two constraints on the lightest neutrino mass (LNM) coming from the `effective' masses ($m_e, m_{ee}$). As we shall see, the parameter space, and especially for the ($m_{ee}$ versus LNM) correlation plots, is well restricted in view of the experimental constraints. Moreover, one of the textures ($M_{\nu 22}=M_{\nu 23}$ with inverted hierarchy), which is not viable except for $3\sigma$ level, can be considered very predictive, in that it foretells, say, that at $3\sigma$-level $\theta_{23}$ lies in the second octant and that $\d$ lies in $[325^o,353^o]$. Finally, the work of \cite{ismael} concerning textures where two independent entries are opposite each other is related to our current work, however they are not equivalent since both works adopt vanishing nonphysical phases where one has no longer phase freedom allowing to relate one form to the other.

One important point that is usually mentioned only in passing  concerns the role of the nonphysical phases in the definition of the texture, in that some studies, e.g. \cite{Dev_2013}, define the texture as the form of $M_\nu$ which satisfies the specific defining constraint for at least one choice of the nonphysical phases, whereas in other studies, e.g. \cite{{Liu_2013}, {Han_2017}}, the texture is defined after fixing the nonphysical phases. The two definitions are not equivalent, and there are advantages/disadvantages in adopting one or the other of the two definitions. In this paper, we shall underline the difference between these two approaches for defining any given texture, contrasting their good and bad points, but for the phenomenology and the analysis, we shall stick to the second definition, putting all the nonphysical phases to zero, because it corresponds to a tighter allowable parameter space, which would increase the predictability power of the texture. Moreover, the specific definition fixing the nonphysical phases to zero makes the computations simpler and unlike the other definition, the corresponding constraint involves normally analytic functions of the matrix entries.

As said earlier, the texture with the constraint of one complex equality (2 real conditions) corresponds to a reduction by $2$ degrees of freedom of the 9 physical parameters of $M_\nu$, so by varying all the seven free parameters ($\theta_{12}$, $\theta_{23}$, $\theta_{13}$, $\delta$, $\rho$, $\sigma$, $\delta m^{2}$) randomly over their allowed experimental ranges, we can reconstruct $M_\nu$ and see whether it meets the remaining experimental bounds getting, thus, allowed regions in the parameter space. We find that out of the  $15$ possible patterns, all of them can accommodate the experimental data in the case of normal hierarchy ($m_{1}<$  $m_{2}<$  $m_{3}$). As to the inverted hierarchy ($m_{3}<$ $m_{1}<$ $m_{2}$), one texture ($M_{\n 33}=M_{\n 23}$) is not viable, and another texture ($M_{\n 22}=M_{\n 23}$, related by $\mu-\tau$-symmetry to the excluded one, is viable but only at the $3\sigma$-level, with a scarce acceptable region in the parameter space, which gives thus a strong predictive power.

Since the current experimental data do not respect the $\mu$-$\tau$ interchange symmetry (e.g. $\theta_{23}$ data is not symmetric around $45^o$), one can not in principle relate automatically the allowed regions of two $\mu$-$\tau$-symmetry-related patterns. Thus, regarding the numerical discussion and correlation plots, we have presented all the possible patterns. As to the analytical expressions, we limited the presentation to nine patterns, as one can deduce, according to $\mu$-$\tau$-symmetry, the corresponding expressions for the remaining related six cases.

We then introduce symmetry realizations for some viable patterns. We start by ``direct'' realizations using $Z_{2}\times Z_{6}$ abelian flavor symmetry within the type-I seesaw scenario and $Z_{2}\times Z_{2}^{'}$ symmetry within the type-II seesaw scenario. The realization within mixed type-(I+II) scenarios by using $Z_2\times Z_{2}^{'}\times U(1)^{3}$ symmetry are also presented where $U(1)^{3}$ represents a separate lepton number symmetry, which is softly broken by the 3-dim operators in the Higgs sector in order to avoid the existence of the Goldstone bosons. We follow by presenting briefly ``indirect'' realizations related to zero-textures

The plan of the paper is as follows: In section 2, we state explicitly the role of the nonphysical phases in the texture definition, whereas in the following section 3 we introduce our notations, adopted conventions, and the experimental bounds of the oscillation and non-oscillation parameters. In section 4, we present the formulae that define the mass ratios ($\frac{m_{1}}{m_{3}}$, $\frac{m_{2}}{m_{3}}$) and the neutrino masses ($m_{1}$, $m_{2}$, $m_{3}$) besides two classifications of all possible texture structures, one classification is based on the $\mu$-$\tau$-interchange symmetry, whereas the other is based on the permutation group of order $3$. In section 5, we present two tables summarizing the different predictions at all $\sigma$-levels, followed by the ($m_{ee}$-LNM) correlation plots at $3\sigma$ for all textures in both types of hierarchy, then we discuss the numerical analysis for all possible textures and include correlation plots at $3\sigma$-level. In section 6, we present symmetry realizations for some viable texture structures in different seesaw schemes. We end up with a summary and conclusion in section 7. In one appendix, we write down the $Z_2\times Z_2^{'}$ invariant terms for the type-II scalar potential in one specific case.

\section{Nonphysical phases and the texture definition}
We contrast in this section two ways used to define a texture given by a certain constraint relating the different entries of $M_\nu$. We assume we are in the flavor basis of the charged leptons which is defined up to a 3-dim phase matrix $P=\mbox{diag}(e^{i\phi_1},e^{i\phi_2},e^{i\phi_3})$, where ($\phi_1, \phi_2, \phi_3$) are the nonphysical charged lepton phases. One can absorb $P$ by redefining equally the ``gauge" neutrino fields, upon which $M_\nu$  is ``phased'' as \footnote{We stress here that the Majorana neutrino mass states are not phased, as the corresponding mass term is not phase invariant, but rather we are redefining both the ``gauge'' neutrino fields and the corresponding mass matrix so that its diagonalizing matrix changes from, say, $V_{\mbox{\tiny PMNS}}$ to $P V_{\mbox{\tiny PMNS}}$.}:
\bea \label{phasing} \left(\nu \rightarrow P \nu\right)  &\Rightarrow& \left( M_\nu \rightarrow M'_\nu=P^* M_\nu P^*\right)\eea

Let's call a change of $P$ a change of basis (or ``gauge''). For a fixed parametrization of the $V_{\mbox{\tiny PMNS}}$ (e.g. the familiar PDG \cite{pdgparam}, or the parametrization that we shall adopt where the third column is real), any $M_\nu$ can be decomposed uniquely as :
\bea \label{diagonalisation} M_\nu &=& (P V_{\mbox{\tiny PMNS}}) M^d (P V_{\mbox{\tiny PMNS}})^T,\eea
where $M^d$ is diagonal with positive masses, so we write
\bea
P^* M_\nu P^* = M^{\mbox{\tiny phys}}_\nu &=& V_{\mbox{\tiny PMNS}} M^d (V_{\mbox{\tiny PMNS}})^T.\eea
 The two matrices $M_\nu$ and $M^{\mbox{\tiny phys}}_\nu $ differ only in the nonphysical phases but have the same physics and no way to distinguish one from the other by physical measurements. Any texture definition should be a characteristic of $M^{\mbox{\tiny phys}}_\nu $, such that two matrices differing only in the nonphysical phases should together, either both satisfy the texture definition or neither does. There are two familiar approaches to guarantee this.

The role of the nonphysical phases in the definition of the texture differs in the two approaches. In the first approach, call it the ``generalized'' definition, the texture is defined by satisfying a given mathematical constraint in at least one basis $P$, whereas in the second approach, call it the ``specific" definition, the given constraint is met in one specific basis, say $P=1$, i.e. it is defined as a constraint on the elements of $M^{\mbox{\tiny phys}}_\nu $ rather than those of $M_\nu$. The generalized definition provides a less stringent constraint leading to an increase of the acceptable points in the correlation plots, which is a disadvantage regarding the predictability of the model. However, this generalized approach optimally leads to a definition which is independent of $P$, acquiring thus the advantage of a basis (or ``gauge'')-independent definition. The systematic way in the generalized approach to reach a ``good'' $P$-invariant definition of the texture is subtle, in that we start from an explicit definition of the texture in one basis -say where $P=1$-, then we replace the entries of $M_\nu$ by the entries of the ``phased'' neutrino mass matrix $M'_\nu=P^* M_\nu P^* $ by introducing the arbitrary phase $\phi_i$, and then subtly try to eliminate $\phi_i$ in order to find, if possible, the $P$-invariant definition of the texture. In contrast, by choosing the basis $P=1$ in the specific approach, one fixes the ``gauge", thus the definition is ``gauge''-dependent.
Let us clarify this point by four examples, where the two definitions coincide in the first two examples whereas they differ in the last two.
 \begin{itemize}
 \item The first example treats the ``zero" texture, and to fix the ideas let us assume that the texture corresponds to one vanishing entry, say $(1,1)$ in the neutrino mass matrix. The specific definition corresponds to a constraint on $M^{\mbox{\tiny phys}}$ entries, so
 \bea \label{SpecificZeroDef} M_\nu(1,1)&=& 0 \mbox{\small  , ``Specific'' Zero texture definition at $P=1$.}\eea  For the generalized definition, we define our texture by \bea M'_\nu(1,1)= 0 &\Rightarrow& e^{-2i(\phi_1)} M_\nu(1,1)= 0.\eea We see now that the two definitions are equivalent and we have  \bea \label{GeneralizedZeroDef} M_\nu(1,1)&=& 0 \mbox{\small  , ``Generalized'' Zero texture  definition.} \eea Note that the zero-texture definition is $P$-invariant and is identical in both the specific and the generalized definition cases.

 \item The vanishing minor texture, and to fix the ideas assume it corresponding to one vanishing minor obtained by deleting the entry $(1,1)$ in $M_\nu$. We thus have
 \bea \label{SpecificMinorDef} M_\nu(2,2) M_\nu(3,3) - M_\nu(3,2) M_\nu(2,3) = 0, \mbox{\small   ``Specific'' Vanishing minor texture definition at $P=1$.}\eea
The generalized definition would apply to any $M_\nu$ where one of the $M'_\nu = P^* M_\nu P^*$  has the corresponding minor equal to zero. So we put
\bea M'_\nu(2,2) M'_\nu(3,3) - M'_\nu(3,2) M'_\nu(2,3) = 0 \Rightarrow \nonumber \\ e^{-2i(\phi_2+\phi_3)} \left[ M_\nu(2,2) M_\nu(3,3) - M_\nu(3,2) M_\nu(2,3)\right] = 0\eea which would give again the same definition
\bea \label{GeneralizedMinorDef} M_\nu(2,2) M_\nu(3,3) - M_\nu(3,2) M_\nu(2,3) = 0, \mbox{\small ``Generalized'' Vanishing minor texture definition.}\eea
The two definitions coincide and the corresponding constraint equation is $P$-invariant.

\item The third example concerns our Equality texture, and to fix the ideas let us assume it corresponds to an equality of the entries $(1,1)$ and $(2,2)$ in the neutrino mass matrix. Thus, we start by the specific definition given in the $P=1$ slice of vanishing nonphysical phases, so we have
 \bea \label{SpecificEqualityDef} M_\nu(1,1)&=& M_\nu(2,2) \mbox{\small  , ``Specific'' Equality texture definition at $P=1$.}\eea  Then, for the generalized definition, we define our texture by \bea M'_\nu(1,1)= M'_\nu(2,2) &\Rightarrow& e^{2i(\phi_2 -\phi_1)} M_\nu(1,1)= M_\nu(2,2).\eea We try now to eliminate the arbitrary phases and find that the generalized definition of the equality texture would amount to  \bea \label{GeneralizedEqualityDef} |M_\nu(1,1)|&=& |M_\nu(2,2)| \mbox{\small  , ``Generalized'' Equality texture  definition.} \eea Note that this generalized definition is again independent of the nonphysical phases and so applies irrespective of which $P$ we choose. However, we see that the generalized definition represents just one ``real" constraint in contrast to the ``complex" constraint in the specific definition.

    \item The specific definition of the $\mu$-$\tau$-symmetry is given simply as
\bea \label{Specific_mutausymmetry} M_\nu(1,2)= M_\nu(1,3) &,& M_\nu(2,2)= M_\nu(3,3), \mbox{   ``Specific''$\mu$-$\tau$-symmetry at $P=1$},\eea and so the generalized $\mu$-$\tau$-symmetry would be given by any neutrino mass matrix $M_\nu$ satisfying
 \bea M'_\nu(1,2)= M'_\nu(1,3) &,& M'_\nu(2,2)= M'_\nu(3,3),\eea for one of the many $M'_\nu$s related to $M_\nu$ by an arbitrary phase $P$.
This leads to \bea e^{-i(\phi_1+\phi_2)} M_\nu(1,2)= e^{-i(\phi_1+\phi_3)}M_\nu(1,3) &,& e^{-2i\phi_2}M_\nu(2,2)=e^{-2i\phi_3} M_\nu(3,3).\eea
Eliminating the arbitrary phases in a clever way in order to reach a definition not showing the nonphysical phases leads to the generalized definition \cite{Grimus_2012} which, compared to the $4$ real constraints of the specific definition of Eq. \ref{Specific_mutausymmetry}, has now only $3$ real constraints:
\bea  \label{generalized_mutausymmetry} |M_\nu(1,2)|= |M_\nu(1,3)| , |M_\nu(2,2)|= |M_\nu(3,3)| &,&  \mbox{``Generalized" $\mu$-$\tau$-symmetry} \nonumber \\ \arg(22)-\arg(33)+2\arg(13)-2\arg(12)=0 &:& \arg(ij)\equiv \arg(M_\nu(ij)) \eea
Note again that the generalized definition is formulated such that the effect of any nonphysical parameter cancels out. Put it differently, in order to check whether a given matrix meets the texture definition, it suffices to restrict to the case $P=1$.
\end{itemize}
Even though the specific definition is a basis(gauge)-dependent definition, however nothing wrong with it in the sense that any mass matrix can be, in a unique way by phasing, put into the slice of vanishing nonphysical phases $P=1$, where the definition is given, and so can be tested whether or not it meets the texture criteria. This specific approach is adopted by all studies (e.g. \cite{Lashin_2012,Lashin_2014,Liu_2013,Han_2017}) which pick the slice $P=1$ for simplicity, but surely the phenomenology will be different from that of the generalized definition.

Note also that it is not always possible to put the generalized definition in a form independent of the nonphysical phases like what we did in Eqs. (\ref{GeneralizedEqualityDef},\ref{generalized_mutausymmetry}), so for a vanishing-trace texture \cite{nasri_2000} defined by ($Tr(M_\nu)=0$) one can not get rid of the phases of $P$ in the generalized definition ($\exists \a,\b \in \mathbb{R}: M_\nu(1,1)+e^{i\a}M_\nu(2,2)+e^{i\b}M_\nu(3,3)=0$). Numerically, the generalized definition corresponds to scanning the parameter space of $P$ with arbitrary values, whereas in the specific definition, $P$ is fixed (namely to the identity matrix) and we do not scan it.

\begin{figure}[hbtp]
\centering
\epsfxsize=8.25cm
\centerline{\epsfbox{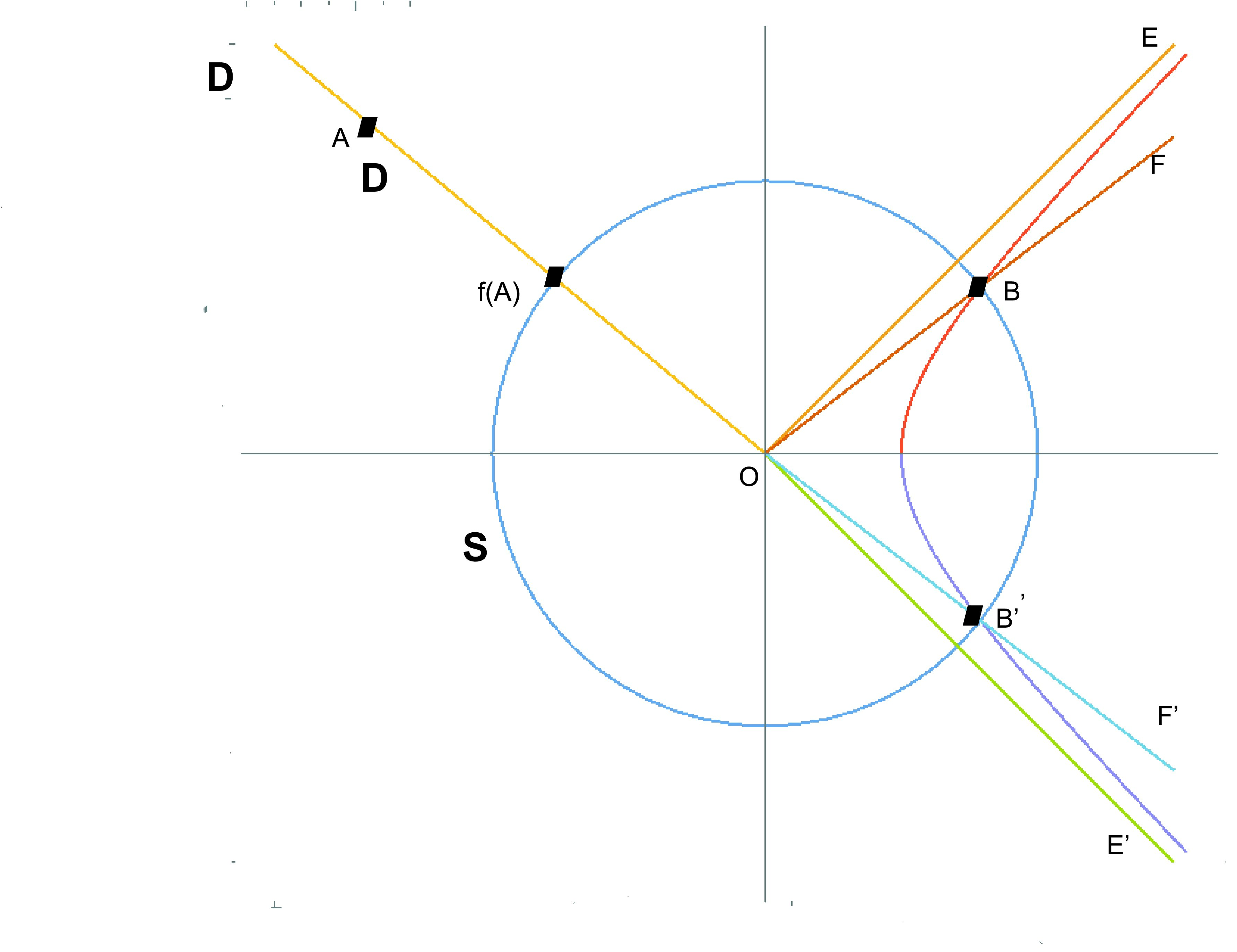}}
\caption{\footnotesize Generalized and Specific Definitions. The defining constraint is met at one point at least in the equivalence class $D$ for the generalized definition, whereas,  for the specific definition, it is met at the intersection of $D$ and the vanishing nonphysical slice $S$.}
\label{FigEquiv}
\end{figure}

Mathematically, we define an equivalence relation on the 12-dim space $\mathcal{M}$ of complex symmetric $3 \times 3$ matrices $A$ by:
\bea \label{equivalence} A \sim A' &\Leftrightarrow \exists  \mbox{\small      phase matrix } P :& A'=P.A.P \eea
and the texture is defined actually on the set of equivalence classes $\mathcal{M}/\sim \equiv \{[M]\}$.
Pictorially, we can represent $\mathcal{M}$ by the plane with polar coordinates, and any 3-dim equivalence class of $A$ is represented by the half-line $D$ from the origin towards $A$ corresponding to a fixed polar angle, whereas the 9-dim slice of $P=1$ is represented by the unit-circle $S$. Thus, changing  $M^{\mbox{\tiny phys}}$ would mean changing the polar angle, whereas changing the nonphysical phases corresponds to changing the polar radius. We see in Fig. \ref{FigEquiv}, that any $D$ intersects $S$ once and only once at a point $f(A)$, which we call the `projection' of $A$ onto $S$. The two ways to define the texture satisfying a certain constraint $C$ given by $g(M)=0$, where $g$ is a vector function of $12$-real variables, is as follows.
\bea \mbox{``Generalized" definition:}&A \in \mbox{texture} \Leftrightarrow& \exists A'\in D: A' \mbox{ satisfies } C: g(A')=0 \\
\mbox{``Specific" definition:} &A \in \mbox{texture} \Leftrightarrow& f(A) \mbox{ satisfies } C: g(f(A))=0
\eea
We see that a ``$P$-invariant definition'' for the texture given by the constraint $g$ should satisfy
\bea \label{P-invariant} g(M)=0 \Rightarrow \forall \mbox{ phase   } P, g(P.M.P)=0 &:& \mbox{``$P$-invariant" constraint $g$}\eea
We see directly that the locus of $g(M)=0$ for a $P$-invariant constraint $g$ would be a union of rays in the pictorial representation, as it does not involve nonphysical phases, and that both the specific and generalized definitions coincide for it\footnote{The converse is not true in general, as one can check that for the trivial non $P$-invariant constraint $g: M_{12}=M_{13}=M_{23}=\arg(11)=\arg(22)=\arg(33)=0$, both the specific and generalized definitions coincide and correspond to the diagonal $3 \times 3$ complex matrices.}.

In order to show pictorially when both definitions coincide, let us take a texture defined by a constraint $C$ represented in the figure by a subspace, say a hyperbole with asymptotes the bisectrices of first and fourth quarters ($OE, OE'$) and which intersects $S$ in the points $B$ and $B'$. We see here that the specific definition applies then to all the matrices ``equivalent'' to $\{B,B'\}$, i.e. to the half-lines ($OF$ and $OF'$). However, if we adopt the generalized definition then all the matrices represented by the sector $EOE'$ satisfy the definition, since any point in this sector has an equivalent point lying on the hyperbole. We see now that both definitions coincide when the constraint $C$ satisfies $f(C)=C\cap S$, such as the constraint represented by the perimeter of the minor circle sector $(OBB')$ which would be satisfied by all the points inside the sector $FOF'$ in both definitions.

In this sense, all the constraints given by Eqs. (\ref{SpecificZeroDef}, \ref{SpecificMinorDef}, \ref{GeneralizedEqualityDef} and \ref{generalized_mutausymmetry}) are $P$-invariant and, a fortiori, lead to the same locus for both definitions, and can be used directly to test any given $M_\nu$ without the need to phase it.

In this work, we shall adopt the specific definition for the Equality texture, and restrict ourselves to the slice of $P=1$, for the following reasons. First, putting $P=1$ simplifies the formulae, and leads to more constrained textures, and so the correlation plots become more clear and nuanced. Second, the specific definition normally involves analytic functions in contrast to the generalized definition (cf. Eqs.{\ref{GeneralizedEqualityDef},\ref{generalized_mutausymmetry}}).%, and so it facilitates finding realization paradigms for the texture. %Third, in the high scale fundamental theory, where a symmetry is imposed in order to get the form of the texture, the neutrino fields are usually coupled to an enriched Higgs sector, and it is difficult to conceive models where the neutrino mass matrices, in the high scale theory, have the phase freedom that they do have in the low scale effective theory.
%Actually,  for the realization models, one can consider the defining symmetry, which leads to the texture form, valid at a particular choice of phase. %Actually, some nonphysical phases may in principle be determinable in a hitherto unknown fundamental theory, and that when integrating out high scale fields a limited effective Lagrangian survives where these phases are unobservable.

However, once we adopt the specific definition we no longer have freedom of phasing the neutrino mass matrix. In this regard, one can see that the work of \cite{ismael} and our current work are not equivalent, in that both adopt the specific definition, so two matrices ($M_{+\nu}$, $M_{-\nu}$) satisfying say ($M^{\mbox{\tiny phys}}_{+\nu}(1,1)=+M^{\mbox{\tiny phys}}_{+\nu}(2,2)$, $M^{\mbox{\tiny phys}}_{-\nu}(1,1)=-M^{\mbox{\tiny phys}}_{-\nu}(2,2)$) are not related and have different physics. Had the generalized definition been adopted then both textures of two-equal-elements or two-opposite-elements would have been equivalent because
\bea M_\nu(a,b)=M_\nu(c,d), P=\mbox{diag}(e^{i\phi_1},e^{i\phi_2},e^{i\phi_3}): \phi_a+\phi_b-\phi_c-\phi_d=\pi &\Rightarrow& M'_\nu(a,b)=-M'_\nu(c,d) \nonumber \\\eea

Note finally that the concept of (nonphysical phases $P$ or ``gauge'') is strongly related to that of (parametrization), i.e. the form of $V_{\mbox{\tiny PMNS}}$ in Eq. (\ref{diagonalisation}), where it is the product $P V_{\mbox{\tiny PMNS}}$ which is parametrization-independent. However, under a change of parametrization, the form of $V_{\mbox{\tiny PMNS}}$ changes, and consequently the nonphysical phases $P$ change as well. Unlike the generalized definition, the specific definition is gauge- and parametrization-dependent since it is defined by fixing first the parametrization followed by defining the texture in the slice of vanishing nonphysical phases within the fixed chosen parametrization \footnote{Again, nothing wrong with the definition being parametrization-dependent. Continuing with the `gauge' analogy, one can think of fixing the parametrization as fixing the ``renormalization scheme" under which computations are carried out.}.

\section{Notations}
As said above, we restrict the study to vanishing nonphysical phases. In the flavor basis, the Majorana neutrino mass matrix $M_{\nu}$ can be diagonalized by a unitary transformation,
\begin{equation}
V_{\mbox{\tiny PMNS}}^{\dagger}M_{\nu}V_{\mbox{\tiny PMNS}}^{*}= \left( \begin {array}{ccc} m_{1}&0&0\\ \noalign{\medskip}0&m_{2}&0
\\ \noalign{\medskip}0&0&m_{3}\end {array} \right),
\end{equation}
with $m_{i}$ (for $i=1,2,3$) are real and positive neutrino masses. We adopt the parameterization where the third column of $V_{\mbox{\tiny PMNS}}$ is real. The three mixing angles ($\theta_{12}$, $\theta_{23}$, $\theta_{13}$) and three phases ($\rho$, $\sigma$, $\delta$) are introduced as:
\begin{align}
P^{\mbox{\tiny Maj.}}=&\nonumber \textrm{diag}(e^{i\rho},e^{i\sigma},1),~U=R_{23}(\theta_{23})~R_{13}(\theta_{13})~\textrm{diag}(1,e^{-i\delta},1)~R_{12}(\theta_{12}),\\
V_{\mbox{\tiny PMNS}}=&UP^{\mbox{\tiny Maj.}}= \left( \begin {array}{ccc} c_{12}c_{13}e^{i\rho}&s_{12}c_{13}e^{i\sigma}&s_{13}\\ \noalign{\medskip}(-c_{12}s_{23}s_{13}-s_{12}c_{23}e^{-i\delta})e^{i\rho}&(-s_{12}s_{23}s_{13}+c_{12}c_{23}e^{-i\delta})e^{i\sigma}&s_{23}c_{13}
\\ \noalign{\medskip}(-c_{12}c_{23}s_{13}+s_{12}s_{23}e^{-i\delta})e^{i\rho}&(-s_{12}c_{23}s_{13}-c_{12}s_{23}e^{-i\delta})e^{i\sigma}&c_{23}c_{13}\end {array} \right),
\end{align}
where $R_{ij}(\theta_{ij})$ is the rotation matrix in the ($i,j$)-plane by angle $\theta_{ij}$ and $c_{12}\equiv \cos\theta_{12}...$.

The advantage of the adopted parametrization is that the Dirac-type CP-violating
phase $\d$ does not appear in the effective mass term of the neutrinoless double beta decay \cite{Fritzsch_2001, Xing_2001}. However, there are simple relations connecting the parameters in the adopted parametrization and those in the standard PDG parametrization. The mixing angles and Dirac CP phase take the same values under the two parametrizations, however the
nonphysical phases and Majorana phases are different. Noting that
\bea
U &=& \textrm{diag}(e^{i\d},1,1) U^{\mbox{\tiny PDG}} \textrm{diag}(e^{-i\d},e^{-i\d},1)
\eea
and assuming the same form for $P^{\mbox{\tiny Maj.}}$ under the two parametrizations, we get\footnote{If the common form $P^{\mbox{\tiny Maj.}}_{\mbox{\tiny PDG}}= \textrm{diag}(1,e^{i\varphi_2/2},e^{i(\varphi_3+2\d)/2})$ then one gets  ($\r = - \frac{\varphi_3}{2}, \s = \frac{\varphi_2-\varphi_3}{2}$) \cite{Lashin_2012}.}
\bea
\phi_1 = \phi_1^{\mbox{\tiny PDG}} - \d, \phi_2 = \phi_2^{\mbox{\tiny PDG}}, \phi_3 = \phi_3^{\mbox{\tiny PDG}}, \r = \r^{\mbox{\tiny PDG}}+\d,
\s = \s^{\mbox{\tiny PDG}}+\d,
\eea

This means that if we take the nonphysical phase $\phi_1$ vanishing in our adopted parametrization, then it is not fixed, and a fortiori not vanishing, under the
standard PDG parametrization. As said earlier, this does not constitute a problem since the texture under study will be defined in the slice of vanishing nonphysical phases after having fixed the parametrization.

We note that the $\mu$-$\tau$ permutation transformation:
\begin{equation} \label{mutau}
T:~~\theta_{23}\rightarrow\frac{\pi}{2}-\theta_{23}~\textrm{and}~\delta\rightarrow\delta\pm \pi,
\end{equation}
interchanges the indices 2 and 3 of $M_{\nu}$ and keeps the index 1 intact:
\begin{align}
M_{\nu11}&\leftrightarrow M_{\nu11}~~M_{\nu12}\leftrightarrow M_{\nu13}\nonumber\\
M_{\nu22}&\leftrightarrow M_{\nu33}~~M_{\nu23}\leftrightarrow M_{\nu23}.\label{T}
\end{align}
The solar and atmospheric neutrino mass-squared differences are defined as:
\begin{equation} \label{DeltaDef}
\delta m^{2}\equiv m_{2}^{2}-m_{1}^{2},~~\Delta m^{2}\equiv\Big| m_{3}^{2}-\frac{1}{2}(m_{1}^{2}+m_{2}^{2})\Big|,
\end{equation}
and data indicate that their ratio, denoted by $R_{\nu}$, is quite small:
\begin{equation}
R_{\nu}\equiv\frac{\delta m^{2}}{\Delta m^{2}}\ll1.
\end{equation}
By studying neutrinoless double-beta decay and beta-decay kinematics, we obtain two non-oscillation parameters which put constraints on the neutrino mass scales: the effective Majorana mass term:
\begin{equation}
m_{ee}=\big| m_{1}~V_{e1}^{2}+m_{2}~V_{e2}^{2}+m_{3}~V_{e3}^{2}\big|=\big| M_{\nu11}\big|,
\end{equation}
and the effective electron-neutrino mass:
\begin{equation}
m_{e}=\sqrt{\sum_{i=1}^{3}(|V_{ei}|^{2}m_{i}^{2})}.
\end{equation}
The \lq{sum}\rq parameter $\Sigma$ is bounded via cosmological observations, and is defined by:
\begin{equation}
\Sigma=\sum_{i=1}^{3}m_{i}.
\end{equation}
The last measurable quantity is Jarlskog rephasing invariant quantity  \cite{PhysRevLett.55.1039}, which measures CP violation in neutrino oscillation and is given by:
\begin{equation}
J=s_{12}~c_{12}~s_{23}~c_{23}~s_{13}~c_{13}^{2}~\sin\delta.
\end{equation}
The experimental bounds for the oscillation parameters ($\theta_{12}$, $\theta_{23}$, $\theta_{13}$, $\delta$, $\delta m^{2}$ and $\Delta m^{2}$ ) are summarized in Table (\ref{TableLisi:as})
\begin{table}[h]
\centering
\scalebox{0.8}{
\begin{tabular}{cccccc}
\toprule
Parameter & Hierarchy & Best fit & $1 \sigma$ & $2 \sigma$ & $3 \sigma$ \\
\toprule
$\delta m^{2}$ $(10^{-5} \text{eV}^{2})$ & NH, IH & 7.50 & [7.30,7.72] & [7.12,7.93] & [6.94,8.14] \\
%%%%%%%%%%%%%%%%%%%%%%%%%%%%%%%%%
\midrule
 \multirow{2}{*}{$\Delta m^{2}$ $(10^{-3} \text{eV}^{2})$} & NH & 2.51 & [2.48,2.53] & [2.45,2.56] & [2.43,2.59] \\
 \cmidrule{2-6}
           & IH & 2.48 & [2.45,2.51] & [2.42,2.54] & [2.40,2.57]\\
\midrule
$\theta_{12}$ ($^{\circ}$) & NH, IH & 34.30 & [33.30,35.30] & [32.30,36.40] & [31.40,37.40] \\
%%%%%%%%%%%%%%%%%%%%%%%%%%%%%%%%%%%%%
\midrule
\multirow{2}{*}{$\theta_{13}$ ($^{\circ}$)}  & NH & 8.53 & [8.41,8.66] & [8.27,8.79] & [8.13,8.92] \\
\cmidrule{2-6}
    & IH & 8.58 & [8.44,8.70] & [8.30,8.83]& [8.17,8.96]\\
%%%%%%%%%%%%%%%%%%%%%%%%%%%%%%%%%%%%%%%%%
\midrule
\multirow{2}{*}{$\theta_{23}$ ($^{\circ}$)}  & NH & 49.26 & [48.47,50.05]& [47.37,50.71] & [41.20,51.33] \\
\cmidrule{2-6}
      & IH & 49.46 & [48.49,50.06]  & [47.35,50.67] & [41.16,51.25]   \\
%%%%%%%%%%%%%%%%%%%%%%%%%%%%%%%%%%%%%%%%%%%%%%
\midrule
\multirow{2}{*}{$\delta$ ($^{\circ}$)}  & NH & 194.00 & [172.00,218.00] & [152.00,255.00] & [128.00,359.00] \\
\cmidrule{2-6}
 & IH & 284.00 & [256.00,310.00] & [226.00,332.00] & [200.00,353.00]   \\
\bottomrule
\end{tabular}}
\caption{\footnotesize The experimental bounds for the oscillation parameters at 1-2-3$\sigma$-levels, taken from the global fit to neutrino oscillation data \cite{de_Salas_2021} (the numerical values of $\Delta m^2$ are different from those in the reference which uses the definition $\Delta m^2 = \Big| \ m_3^2 - m_1^2 \Big|$ instead of Eq. \ref{DeltaDef}). Normal and Inverted Hierarchies are
respectively denoted by NH and IH}.
\label{TableLisi:as}
\end{table}

For the non-oscillation parameters $\Sigma$ and $m_{ee}$, we adopt the ranges mentioned in the recent reference \cite{Salas_2018}\footnote{We adopt for $\Sigma$ the  results of Planck 2018 \cite{Planck_2018} from temperature information with low energy, but without using the simulator SimLOW. Including the latter simulator, one gets $\Sigma < 0.54$. However, we have checked that for all patterns, most of the accepted points satisfy this tighter bound, or others found in \cite{vagnozziPrd2016,vagnozziPrd2017}, such that the plots -as well as the allowed ranges- change very little whereas all the conclusions about the textures viability remain the same}, while for $m_{e}$ we adopt values found earlier \cite{Andreotti:2010vj}:
\begin{equation}\label{non-osc-cons}
\begin{aligned}
\Sigma~~~~~&<0.7~\textrm{eV},\\
 m_{ee}~~&<0.3~\textrm{eV},\\
 m_{e}~~~&<1.8~\textrm{eV}.
\end{aligned}
\end{equation}

\section{Texture of an equality between two matrix elements}
Because $M_{\nu}$ is a symmetric matrix and has 6 independent elements, we have 15 possibilities ($^{6}C_{2}=15$) for equality between two independent matrix elements. The texture structures can be classified into 7 categories, each -except the last- has 3 elements:
\begin{align}
A_{1}:&\left(\begin{array}{ccc}
a&a&c\\
a&b&d\\
c&d&e
\end{array}\right),~A_{2}:\left(\begin{array}{ccc}
a&b&a\\
b&c&d\\
a&d&e
\end{array}\right);\nonumber\\
B_{1}:&\left(\begin{array}{ccc}
a&b&c\\
b&b&d\\
c&d&e
\end{array}\right),~B_{2}:\left(\begin{array}{ccc}
a&b&c\\
b&d&e\\
c&e&c
\end{array}\right);\nonumber\\
C_{1}:&\left(\begin{array}{ccc}
a&b&c\\
b&d&e\\
c&e&e
\end{array}\right),~C_{2}:\left(\begin{array}{ccc}
a&b&c\\
b&d&d\\
c&d&e
\end{array}\right);\nonumber\\
D_{1}:&\left(\begin{array}{ccc}
a&b&c\\
b&c&d\\
c&d&e
\end{array}\right),~D_{2}:\left(\begin{array}{ccc}
a&b&c\\
b&d&e\\
c&e&b
\end{array}\right);\nonumber\\
E_{1}:&\left(\begin{array}{ccc}
a&b&c\\
b&a&d\\
c&d&e
\end{array}\right),~E_{2}:\left(\begin{array}{ccc}
a&b&c\\
b&d&e\\
c&e&a
\end{array}\right);\nonumber\\
F_{1}:&\left(\begin{array}{ccc}
a&b&c\\
b&d&b\\
c&b&e
\end{array}\right),~F_{2}:\left(\begin{array}{ccc}
a&b&c\\
b&d&c\\
c&c&e
\end{array}\right);\nonumber\\
G_{1}:&\left(\begin{array}{ccc}
a&b&c\\
b&d&a\\
c&a&e
\end{array}\right),~G_{2}:\left(\begin{array}{ccc}
a&b&c\\
b&d&e\\
c&e&d
\end{array}\right),~G_{3}:\left(\begin{array}{ccc}
a&b&b\\
b&c&d\\
b&d&e
\end{array}\right).
\end{align}

We see that, apart from the G-category, the two elements in each category are transformed into each other upon carrying out a $\mu$-$\tau$ interchange symmetry. Each of the three elements in the G-category is invariant under $\mu$-$\tau$ transformation. Therefore, there are 9 independent patterns out of the 15 possible ones.

Actually there is a further classification of the one-equality texture based on the permutation group $S_3$. We know that $S_3$ is generated by the transpositions $(T_{ij}):i\neq j \in \{1,2,3\}$, where if we identify the elements of $\{1,2,3\}$ with three basis elements of a 3-dim vector space, then each transposition $T_{ij}$ can be represented  by a $3\times 3$ matrix formed by switching the $i^{th}$ and $j^{th}$ columns in the $3 \times 3$ identity matrix. More generally, $M.T_{ij}(T_{ij}.M)$ is formed by exchanging the  $i^{th}$ and $j^{th}$ columns (rows) in $M$, whereas the net effect of applying $T_{ij}$ on both left and right $(T_{ij}.M.T_{ij})$ is to switch the indices $i\leftrightarrow j$ in $M$. Let us denote the set of one-equality textures by $V = \cup V_i$ where $V_i$ is one of the 15 10-dim vector spaces representing one of the classes mentioned above. Let us define the following `adjoint' action of $S_3$ on $V$\footnote{\label{rep} Embedding $V$ as a subset of the 12-dim vectorspace $\mathbb{M}$ of complex symmetric matrices, and embedding $S_3$ as a subgroup of $U(3)$ the group of unitary $3 \times 3$ matrices, we have the `adjoint' representation $\mathcal{D}$ from $U(3)$ on $\mathbb{M}$ defined such that $\mathcal{D}_U(M)=U^*.M.U^\dagger$. Note that $\mathcal{D}$ is slightly different from the usual adjoint representation defined by ($\mbox{adj}_{U}(M)=U.M.U^{-1}$), but the two definitions coincide on the subgroup of real orthogonal matrices.}: \bea \a : S_3 \times V \rightarrow V &:& g(M) \equiv \a(g,M) = g.M.g^T = g.M.g^{-1}\eea
We see that $V$ can be written as a union of four invariant subsets $Y$ of $V$, such that $S_3.Y \equiv \a(S_3,Y)  = Y$, as follows.
\begin{align}
Y_{1}= E_1 \cup E_2 \cup G_2  &:& M_{ii}=M_{jj}, i \neq j; \nonumber\\
Y_{2}= F_1 \cup F_2 \cup G_3  &:& M_{ij}=M_{jk}, i \neq j \neq k; \nonumber\\
Y_{3}= G_1 \cup D_1 \cup D_2  &:& M_{ii}=M_{jk}, i \neq j \neq k;\nonumber\\
Y_{4}= A_1 \cup A_2\cup B_1 \cup B_2 \cup C_1  \cup C_2  &:& M_{ii}=M_{ij}, i \neq j ;
\end{align}
We see that each invariant class $Y_i$ is a union of orbits \footnote{The orbit of an element $x \in V$ is ($S_3.x \equiv \a(S_3,x)$) the image of the action of $S_3$ on this element.}. Each orbit in ($Y_i, i=1,2,3$) counts three members, whereas each orbit in $Y_4$ counts 6 members. This classification is important because, as we shall see, only elements in $(Y_1 \cup Y_2)$ can  be related to one-zero textures, and so we can find realization of $(Y_1 \cup Y_2)$ starting from known realizations of zero-textures. The classes $(Y_1 \cup Y_2)$ related to one-zero textures are characterized also as being the classes to which belong the `perturbed' $\mu$-$\tau$-symmetry patterns.

The equality condition is written as:
\begin{equation}
M_{\nu~ab}-M_{\nu~cd}=0,
\end{equation}
then we get
\begin{equation}
\sum_{k=1}^{3}(U_{ak}U_{bk}-U_{ck}U_{dk})\lambda_{k}=0,
\label{cond}
\end{equation}
where $\lambda_{1}=m_{1}e^{2i\rho}$, $\lambda_{2}=m_{2}e^{2i\sigma}$ and $\lambda_{3}=m_{3}$. By solving Eq. (\ref{cond}), we get the mass ratios as functions of the mixing and phase angles:
\begin{align} \label{m1-m3}
\frac{m_{1}}{m_{3}}=&\frac{\Re(A_{3})\Im(A_{2}e^{2i\sigma})-\Re(A_{2}e^{2i\sigma})\Im(A_{3})}{\Re(A_{2}e^{2i\sigma})\Im(A_{1}e^{2i\rho})-\Re(A_{1}e^{2i\rho})\Im(A_{2}e^{2i\sigma})},\\
\label{m2-m3} \frac{m_{2}}{m_{3}}=&\frac{\Re(A_{1}e^{2i\rho})\Im(A_{3})-\Re(A_{3})\Im(A_{1}e^{2i\rho})}{\Re(A_{2}e^{2i\sigma})\Im(A_{1}e^{2i\rho})-\Re(A_{1}e^{2i\rho})\Im(A_{2}e^{2i\sigma})},
\end{align}
where the coefficients $A_{k}$ is defined as,
\begin{equation}
A_{k}=U_{ak}U_{bk}-U_{ck}U_{dk},~~~k=1,2,3.\label{coeff}
\end{equation}
The neutrino masses are written as,
\begin{equation}
m_{3}=\sqrt{\frac{\delta m^{2}}{(\frac{m_{2}}{m_{3}})^{2}-(\frac{m_{1}}{m_{3}})^{2}}},~~~m_{1}=m_{3}\times\frac{m_{1}}{m_{3}},~~~m_{2}=m_{3}\times\frac{m_{2}}{m_{3}}.
\end{equation}

As we see, there are seven input parameters ($\theta_{12}$, $\theta_{23}$, $\theta_{13}$, $\delta$, $\rho$, $\sigma$, $\delta m^{2}$) which together with the two real conditions in Eq. (\ref{cond}) permit us to determine the nine degrees of freedom of $M_{\nu}$. The strategy we follow consists of spanning ($\theta_{12}$, $\theta_{23}$, $\theta_{13}$, $\delta$, $\delta m^{2}$) over their experimental ranges at the 1-2-3$\sigma$-levels, while two Majorana-phases ($\rho$ and $\sigma$) are covered over their full ranges. Thus, we can determine the allowed regions in the 7-dim space at the 1-2-3$\sigma$-levels satisfying the other experimental constraints of $\Delta m^{2}$ together with those of Eq. (\ref{non-osc-cons}). Moreover, the correlations between any two of the neutrino physical parameters ($\theta_{12}$, $\theta_{23}$, $\theta_{13}$, $\delta$, $\rho$, $\sigma$, $m_{1}$, $m_{2}$, $m_{3}$, $m_{ee}$, $m_{e}$, $J$) can be studied graphically.

\section{Numerical discussion}
In this section, we state the results of our numerical analysis for the fifteen possible textures. For each one of the nine independent patterns, i.e. not related by $\mu-\tau$-symmetry to any other independent patten, we introduce the coefficients $A's$ of Eq. (\ref{coeff}), and evaluate the analytical expressions for the mass ratios and other neutrino physical parameters, which, however, are found to be too cumbersome to be presented. Therefore, we give some of them in their approximate form. For the remaining six cases, the expressions for $A's$ coefficients and the approximate analytical formulas for the mass ratios can be obtained by using the transformations in Eq. \ref{mutau}. The allowed ranges of the neutrino physical parameters ($\theta_{12}$, $\theta_{23}$, $\theta_{13}$, $\delta$, $\rho$, $\sigma$, $m_{1}$, $m_{2}$, $m_{3}$, $m_{ee}$, $m_{e}$ and $J$) at all statistical levels for each type of hierarchy are listed in Tables (\ref{numerical1},\ref{numerical2}). We present 9 or 12 correlation plots \footnote{Actually, the parameters $\theta_{13}$ and $\theta_{12}$ are relatively well measured so the corresponding correlation plots were omitted. Those involving the parameter $\theta_{23}$ in general do not show discernable acceptable regions, and thus were dropped except in the patterns $C_1, C_2, G_2$ and $G_3$ where they show distinguishable features.}   for each pattern for both normal and inverted ordering generated from the accepted points of the neutrino physical parameters at the 3-$\sigma$ level. Moreover, we state the neutrino mass matrix $M_{\nu}$ for each texture with either hierarchy type at one representative acceptable point in the parameter space at the 3-$\sigma$ level. We chose this representative point to be as close as possible to the best fit values for mixing and Dirac phase angles.

For each point, out of $N$ of order ($10^{6}$-$10^{10}$) thrown randomly and uniformly in the 7-dim space of the parameters ($\theta_{12}$, $\theta_{23}$, $\theta_{13}$,  $\delta$, $\rho$, $\sigma$, $\delta m^{2}$), we test the other experimental constraints of $\Delta m^{2}$ together with those of Eq. (\ref{non-osc-cons}) in order to determine the acceptable regions in the parameter space. We find that among the presented patterns, all are viable except the $C_{1}$ texture  of inverted ordering.

From Tables (\ref{numerical1},\ref{numerical2}), we see that the first neutrino mass $m_{1}$ can reach a vanishing value, in normal ordering, for $A_{1}$, $A_{2}$ and $C_{2}$ patterns at the 2-3-$\sigma$ levels as well as for $G_{2}$ pattern at the 3-$\sigma$-level. However, the third neutrino mass $m_{3}$ can reach zero value, in inverted ordering, for $A_{2},~B_{1},~B_{2},~D_{1},~D_{2},~F_{2},~G_{2},~G_{3}$ patterns at all $\sigma$-levels in addition to  $A_{1}$, $F_{1}$ at the 2-3-$\sigma$ levels. Thus, out of the fifteen possible textures, four textures are predicted to allow for singular (non-invertible) neutrino mass matrix for normal ordering, whereas ten textures can become singular in the case of inverted ordering. The predictions for the singular textures are consistent with the results found in \cite{Han_2017} except, due to new experimental data unavailable in the latter reference, for the $C_{1}$ texture. The effective Majorana mass ($m_{ee}$) can reach zero value in the case of normal ordering in the patterns ($A_{1}$, $A_{2}$) at all $\sigma$-levels, and in the patterns ($C_1$, $C_2$) at the 2-3-$\sigma$ levels, whereas it reaches zero in the pattern ($G_{2}$) only at the 3-$\sigma$ level. %The allowed values of the $J$ parameter at all $\sigma$-levels for inverted ordering are negative in all patterns, which means that $\delta$ must belong to the third and fourth quarters.
For both normal and inverted ordering in all patterns, the correlation between (J, $\delta$) is sinusoidal because of tiny variation of the mixing angles compared to the Dirac phase, thus $J\propto \sin\delta$. The appearing sinusoidal envelope is not covering a full sine curve, but rather the portion corresponding to the acceptable range of $\delta$. We also find a quasi degeneracy characterized by $m_{1}\approx m_{2}$ for all patterns in inverted type.

In the next subsections, we discuss the numerical results for all possible patterns supplemented by the correlation plots, at $3\sigma$ level for each hierarchy type. For each pattern apart from $C_{1}$, we present one single large figure where the red (blue) sub-figures represent the correlation plots for normal (inverted) hierarchy.

One remark is in order here in that the analytic expressions, albeit approximative, of the mass ratios $(\frac{m_1}{m_3}, \frac{m_2}{m_3})$ (Eqs. \ref{m1-m3}, \ref{m2-m3}) lead to an approximate formula of $\frac{m_2}{m_1}$ in terms of the mixing and phase angles. Imposing the ordering $m_2 > m_1$, and fixing the mixing angles around their central, or best fit, values due to the smallness of their allowed ranges compared to those of the phase angles, leads to a correlation amidst the phase angles which can be checked in the corresponding correlation plots.

\newpage
\clearpage
\begin{landscape}
\begin{table}[h]
 \begin{center}
\scalebox{0.52}{
{\scriptsize
 \begin{tabular}{c|c|c|c|c|c|c|c|c|c|c|c|c}
  \hline
 \hline
  \multicolumn{13}{c}{\mbox{Pattern} $A_1 \equiv M_{\nu~11}=M_{\nu~12}$} \\
\hline
\hline
  \mbox{quantity} & $\theta_{12}^{\circ}$ & $\theta_{23}^{\circ}$& $\theta_{13}^{\circ}$ & $\delta^{\circ}$ & $\rho^{\circ}$ & $\sigma^{\circ}$ & $m_{1}$ $(10^{-1} \text{eV})$ & $m_{2}$ $(10^{-1} \text{eV})$ & $m_{3}$ $(10^{-1} \text{eV})$ & $m_{ee}$ $(10^{-1} \text{eV})$
 & $m_{e}$ $(10^{-1} \text{eV})$ & $J$ $(10^{-1})$\\
 \hline
  \multicolumn{13}{c}{\mbox{Normal  Hierarchy}} \\
 \cline{1-13}
 $1~\sigma$ &$33.30 - 35.30 $& $48.47 - 50.05$ &$8.41 - 8.66$ &$172.11 - 217.99$ &$41.86 - 122.87$ &$0.00 - 33.37 \cup 163.60 - 179.99$&
  $0.01 - 0.55$& $0.08 - 0.56$ &$0.50 - 0.75$ & $2.97\times10^{-3} - 0.22$ &$0.09 - 0.56$ & $-0.21 -  0.04$  \\
 \hline
 $2~\sigma$ &$32.30 - 36.39 $& $47.37 - 50.71$ &$8.27 - 8.79$ &$152.02 - 254.99$ &$1.18 - 164.50 \cup 173.50 - 179.43$ &
  $0.01 - 87.01 \cup 116.00 - 179.99$& $6.18\times10^{-6} - 0.97$ &$0.08 - 0.97$ & $0.50 - 1.09$ & $4.87\times10^{-4} - 0.51$ & $0.08 - 0.97$ & $-0.33 -  0.16$  \\
 \hline
 $3~\sigma$ &$31.40 - 37.39 $& $41.21 - 51.33$ &$8.13 - 8.92$ &$128.05 -358.87$ &$0.18 - 179.88$ &
  $0.02 - 179.99$& $5.65\times10^{-5} - 0.38$ &$0.08 - 0.39$ & $0.49 - 0.63$ &$8.16\times10^{-4} -  0.22$ &$0.08  -  0.39$ & $-0.36 - 0.27$  \\
 \hline
 \multicolumn{13}{c}{\mbox{Inverted  Hierarchy}} \\
 \cline{1-13}
 $1~\sigma$ &$\times$ & $\times$ & $\times$ & $\times$ & $\times$ & $\times$ & $\times$ & $\times$ & $\times$ & $\times$ & $\times$ & $\times$ \\
 \hline
 $2~\sigma$ &$32.37 - 36.40$ & $47.35 -50.67$ & $8.30 - 8.83$ & $226.00 - 247.24$ & $0.16 - 76.05 \cup 80.83 - 179.90$ & $0.18 -  179.95$ & $0.49 - 2.11$ & $0.49 - 2.11$ & $1.84\times10^{-4} - 2.05$ & $0.23 - 1.16$& $ 0.48 - 2.11$ & $-0.32 - -0.23$  \\
 \hline
 $3~\sigma$ & $31.40 - 37.39$ & $41.16 - 51.24$ & $8.17 - 8.96$ & $200.48 - 250.52$ & $0.05 -  179.70$ & $0.14  - 179.84$ & $0.48 - 2.24$ & $0.49 - 2.25$ & $2.11\times10^{-4} - 2.19$ & $0.23 - 1.24$ & $0.48 -  2.24$ & $-0.34 - -0.11$  \\
 \hline
 \hline
 %%%%%%%%%%%%%%%%%%%%%%%%%%%%%%%%%%%%%%%%%%%%%%%%%%%%%%%%%%%%%%%%%%%%%%%%%%%%%%%
  %%%%%%%%%%%%%%%%%%%%%%%%%%%%%%%%%%%%%%%%%%%%%%%%%%%%%%%%%%%%%%%%%%%%%%%%%%%%%%%%%%%%%%%%%%%%%%%%%
 %%%%%%%%%%%%%%%%%%%%%%%%%%%%%%%%%%%%%%%%%%%%%%%%%%%%%%%%%%%%%%%%%%%%
 %%%%%%%%%%%%%%%%%%%%%%%%%%%%%%%%%%%%%%%%%%%%%%%%%%%%%%%%%%%%%%%%%%%%%%%%%%%%%%%%%
 \multicolumn{13}{c}{\mbox{Pattern} $A_2 \equiv M_{\nu~11}=M_{\nu~13}$} \\
\hline
\hline
  \mbox{quantity} & $\theta_{12}^{\circ}$ & $\theta_{23}^{\circ}$& $\theta_{13}^{\circ}$ & $\delta^{\circ}$ & $\rho^{\circ}$ & $\sigma^{\circ}$ & $m_{1}$ $(10^{-1} \text{eV})$ & $m_{2}$ $(10^{-1} \text{eV})$ & $m_{3}$ $(10^{-1} \text{eV})$ & $m_{ee}$ $(10^{-1} \text{eV})$
 & $m_{e}$ $(10^{-1} \text{eV})$ & $J$ $(10^{-1})$\\
 \hline
  \multicolumn{13}{c}{\mbox{Normal  Hierarchy}} \\
 \cline{1-13}
 $1~\sigma$ &$33.30 - 35.30 $& $48.47 - 50.05$ &$8.41 - 8.66$ &$172.04 - 217.97$ &$0.00 - 23.43 \cup 168.80 - 179.98$ &$0.01 -  179.99$& $0.01 - 0.06$ & $0.08 - 0.10$ &$0.50 - 0.51$ & $2.15\times10^{-3} - 0.07$ &$0.09 - 0.11$ & $-0.21 -  0.04$  \\
 \hline
 $2~\sigma$ &$32.30 - 36.40 $& $47.37 - 50.71$ &$8.27 - 8.79$ &$152.00 - 176.30 \cup 178.60 - 254.99$ &$0.01 - 79.83 \cup 90.10 - 90.82 \cup 125.60 - 179.95$ & $0.07 - 179.99$& $1.23\times10^{-5} - 0.11$ &$0.08 - 0.14$ & $0.50 - 0.52$ & $1.04\times10^{-3} - 0.08$ & $0.08 - 0.14$ & $-0.34 -  0.16$  \\
 \hline
 $3~\sigma$ &$31.40 - 37.40 $& $41.20 - 51.33$ &$8.13 - 8.92$ &$128.07 -358.92$ &$0.03 - 179.96$ & $0.07 - 179.89$& $7.47\times10^{-6} - 0.81$ &$0.08 - 0.81$ & $0.49 - 0.95$ &$6.74\times10^{-4} -  0.42$ &$0.08  -  0.81$ & $-0.36 - 0.27$  \\
 \hline
 \multicolumn{13}{c}{\mbox{Inverted  Hierarchy}} \\
 \cline{1-13}
 $1~\sigma$ &$33.30 - 35.30$ & $48.49 - 50.06$ & $8.44 - 8.70$ & $293.06 - 309.99$ & $0.03 - 179.94$ & $0.18 - 179.94 $ & $0.49 - 2.20$ & $0.50 - 2.20$ & $4.45\times 10^{-6} - 2.14$ & $0.27 - 1.27$ & $0.49 - 2.20$ & $-0.31 - -0.25$ \\
 \hline
 $2~\sigma$ &$32.30 - 36.39$ & $47.35 -50.66$ & $8.30 - 8.83$ & $290.70 - 320.00$ & $0.23 - 12.04 \cup 15.55 - 179.93$ & $0.09 -    179.83$ & $0.49 - 1.78$ & $0.49 - 1.78$ & $1.00\times10^{-4} - 1.71$ & $0.26 - 1.05$& $ 0.48 - 1.78$ & $-0.32 - -0.21$  \\
 \hline
 $3~\sigma$ & $31.40 - 37.39$ & $41.16 - 51.24$ & $8.17 - 8.96$ & $287.81 - 342.45$ & $0.01 - 10.37 \cup 13.23 - 179.91$ & $0.11  - 179.96$ & $0.48 - 2.10$ & $0.49 - 2.10$ & $2.61\times10^{-5} - 2.04$ & $0.24 - 1.25$ & $0.48 -  2.10$ & $-0.34 - -0.10$  \\
 \hline
 \hline
 \multicolumn{13}{c}{\mbox{Pattern} $B_{1} \equiv M_{\nu~12} = M_{\nu~22} $} \\
\hline
\hline
  \mbox{quantity} & $\theta{12}^{\circ}$ & $\theta_{23}^{\circ}$& $\theta_{13}^{\circ}$ & $\delta^{\circ}$ & $\rho^{\circ}$ & $\sigma^{\circ}$ & $m_{1}$ $(10^{-1} \text{eV})$ & $m_{2}$ $(10^{-1} \text{eV})$ & $m_{3}$ $(10^{-1} \text{eV})$ & $m_{ee}$ $(10^{-1} \text{eV})$
 & $m_{e}$ $(10^{-1} \text{eV})$ & $J$ $(10^{-1})$\\
 \hline
 %%%%%%%%%%%%%%%%%%%%%%%%%%%%%%%%%%%%%%%%%%%%%%%%%%%%%%%%%%%%%%%%%
 \multicolumn{13}{c}{\mbox{Normal  Hierarchy}} \\
 \cline{1-13}
 $1~\sigma$ &$33.30 - 35.30$ & $48.47 - 50.04$ & $8.41- 8.66$ &$172.01 - 217.99$ &$0.07 - 91.83 \cup 96.65 - 179.97$ & $78.64 - 124.38$ & $0.28 - 1.21$ & $0.29 - 1.21$ & $0.57 - 1.31$ & $0.10 - 1.16$ & $0.29 - 1.21$ & $-0.21   - 0.05$  \\
 \hline
 $2~\sigma$ & $32.30 - 36.40$ & $47.37 - 50.70$ & $8.27 - 8.79$ & $152.03 - 254.95$ &$0.09 - 179.98$ & $63.79 - 158.95$ & $0.25 - 1.77$ & $0.27 - 1.77$ & $0.56 - 1.84$ &$0.08 -  1.70$ & $0.27 - 1.77$ & $-0.33  - 0.16$  \\
 \hline
 $3~\sigma$ & $31.40 - 37.40$ & $41.20 - 51.32$ & $8.13 - 8.92$ &$128.03 - 358.71$ &$0.02 - 179.97$ &
  $0.15 - 179.96$& $0.15 - 1.78$ & $0.17 - 1.78$ & $0.52 - 1.85$ & $0.04 - 1.63$ & $0.17 - 1.78$ & $-0.36 - 0.27$  \\
 \hline
 %%%%%%%%%%%%%%%%%%%%%%%%%%%%%%%%%%%%%%%%%%%%%%%%%%%%%%%%%%%%%%%%%%%%%%%%
 \multicolumn{13}{c}{\mbox{Inverted  Hierarchy}} \\
 \cline{1-13}
 $1~\sigma$ &$33.30 - 35.30$ & $48.49 - 50.06$ & $8.44 - 8.70$ & $ 256.00 - 309.98$ & $0.01  - 179.90$ & $0.01 - 179.99$ & $0.49 - 2.07$ & $0.50 - 2.07$ & $1.90\times10^{-4} - 2.01$ & $0.43 -  2.02$& $ 0.49 - 2.07$ & $-0.34 - -0.25$  \\
 \hline
 $2~\sigma$ &$32.30 - 36.40$ & $ 47.35 - 50.67$ & $8.30 -  8.83$ &$ 227.00 - 330.74$ & $0.13 - 179.97$ & $0.00 -  179.95$ & $0.48 - 2.25$ & $0.49 - 2.25$ & $ 4.91\times10^{-4} - 2.20$ & $0.43 -  2.10$& $0.48 - 2.25$ & $-0.35 - -0.16$  \\
 \hline
 $3~\sigma$ & $31.41 - 37.40$ & $41.17 - 51.25$ & $8.17 - 8.96$ & $224.88 - 325.14$ & $ 0.02 - 179.99$ & $0.00 - 179.95$ & $0.48  - 2.09$ & $0.49 - 2.09$ & $1.03\times10^{-4} - 2.03$ & $0.42 - 1.70$ & $0.48 - 2.09$ & $-0.36 - -0.18$  \\
 \hline
%%%%%%%%%%%%%%%%%%%%%%%%%%%%%%%%%%%%%%%%%%%%%%%%%%%%%%%%%%%%%%%%%%%%%%%%%%%%%%%%%%%%%%%
 %%%%%%%%%%%%%%%%%%%%%%%%%%%%%%%%%%%%%%%%%%%%%%%%%%%%%%%%%%%%%%%%%%%%%%%%%%%%%%%%%
  %%%%%%%%%%%%%%%%%%%%%%%%%%%%%%%%%%%%%%%%%%%%%%%%%%%%%%%%%%%%%%%%%%%%%%%%%%%%%%%%%%%%%%%%%%%%%%%%%
 %%%%%%%%%%%%%%%%%%%%%%%%%%%%%%%%%%%%%%%%%%%%%%%%%%%%%%%%%%%%%%%%%%%%%%%%%%%%%%%%%
 \hline
 \multicolumn{13}{c}{\mbox{Pattern} $B_{2} \equiv M_{\nu~13} = M_{\nu~33} $} \\
\hline
\hline
  \mbox{quantity} & $\theta{12}^{\circ}$ & $\theta_{23}^{\circ}$& $\theta_{13}^{\circ}$ & $\delta^{\circ}$ & $\rho^{\circ}$ & $\sigma^{\circ}$ & $m_{1}$ $(10^{-1} \text{eV})$ & $m_{2}$ $(10^{-1} \text{eV})$ & $m_{3}$ $(10^{-1} \text{eV})$ & $m_{ee}$ $(10^{-1} \text{eV})$
 & $m_{e}$ $(10^{-1} \text{eV})$ & $J$ $(10^{-1})$\\
 \hline
 %%%%%%%%%%%%%%%%%%%%%%%%%%%%%%%%%%%%%%%%%%%%%%%%%%%%%%%%%%%%%%%%%
 \multicolumn{13}{c}{\mbox{Normal  Hierarchy}} \\
 \cline{1-13}
 $1~\sigma$ &$33.30 - 35.30$ & $48.47 - 50.05$ & $8.41- 8.66$ &$172.01 - 217.99$ &$83.13 - 123.54$ & $0.01 - 179.96$ & $0.19 - 0.63$ & $0.21 - 0.64$ & $0.54 - 0.81$ & $0.06 - 0.54$ & $0.21 - 0.64$ & $-0.21 - 0.04$  \\
 \hline
 $2~\sigma$ & $32.30 - 36.40$ & $47.37 - 50.70$ & $8.27 - 8.79$ & $152.02 - 254.99$ &$66.08 - 165.04$ & $0.01 - 179.91$ & $0.15 - 1.43$ & $0.17 - 1.44$ & $0.52 - 1.52$ &$0.04 -  1.27$ & $0.17 - 1.44$ & $-0.33  - 0.16$  \\
 \hline
 $3~\sigma$ & $31.40 - 37.40$ & $41.20 - 51.32$ & $8.13 - 8.92$ &$128.05 - 177.70 \cup 182.10 - 359.93$ &$0.04 - 179.91$ &
  $0.57 - 179.94$& $0.12 - 2.11$ & $0.14 - 2.11$ & $0.51 - 2.17$ & $0.03 - 2.04$ & $0.15 - 2.11$ & $-0.36 - 0.27$  \\
 \hline
 %%%%%%%%%%%%%%%%%%%%%%%%%%%%%%%%%%%%%%%%%%%%%%%%%%%%%%%%%%%%%%%%%%%%%%%%
 \multicolumn{13}{c}{\mbox{Inverted  Hierarchy}} \\
 \cline{1-13}
 $1~\sigma$ &$33.30 - 35.30$ & $48.49 - 50.06$ & $8.44 - 8.70$ & $ 256.00 - 289.12$ & $0.01  -  179.98$ & $0.00 - 179.96$ & $0.49 - 2.27$ & $0.50 - 2.28$ & $1.13\times10^{-6} - 2.22$ & $0.41 -  2.18$ & $0.49 - 2.27$ & $-0.34 - -0.31$  \\
 \hline
 $2~\sigma$ &$32.30 - 36.40$ & $47.35 - 50.66$ & $8.30 - 8.83$ & $226.41 - 291.28$ & $0.01 - 179.98$ & $0.03 -  179.91$ & $0.48 - 2.15$ & $0.49 - 2.15$ & $ 5.51\times10^{-4} - 2.09$ & $0.41 -  1.79$ & $0.48 - 2.15$ & $-0.35 - -0.23$  \\
 \hline
 $3~\sigma$ & $31.40 - 37.39$ & $41.16 - 51.24$ & $8.17 - 8.96$ & $214.08 - 311.63$ & $ 0.22 - 179.99$ & $0.12 - 179.99$ & $0.48  - 2.04$ & $0.49 - 2.04$ & $6.78\times10^{-5} - 1.97$ & $0.41 - 2.02$ & $0.48 - 2.04$ & $-0.36 - -0.18$  \\
 \hline
%%%%%%%%%%%%%%%%%%%%%%%%%%%%%%%%%%%%%%%%%%%%%%%%%%%%%%%%%%%%%%%%%%%%%%%%%%%%%%%%%%%%%%%
 %%%%%%%%%%%%%%%%%%%%%%%%%%%%%%%%%%%%%%%%%%%%%%%%%%%%%%%%%%%%%%%%%%%%%%%%%%%%%%%%%
  %%%%%%%%%%%%%%%%%%%%%%%%%%%%%%%%%%%%%%%%%%%%%%%%%%%%%%%%%%%%%%%%%%%%%%%%%%%%%%%%%%%%%%%%%%%%%%%%%
 %%%%%%%%%%%%%%%%%%%%%%%%%%%%%%%%%%%%%%%%%%%%%%%%%%%%%%%%%%%%%%%%%%%%%%%%%%%%%%%%%
 \hline
 \multicolumn{13}{c}{\mbox{Pattern} $C_1 \equiv M_{\nu~23} = M_{\nu~33} $} \\
\hline
\hline
 \mbox{quantity} & $\theta_{12}^{\circ}$ & $\theta_{23}^{\circ}$& $\theta_{13}^{\circ}$ & $\delta$ & $\rho$ & $\sigma$ & $m_{1}$ $(10^{-1} \text{eV})$ & $m_{2}$ $(10^{-1} \text{eV})$ & $m_{3}$ $(10^{-1} \text{eV})$ & $m_{ee}$ $(10^{-1} \text{eV})$
 & $m_{e}$ $(10^{-1} \text{eV})$ & $J$ $(10^{-1})$\\
 \hline
 %%%%%%%%%%%%%%%%%%%%%%%%%%%%%%%%%%%%%%%%%%%%%%%%%%%%%%%%%%%%%%%%%
 \multicolumn{13}{c}{\mbox{Normal  Hierarchy}} \\
 \cline{1-13}
 $1~\sigma$ &$\times$ & $\times$ & $\times$ & $\times$ & $\times$ & $\times$ & $\times$ & $\times$ & $\times$ & $\times$ & $\times$ & $\times$  \\
 \hline
 $2~\sigma$ & $35.39 -  36.40$ & $50.06 - 50.70$ & $8.27 - 8.79$ & $152.02 - 227.99$ & $49.49 - 137.33$ & $0.05 - 48.34 \cup 145.00 - 179.98$ &  $0.04 - 0.13$ & $0.09 - 0.16$ & $0.50 - 0.52$ & $8.82\times10^{-5} -  0.02$ &  $0.09 - 0.16$ & $-0.24 -  0.16$  \\
 \hline
 $3~\sigma$ &$34.49 - 37.40$ & $41.20 - 42.15 \cup 49.53 - 51.33$ &$8.13 - 8.92$ & $128.05 - 257.64$ & $0.01 - 179.92$ & $0.49 - 179.97$ &$0.01 - 0.29$ & $0.08 - 0.30$ & $0.50 - 0.57$ & $4.64\times10^{-4} - 0.09$ & $0.09 - 0.30$ & $-0.34 - 0.27$  \\
 \hline
 %%%%%%%%%%%%%%%%%%%%%%%%%%%%%%%%%%%%%%%%%%%%%%%%%%%%%%%%%%%%%%%%%%%%%%%%
 \multicolumn{13}{c}{\mbox{Inverted  Hierarchy}} \\
 \cline{1-13}
 $1~\sigma$ &$\times$& $\times$ &$\times$ &$\times$ &$\times$ &
  $\times$& $\times$ &$\times$ & $\times$ &$\times$ &
  $\times$ & $\times$ \\
 \hline
 $2~\sigma$ &$\times$& $\times$ &$\times$ &$\times$ &$\times$ &
  $\times$& $\times$ &$\times$ & $\times$ &$\times$ &
  $\times$ & $\times$  \\
 \hline
  $3~\sigma$ &$\times$& $\times$ &$\times$ &$\times$ &$\times$ &
  $\times$& $\times$ &$\times$ & $\times$ &$\times$ &
  $\times$ & $\times$  \\
 \hline
 %%%%%%%%%%%%%%%%%%%%%%%%%%%%%%%%%%%%%%%%%%%%%%%%%%%%%%%%%%%%%%%%%%%%%%%%%%%%%%%%%%%%%%%
 %%%%%%%%%%%%%%%%%%%%%%%%%%%%%%%%%%%%%%%%%%%%%%%%%%%%%%%%%%%%%%%%%%%%%%%%%%%%%%%%%
  %%%%%%%%%%%%%%%%%%%%%%%%%%%%%%%%%%%%%%%%%%%%%%%%%%%%%%%%%%%%%%%%%%%%%%%%%%%%%%%%%%%%%%%
 %%%%%%%%%%%%%%%%%%%%%%%%%%%%%%%%%%%%%%%%%%%%%%%%%%%%%%%%%%%%%%%%%%%%%%%%%%%%%%%%%
 \hline
 \multicolumn{13}{c}{\mbox{Pattern} $C_2 \equiv M_{\nu~23} = M_{\nu~22} $} \\
\hline
\hline
 \mbox{quantity} & $\theta_{12}^{\circ}$ & $\theta_{23}^{\circ}$& $\theta_{13}^{\circ}$ & $\delta$ & $\rho$ & $\sigma$ & $m_{1}$ $(10^{-1} \text{eV})$ & $m_{2}$ $(10^{-1} \text{eV})$ & $m_{3}$ $(10^{-1} \text{eV})$ & $m_{ee}$ $(10^{-1} \text{eV})$
 & $m_{e}$ $(10^{-1} \text{eV})$ & $J$ $(10^{-1})$\\
 \hline
 %%%%%%%%%%%%%%%%%%%%%%%%%%%%%%%%%%%%%%%%%%%%%%%%%%%%%%%%%%%%%%%%%
 \multicolumn{13}{c}{\mbox{Normal  Hierarchy}} \\
 \cline{1-13}
 $1~\sigma$ &$\times$ & $\times$ & $\times$ & $\times$ & $\times$ & $\times$ & $\times$ & $\times$ & $\times$ & $\times$ & $\times$ & $\times$  \\
 \hline
 $2~\sigma$ & $32.41 -  36.40$ & $49.98 - 50.00 \cup 50.09 - 50.71$ & $8.27 - 8.79$ & $152.02 - 181.40 \cup 189.40 - 254.99$ & $0.42 - 177.92$ & $64.25 - 82.74 \cup 87.24 - 92.01 \cup 98.08 - 167.40$ &  $1.17\times 10^{-6} - 0.10$ & $0.08 - 0.13$ & $0.50 - 0.52$ & $5.54\times10^{-4} -  0.04$ &  $0.08 - 0.13$ & $-0.33 -  0.15$  \\
 \hline
 $3~\sigma$ &$31.40 - 37.40$ & $48.15 - 51.33$ &$8.13 - 8.92$ & $128.05 - 359.00$ & $0.03 - 179.93$ & $0.12 - 179.97$ &$1.09\times 10^{-5} - 0.60$ & $0.08 - 0.61$ & $0.49 - 0.79$ & $1.32\times10^{-3} - 0.17$ & $0.08 - 0.61$ & $-0.35 - 0.27$  \\
 \hline
 %%%%%%%%%%%%%%%%%%%%%%%%%%%%%%%%%%%%%%%%%%%%%%%%%%%%%%%%%%%%%%%%%%%%%%%%
 \multicolumn{13}{c}{\mbox{Inverted  Hierarchy}} \\
 \cline{1-13}
 $1~\sigma$ &$\times$& $\times$ &$\times$ &$\times$ &$\times$ &
  $\times$& $\times$ &$\times$ & $\times$ &$\times$ &
  $\times$ & $\times$ \\
 \hline
 $2~\sigma$ &$\times$& $\times$ &$\times$ &$\times$ &$\times$ &
  $\times$& $\times$ &$\times$ & $\times$ &$\times$ &
  $\times$ & $\times$  \\
 \hline
  $3~\sigma$ &$36.42 - 37.39$& $50.14 - 51.24$ &$8.17 - 8.95$ & $325.11 - 352.72$ & $2.28 - 13.85 \cup 136.30 - 179.73$ &
  $44.22 - 101.30$& $0.63 - 1.73$ &$0.63 - 1.73$ & $0.39 - 1.66$ &$0.17 - 0.47$ & $0.62 - 1.73$ & $-0.20 - -0.04$  \\
 \hline
 %%%%%%%%%%%%%%%%%%%%%%%%%%%%%%%%%%%%%%%%%%%%%%%%%%%%%%%%%%%%%%%%%%%%%%%%%%%%%%%%%%%%%%%
 %%%%%%%%%%%%%%%%%%%%%%%%%%%%%%%%%%%%%%%%%%%%%%%%%%%%%%%%%%%%%%%%%%%%%%%%%%%%%%%%%
  %%%%%%%%%%%%%%%%%%%%%%%%%%%%%%%%%%%%%%%%%%%%%%%%%%%%%%%%%%%%%%%%%%%%%%%%%%%%%%%%%%%%%%%
 %%%%%%%%%%%%%%%%%%%%%%%%%%%%%%%%%%%%%%%%%%%%%%%%%%%%%%%%%%%%%%%%%%%%%%%%%%%%%%%%%
 \hline
 \multicolumn{13}{c}{\mbox{Pattern} $D_1 \equiv M_{\nu~13} = M_{\nu~22} $} \\
\hline
\hline
 \mbox{quantity} & $\theta_{12}^{\circ}$ & $\theta_{23}^{\circ}$& $\theta_{13}^{\circ}$ & $\delta^{\circ}$ & $\rho^{\circ}$ & $\sigma^{\circ}$  & $m_{1}$ $(10^{-1} \text{eV})$ & $m_{2}$ $(10^{-1} \text{eV})$ & $m_{3}$ $(10^{-1} \text{eV})$ & $m_{ee}$ $(10^{-1} \text{eV})$
 & $m_{e}$ $(10^{-1} \text{eV})$ & $J$ $(10^{-1})$\\
 \hline
 %%%%%%%%%%%%%%%%%%%%%%%%%%%%%%%%%%%%%%%%%%%%%%%%%%%%%%%%%%%%%%%%%
 \multicolumn{13}{c}{\mbox{Normal  Hierarchy}} \\
 \cline{1-13}
 $1~\sigma$ & $33.30 - 35.29$ & $48.47 - 50.05$ &$8.41 - 8.66$ & $172.11 - 217.97$ &$82.42 - 119.94$ & $0.02 - 13.70 \cup 23.70 - 179.80$ & $0.56 - 2.21$ & $0.56 - 2.21$ & $0.75 - 2.27$ & $0.27 - 1.91$ & $0.56 - 2.22$ & $-0.21 - 0.04$  \\
 \hline
 $2~\sigma$ & $32.30 - 36.40$ & $47.37 - 50.70$ & $8.27 - 8.79$ & $152.04 -254.99 $ &$66.45  - 152.53$ & $0.00 -72.05 \cup 75.26 - 179.99$ & $0.44 - 2.21$ & $0.45 - 2.21$ & $0.67 - 2.27$  & $0.17 - 2.14$ & $0.45 - 2.21$ & $-0.33 -  0.16$  \\
 \hline
 $3~\sigma$ &$31.40 - 37.40$& $41.20 - 51.32$ &$8.13 - 8.92$ & $128.07  - 358.99$ & $0.12 - 179.93$ & $0.02 - 179.99$ & $0.24 - 2.17$ & $0.25 - 2.17$ & $0.56 - 2.23$ & $0.08 -  1.86$ & $0.25 - 2.17$ & $-0.36 -  0.28$  \\
 \hline
 %%%%%%%%%%%%%%%%%%%%%%%%%%%%%%%%%%%%%%%%%%%%%%%%%%%%%%%%%%%%%%%%%%%%%%%%
 \multicolumn{13}{c}{\mbox{Inverted  Hierarchy}} \\
 \cline{1-13}
 $1~\sigma$ & $33.30 - 35.30$& $48.49 -50.06$ & $8.44 - 8.70$ & $256.00 - 309.11$ &$0.02 - 134.30 \cup 141.90 - 143.50 \cup 148.00 - 179.98$ & $0.05 - 90.42 \cup 95.31 - 105.10 \cup 111.90 - 179.90$ & $0.49 - 1.82$ &$0.50 - 1.83$ & $1.29\times10^{-4} - 1.76$ & $0.37 - 1.64$ & $0.49 - 1.82$ & $-0.34 - -0.26$  \\
 \hline
  $2~\sigma$ &$32.30 -36.40$& $47.35  - 50.67 $ &$8.30 - 8.83$ &$226.05- 317.97$ & $0.01 -  179.98$ & $0.00 - 179.97$ & $0.48 - 2.14$ & $0.49 - 2.14$ & $3.54\times10^{-5} - 2.08$ & $0.35 - 2.05$ & $0.48 - 2.14$ & $-0.35 - -0.22$    \\
 \hline
 $3~\sigma$ & $31.40 - 37.39$ & $41.17 - 51.24$ & $8.17 - 8.96$ & $200.18 - 316.77 $ & $0.08  - 179.91$ & $0.02 - 179.92$ & $0.48 - 2.07$ & $0.49 -  2.07$ & $4.30\times10^{-4} - 2.01$ & $0.31 - 1.61$ & $0.48 - 2.07$ & $-0.36  - -0.11$  \\
 \hline
 \hline
  %%%%%%%%%%%%%%%%%%%%%%%%%%%%%%%%%%%%%%%%%%%%%%%%%%%%%%%%%%%%%%%%%%%%%%%%%%%%%%%
 \multicolumn{13}{c}{\mbox{Pattern} $D_2 \equiv M_{\nu~12} = M_{\nu~33} $} \\
\hline
\hline
 \mbox{quantity} & $\theta_{12}^{\circ}$ & $\theta_{23}^{\circ}$& $\theta_{13}^{\circ}$ & $\delta^{\circ}$ & $\rho^{\circ}$ & $\sigma^{\circ}$  & $m_{1}$ $(10^{-1} \text{eV})$ & $m_{2}$ $(10^{-1} \text{eV})$ & $m_{3}$ $(10^{-1} \text{eV})$ & $m_{ee}$ $(10^{-1} \text{eV})$
 & $m_{e}$ $(10^{-1} \text{eV})$ & $J$ $(10^{-1})$\\
 \hline
 %%%%%%%%%%%%%%%%%%%%%%%%%%%%%%%%%%%%%%%%%%%%%%%%%%%%%%%%%%%%%%%%%
 \multicolumn{13}{c}{\mbox{Normal  Hierarchy}} \\
 \cline{1-13}
 $1~\sigma$ & $33.30 - 35.30$ & $48.47 - 50.05$ &$8.41 - 8.66$ & $172.03 - 218.00$ &$0.07 - 179.67$ & $82.32  - 122.38$ & $0.20 - 0.34$ & $0.22 - 0.35$ & $0.54 - 0.61$ & $0.09 - 0.29$ & $0.22 - 0.35$ & $-0.21 - 0.04$  \\
 \hline
 $2~\sigma$ & $32.30 - 36.39$ & $47.37 - 50.70$ & $8.27 - 8.79$ & $152.04 -254.98 $ &$0.02  - 179.96$ & $66.96 - 161.38$ & $0.19 - 2.20$ & $0.21 - 2.20$ & $0.53 - 2.26$  & $0.08 - 1.40$ & $0.21 - 2.20$ & $-0.33 -  0.16$  \\
 \hline
 $3~\sigma$ &$31.40 - 37.39$& $41.20 - 51.32$ &$8.13 - 8.92$ & $128.02  - 358.96$ & $0.08 - 179.87$ & $0.03 - 179.95$ & $0.18 - 2.02$ & $0.20 - 2.02$ & $0.53 - 2.08$ & $0.07 -  1.84$ & $0.20 - 2.02$ & $-0.35 -  0.27$  \\
 \hline
 %%%%%%%%%%%%%%%%%%%%%%%%%%%%%%%%%%%%%%%%%%%%%%%%%%%%%%%%%%%%%%%%%%%%%%%%
 \multicolumn{13}{c}{\mbox{Inverted  Hierarchy}} \\
 \cline{1-13}
 $1~\sigma$ & $33.30 - 35.30$& $48.49 -50.06$ & $8.44 - 8.70$ & $256.25 - 309.97$ &$0.09 - 179.93$ & $0.09 - 179.92$ & $0.49 - 2.03$ &$0.50 - 2.03$ & $3.06\times10^{-4} - 1.97$ & $0.30 - 1.77$ & $0.48 - 2.03$ & $-0.34 - -0.25$  \\
 \hline
  $2~\sigma$ &$32.30 -36.39$& $47.35  - 50.67 $ &$8.30 - 8.83$ &$254.05- 330.65$ & $0.01 - 179.91$ & $0.35 - 179.98$ & $0.48 - 2.25$ & $0.49 - 2.25$ & $6.61\times10^{-5} - 2.20$ & $0.30 - 1.51$ & $0.48 - 2.25$ & $-0.35 - -0.15$    \\
 \hline
 $3~\sigma$ & $31.40 - 37.39$ & $41.16 - 51.23$ & $8.17 - 8.96$ & $242.70 - 339.20 \cup 347.00 - 350.05 $ & $0.03  - 179.90$ & $0.18 - 179.98$ & $0.48 - 2.34$ & $0.49 - 2.34$ & $2.15\times10^{-5} - 2.29$ & $0.29 - 1.73$ & $0.48 - 2.34$ & $-0.36  - -0.06$  \\
 \hline
 \hline
  %%%%%%%%%%%%%%%%%%%%%%%%%%%%%%%%%%%%%%%%%%%%%%%%%%%%%%%%%%%%%%%%%%%%%%%%%%%%%%%
 \end{tabular}
 }}
 \end{center}
 \vspace{-1em} \caption{\footnotesize  The 1-2-3 $\sigma$ predictions for the neutrino physical parameters ($\theta_{12}$, $\theta_{23}$, $\theta_{13}$, $\delta$, $\rho$, $\sigma$, $m_{1}$, $m_{2}$, $m_{3}$, $m_{ee}$, $m_{e}$ and $J$) with regard to A, B, C and D categories.}
  \label{numerical1}
 \end{table}
\end{landscape}
%%%%%%%%%%%%%%%%%%%%%%%%%%%%%%%%%%%%%%%%%%%%%%%%%%%%%%%%%%%%%%%%%%%%%%%%%%%%%%%%%%%%%%%%%%%%%%%%%%%%%%%%%%%%%%%%%%%%%%%%%%%%%%%%%%%%%%%%%%%%%%%%%%%%%%%%%%%%%%%%%%%%%%%%%%%%%%%%%%%%%%%%%%%%%%%%%%%%%%%%%%%%%%%%table1 end%%%%%%%%%%%%%%%%%%%%%%%%%%%%%%%%%%%%%%%%%%%%%%%%%%%%%%%%%%
%%%%%%%%%%%%%%%%%%%%%%%%%%%%%%%%%%%%%%%%%%%%%%%%%%%%%%%%%%%%%%%%%%%%%%%%%%%%%%%%%%%%%%%%%%%%%%%%%%%%%%%%%%%%%%%%%%%%%%%%%%%%%%%%%%%%%%%%%%%%%%%%%%%%%%%%%%%%%%%%%%%%%%%%%%%%%%%%%%%%%%%%%%%%%%%%%%%%%%%%%%%%%%%%table1 end%%%%%%%%%%%%%%%%%%%%%%%%%%%%%%%%%%%%%%%%%%%%%%%%%%%%%%%%%%

\newpage
\clearpage
\begin{landscape}
\begin{table}[h]
 \begin{center}
\scalebox{0.54}{
{\scriptsize
 \begin{tabular}{c|c|c|c|c|c|c|c|c|c|c|c|c}
  \hline
 \hline
  \multicolumn{13}{c}{\mbox{Pattern} $E_1 \equiv M_{\nu~11} = M_{\nu~22}$} \\
\hline
\hline
  \mbox{quantity} & $\theta_{12}^{\circ}$ & $\theta_{23}^{\circ}$& $\theta_{13}^{\circ}$ & $\delta^{\circ}$ & $\rho^{\circ}$ & $\sigma^{\circ}$ & $m_{1}$ $(10^{-1} \text{eV})$ & $m_{2}$ $(10^{-1} \text{eV})$ & $m_{3}$ $(10^{-1} \text{eV})$ & $m_{ee}$ $(10^{-1} \text{eV})$
 & $m_{e}$ $(10^{-1} \text{eV})$ & $J$ $(10^{-1})$\\
 \hline
 %%%%%%%%%%%%%%%%%%%%%%%%%%%%%%%%%%%%%%%%%%%%%%%%%%%%%%%%%%%%%%%%%
 \multicolumn{13}{c}{\mbox{Normal  Hierarchy}} \\
 \cline{1-13}
 $1~\sigma$ & $33.30 - 35.30$ & $48.47 - 50.05$ & $8.41 - 8.66$ & $172.00 - 218.00$ &$0.02 - 12.72 \cup 159.20 - 179.99$ & $0.01 - 179.96$ & $0.29 - 2.18$ & $0.30 - 2.18$ & $0.58 - 2.24 $ & $ 0.21 - 1.86$ & $0.30 - 2.18$ & $-0.21 -  0.04$  \\
 \hline
 $2~\sigma$ & $32.30 - 36.40$ & $47.37 - 50.71$ & $8.27 - 8.79$ & $152.00 - 255.00$ &$0.05 - 24.39 \cup 153.90 - 179.90$ & $0.03  - 179.98$ & $0.19 - 2.17$ & $ 0.20 - 2.17 $ & $0.53 - 2.23 $& $0.18 - 1.34$ & $0.21 - 2.17$ & $-0.33 -  0.16$  \\
 \hline
 $3~ \sigma$ &$ 31.40 - 37.39$ & $41.20 - 51.32$ & $8.13 - 8.92$ & $128.03  - 352.15$ & $0.24 - 38.77 \cup 149.90 - 179.90$ & $0.01 - 179.93$ & $0.12 - 2.09 $ & $0.15 -2.10$ & $0.51 - 2.15$ & $ 0.13 - 1.61$ & $0.15 - 2.10$ & $-0.36 - 0.27$ \\
 \hline
 %%%%%%%%%%%%%%%%%%%%%%%%%%%%%%%%%%%%%%%%%%%%%%%%%%%%%%%%%%%%%%%%%%%%%%%%
 \multicolumn{13}{c}{\mbox{Inverted  Hierarchy}} \\
 \cline{1-13}
 $1~\sigma$ & $33.30 - 35.30$  &$48.49 - 50.06$ &$8.44 - 8.70$ &$256.03 - 309.99$ &$0.01 - 32.37 \cup 156.20 - 179.90$ & $36.32 - 150.39$ &$0.51 - 2.27$ &$0.52 - 2.27$ &$0.15 - 2.21$ &$0.17 - 1.17$& $0.51 - 2.27$ &$-0.34 - -0.25$  \\
 \hline
 $2~\sigma$ &$32.30 - 36.40$ & $47.35 - 50.67$ & $8.30 - 8.83$ & $226.02 - 331.98$ & $0.14 - 34.48 \cup 153.60 - 179.91$ & $26.78 - 158.93$ & $0.51 - 2.00$ & $0.52 - 2.00$ & $0.13 - 1.94$ & $0.15 - 1.31$ & $ 0.51 - 2.00$ & $-0.35 - - 0.15$  \\
 \hline
 $3~\sigma$ & $31.40 - 37.40$ & $41.16 - 51.25$ & $8.17 - 8.96$ & $200.64 - 352.63$ & $0.03 - 38.36 \cup 149.80 - 179.97$ & $16.70 - 164.72$ & $0.50 -2.22$ & $0.51 - 2.22$ &$0.12 - 2.16$ & $0.13 - 1.49$ & $0.50 - 2.22$& $-0.36 -  -0.04$  \\
 \hline
 %%%%%%%%%%%%%%%%%%%%%%%%%%%%%%%%%%%%%%%%%%%%%%%%%%%%%%%%%%%%%%%%%%%%%%%%%%%%%%%%%%%%%%%%%%%%%%%%%
 %%%%%%%%%%%%%%%%%%%%%%%%%%%%%%%%%%%%%%%%%%%%%%%%%%%%%%%%%%%%%%%%%%%%
 %%%%%%%%%%%%%%%%%%%%%%%%%%%%%%%%%%%%%%%%%%%%%%%%%%%%%%%%%%%%%%%%%%%%%%%%%%%%%%%%%
 %%%%%%%%%%%%%%%%%%%%%%%%%%%%%%%%%%%%%%%%%%%%%%%%%%%%%%%%%%%%%%%%%%%%%%%%%%%%%%%%%%%%%%%
 %%%%%%%%%%%%%%%%%%%%%%%%%%%%%%%%%%%%%%%%%%%%%%%%%%%%%%%%%%%%%%%%%%%%%%%%%%%%%%%%%
 \hline
 \multicolumn{13}{c}{\mbox{Pattern} $E_2 \equiv M_{\nu~11} = M_{\nu~33}$} \\
\hline
\hline
  \mbox{quantity} & $\theta_{12}^{\circ}$ & $\theta_{23}^{\circ}$& $\theta_{13}^{\circ}$ & $\delta^{\circ}$ & $\rho^{\circ}$ & $\sigma^{\circ}$ & $m_{1}$ $(10^{-1} \text{eV})$ & $m_{2}$ $(10^{-1} \text{eV})$ & $m_{3}$ $(10^{-1} \text{eV})$ & $m_{ee}$ $(10^{-1} \text{eV})$
 & $m_{e}$ $(10^{-1} \text{eV})$ & $J$ $(10^{-1})$\\
 \hline
 %%%%%%%%%%%%%%%%%%%%%%%%%%%%%%%%%%%%%%%%%%%%%%%%%%%%%%%%%%%%%%%%%
 \multicolumn{13}{c}{\mbox{Normal  Hierarchy}} \\
 \cline{1-13}
 $1~\sigma$ & $33.30 - 35.29$ & $48.47 - 50.04$ & $8.41 - 8.66$ & $172.40 - 177.10 \cup 180.10 - 217.99$ &$0.01 - 12.17 \cup 145.90 - 179.98$ & $0.00 - 43.26 \cup 63.72 - 179.97$ & $0.20 - 2.20$ & $0.21 - 2.21$ & $0.54 - 2.26 $ & $ 0.17 - 1.57$ & $0.22 - 2.20$ & $-0.21 -  0.04$  \\
 \hline
 $2~\sigma$ & $32.30 - 36.39$ & $47.37 - 50.70$ & $8.27 - 8.79$ & $152.05 - 173.40 \cup 176.20 - 182.80 \cup 185.90 - 255.00$ &$0.01 - 29.35 \cup 141.80 - 179.97$ & $0.00 - 75.32 \cup 78.59 - 89.79 \cup 94.57 - 179.99$ & $0.12 - 1.58$ & $ 0.14 - 1.58 $ & $0.51 - 1.66 $& $0.13 - 1.56$ & $0.14 - 1.58$ & $-0.34 -  0.15$  \\
 \hline
 $3~ \sigma$ & $31.40 - 37.39$ & $41.20 - 51.32$ & $8.13 - 8.92$ & $128.05 - 170.90 \cup 183.40 - 358.62$ & $0.00 - 37.68 \cup 142.30 - 179.97$ & $0.02 - 73.92 \cup 76.28 -102.20 \cup 105.70 - 179.98$ & $0.10 - 2.29 $ & $0.13 -2.29$ & $0.51 - 2.34$ & $ 0.11 - 1.17$ & $0.13 - 2.29$ & $-0.36 - 0.27$ \\
 \hline
 %%%%%%%%%%%%%%%%%%%%%%%%%%%%%%%%%%%%%%%%%%%%%%%%%%%%%%%%%%%%%%%%%%%%%%%%
 \multicolumn{13}{c}{\mbox{Inverted  Hierarchy}} \\
 \cline{1-13}
 $1~\sigma$ & $33.30 - 35.29$  &$48.49 - 50.06$ & $8.44 - 8.70$ & $256.01 - 310.00$ &$0.03 - 28.84 \cup 147.40 - 179.63$ & $46.59 - 151.70$ & $0.51 - 2.32$ & $0.52 - 2.33$ & $0.15 - 2.27$ & $0.17 - 1.48$& $0.51 - 2.32$ & $-0.34 - -0.25$  \\
 \hline
 $2~\sigma$ & $32.30 - 36.39$ & $47.35 - 50.66$ & $8.30 - 8.83$ & $226.16 - 331.71$ & $0.21 - 30.55 \cup 141.40 - 179.74$ & $27.99 - 161.25$ & $0.51 - 2.25$ & $0.51 - 2.25$ & $0.13 - 2.20$ & $0.15 - 1.16$ & $ 0.50 - 2.25$ & $-0.35 - -0.15$  \\
 \hline
 $3~\sigma$ & $31.41 - 37.39$ & $41.18 - 51.24$ & $8.17 - 8.96$ & $200.02 - 346.13$ & $0.01 - 30.61 \cup 139.20 - 179.93$ & $16.60 - 167.34$ & $0.50 -2.01$ & $0.51 - 2.02$ &$0.11 - 1.95$ & $0.13 - 1.82$ & $0.50 - 2.01$& $-0.36 -  -0.08$  \\
 \hline
 %%%%%%%%%%%%%%%%%%%%%%%%%%%%%%%%%%%%%%%%%%%%%%%%%%%%%%%%%%%%%%%%%%%%%%%%%%%%%%%%%%%%%%%%%%%%%%%%%%%%%%%%%%%%%%%%%%%%%%%%%%%%%%%%%%%%%%%%%%%%%%%%%%%%%%%%%%%%%%%%%%%%%%%%%%%%%%%%%%%%%%%%%%%%%%5
 \hline
 \multicolumn{13}{c}{\mbox{Pattern} $F_1 \equiv M_{\nu~12} = M_{\nu~23} $} \\
\hline
\hline
 \mbox{quantity} & $\theta_{12}^{\circ}$ & $\theta_{23}^{\circ}$& $\theta_{13}^{\circ}$ & $\delta$ & $\rho$ & $\sigma$ & $m_{1}$ $(10^{-1} \text{eV})$ & $m_{2}$ $(10^{-1} \text{eV})$ & $m_{3}$ $(10^{-1} \text{eV})$ & $m_{ee}$ $(10^{-1} \text{eV})$
 & $m_{e}$ $(10^{-1} \text{eV})$ & $J$ $(10^{-1})$\\
 \hline
 %%%%%%%%%%%%%%%%%%%%%%%%%%%%%%%%%%%%%%%%%%%%%%%%%%%%%%%%%%%%%%%%%
 \multicolumn{13}{c}{\mbox{Normal  Hierarchy}} \\
 \cline{1-13}
 $1~\sigma$ & $33.30 - 35.30$ & $48.47 - 50.05$ & $8.41 - 8.66$ & $172.19 - 178.20 \cup 183.90 - 218.00$ & $2.42 - 47.27 \cup 172.60 - 178.57$ & $14.02  - 166.02$ & $0.37 - 2.27$ & $0.38 - 2.27$ & $0.62 - 2.32$ & $0.21 - 2.24$ & $0.38 - 2.27$ & $-0.21 -  - 0.04$  \\
 \hline
 $2~\sigma$ & $32.30 -  36.40$ & $47.37 - 50.71$ & $8.27 - 8.79$ & $152.05 - 172.00 \cup 180.70 - 254.99$ & $1.32 - 87.92 \cup 147.80 - 177.53$ & $23.13 - 158.48$ & $0.25 - 2.27$ & $0.26 - 2.28$ & $0.56  - 2.33$ & $0.11 - 2.28$ & $0.26 - 2.28$ & $-0.34 -  0.16$  \\
 \hline
 $3~\sigma$ &$31.40 - 37.40$& $41.20 - 51.32$ &$8.13 - 8.92$ & $128.04 - 179.00 \cup 184.70 - 358.84$ & $ 2.46 - 178.62$ & $0.15 - 9.87 \cup 27.30 - 179.40$ & $0.21 - 1.78$ & $0.22 - 1.78$ & $0.54 - 1.85$ & $0.04 - 1.71$ & $0.22 - 1.78$ & $-0.35 - 0.27$  \\
 \hline
 %%%%%%%%%%%%%%%%%%%%%%%%%%%%%%%%%%%%%%%%%%%%%%%%%%%%%%%%%%%%%%%%%%%%%%%%
 \multicolumn{13}{c}{\mbox{Inverted  Hierarchy}} \\
 \cline{1-13}
 $1~\sigma$ & $33.30 - 35.30$ & $48.49 - 50.06$ & $8.44 - 8.70$ & $ 256.00 - 309.93$ & $0.01 - 26.03 \cup 92.06 - 179.94$ & $66.41 - 152.94$ & $0.49 - 2.25$ & $0.50 - 2.25$ & $0.05 - 2.19$ & $0.32 - 2.07$ & $0.49 - 2.25$ & $-0.34 - -0.25$  \\
 \hline
 $2~\sigma$ & $32.30 - 36.39$ & $ 47.35 - 50.67$ & $8.30 - 8.83$ & $226.00 - 314.40 \cup 317.30 - 321.10 \cup 324.20 - 329.80$ & $0.03 - 179.99$ & $0.17 - 178.89$ & $0.48 - 2.17$ & $0.49 - 2.18$ & $3.99\times10^{-5} - 2.12$ & $0.30 - 2.03$ & $0.48 - 2.17$ & $-0.35 - -0.16$  \\
 \hline
 $3~\sigma$ & $31.40 - 37.39$ & $41.16  - 51.25$ & $8.17 - 8.96$ & $200.23 - 321.60 \cup 325.90 - 327.30 \cup 332.60 - 337.47$ & $0.00 - 179.98$ & $0.16 - 179.43$ & $0.48 - 2.29$ &
$0.49 - 2.29$ & $2.51\times10^{-5} - 2.23$ & $0.29 - 1.85$ & $0.48 - 2.29$ & $-0.36 -  -0.17$  \\
 \hline
 %%%%%%%%%%%%%%%%%%%%%%%%%%%%%%%%%%%%%%%%%%%%%%%%%%%%%%%%%%%%%%%%%%%%%%%%%%%%%%%%%%%%%%%
 %%%%%%%%%%%%%%%%%%%%%%%%%%%%%%%%%%%%%%%%%%%%%%%%%%%%%%%%%%%%%%%%%%%%%%%%%%%%%%%%%
  %%%%%%%%%%%%%%%%%%%%%%%%%%%%%%%%%%%%%%%%%%%%%%%%%%%%%%%%%%%%%%%%%%%%%%%%%%%%%%%%%%%%%%
 %%%%%%%%%%%%%%%%%%%%%%%%%%%%%%%%%%%%%%%%%%%%%%%%%%%%%%%%%%%%%%%%%%%%%%%%%%%%%%%%%
 \hline
 \multicolumn{13}{c}{\mbox{Pattern} $F_2 \equiv M_{\nu~13} = M_{\nu~23} $} \\
\hline
\hline
 \mbox{quantity} & $\theta_{12}^{\circ}$ & $\theta_{23}^{\circ}$& $\theta_{13}^{\circ}$ & $\delta$ & $\rho$ & $\sigma$ & $m_{1}$ $(10^{-1} \text{eV})$ & $m_{2}$ $(10^{-1} \text{eV})$ & $m_{3}$ $(10^{-1} \text{eV})$ & $m_{ee}$ $(10^{-1} \text{eV})$
 & $m_{e}$ $(10^{-1} \text{eV})$ & $J$ $(10^{-1})$\\
 \hline
 %%%%%%%%%%%%%%%%%%%%%%%%%%%%%%%%%%%%%%%%%%%%%%%%%%%%%%%%%%%%%%%%%
 \multicolumn{13}{c}{\mbox{Normal  Hierarchy}} \\
 \cline{1-13}
 $1~\sigma$ & $33.30 - 35.30$ & $48.47 - 50.05$ & $8.41 - 8.66$ & $172.03 - 217.99$ & $13.35 - 165.56$ & $0.03 - 43.80 \cup 161.80 - 179.95$ & $0.20 - 1.86$ & $0.22 - 1.87$ & $0.54 - 1.93$ & $0.05 - 1.85$ & $0.22 - 1.87$ & $-0.21 - 0.04$  \\
 \hline
 $2~\sigma$ & $32.30 - 36.39$ & $47.37 - 50.70$ & $8.27 - 8.79$ & $152.03 - 254.92$ & $14.20 - 162.02$ & $0.00 -79.30 \cup 145.90 - 179.99$ & $0.20 - 2.20$ & $0.22 - 2.20$ & $0.54 - 2.26$ & $0.05 - 1.61$ & $0.22 - 2.20$ & $-0.34 - 0.16$  \\
 \hline
 $3~\sigma$ &$31.40 - 37.40$& $41.20 - 51.32$ & $8.13 - 8.92$ & $128.05 - 357.19$ & $ 15.56 - 179.25$ & $0.03 - 179.98$ & $0.20 - 2.30$ & $0.22 - 2.31$ & $0.54 - 2.36$ & $0.04 - 2.20$ & $0.22 - 2.31$ & $-0.36 - 0.28$  \\
 \hline
 %%%%%%%%%%%%%%%%%%%%%%%%%%%%%%%%%%%%%%%%%%%%%%%%%%%%%%%%%%%%%%%%%%%%%%%%
 \multicolumn{13}{c}{\mbox{Inverted  Hierarchy}} \\
 \cline{1-13}
 $1~\sigma$ & $33.30 - 35.30$ & $48.49 - 50.06$ & $8.44 - 8.70$ & $256.10 - 309.90$ & $0.04 - 179.99$ & $0.31 - 179.52$ & $0.49 - 1.95$ & $0.50 - 1.95$ & $3.15\times 10^{-5} - 2.19$ & $0.35 - 1.60$ & $0.49 - 1.95$ & $-0.34 - -0.25$  \\
 \hline
 $2~\sigma$ & $32.30 - 36.40$ & $ 47.35 - 50.67$ & $8.30 - 8.83$ & $226.03 - 227.40 \cup 228.90 - 331.92$ & $0.06 - 179.96$ & $0.73 - 178.68$ & $0.48 - 2.24$ & $0.49 - 2.25$ & $8.45\times10^{-5} - 2.19$ & $0.35 - 2.15$ & $0.48 - 2.24$ & $-0.35 - -0.16$  \\
 \hline
 $3~\sigma$ & $31.40 - 37.40$ & $41.16  - 51.24$ & $8.17 - 8.96$ & $205.50 - 213.00 \cup 217.70 - 351.90$ & $0.01 - 179.83$ & $0.05 - 179.43$ & $0.48 - 1.94$ & $0.49 - 1.95$ & $5.95\times10^{-5} - 1.88$ & $0.31 - 1.91$ & $0.48 - 1.94$ & $-0.36 - -0.04$  \\
 \hline
 %%%%%%%%%%%%%%%%%%%%%%%%%%%%%%%%%%%%%%%%%%%%%%%%%%%%%%%%%%%%%%%%%%%%%%%%%%%%%%%%%%%%%%%
 %%%%%%%%%%%%%%%%%%%%%%%%%%%%%%%%%%%%%%%%%%%%%%%%%%%%%%%%%%%%%%%%%%%%%%%%%%%%%%%%%
  %%%%%%%%%%%%%%%%%%%%%%%%%%%%%%%%%%%%%%%%%%%%%%%%%%%%%%%%%%%%%%%%%%%%%%%%%%%%%%%%%%%%%%
 %%%%%%%%%%%%%%%%%%%%%%%%%%%%%%%%%%%%%%%%%%%%%%%%%%%%%%%%%%%%%%%%%%%%%%%%%%%%%%%%%
 \hline
 \multicolumn{13}{c}{\mbox{Pattern} $G_{1}\equiv M_{\nu~11} = M_{\nu~23} $} \\
\hline
\hline
 \mbox{quantity} & $\theta_{12}^{\circ}$ & $\theta_{23}^{\circ}$& $\theta_{13}^{\circ}$ & $\delta^{\circ}$ & $\rho^{\circ}$ & $\sigma^{\circ}$  & $m_{1}$ $(10^{-1} \text{eV})$ & $m_{2}$ $(10^{-1} \text{eV})$ & $m_{3}$ $(10^{-1} \text{eV})$ & $m_{ee}$ $(10^{-1} \text{eV})$
 & $m_{e}$ $(10^{-1} \text{eV})$ & $J$ $(10^{-1})$\\
 \hline
 %%%%%%%%%%%%%%%%%%%%%%%%%%%%%%%%%%%%%%%%%%%%%%%%%%%%%%%%%%%%%%%%%
 \multicolumn{13}{c}{\mbox{Normal  Hierarchy}} \\
 \cline{1-13}
 $1~\sigma$ & $33.30 - 35.30$ & $48.47 - 50.05$ &$8.41 - 8.66$ & $172.02 - 217.88$ &$0.06 - 30.64 \cup 151.60 - 179.91$ & $0.05 -  60.61 \cup 134.00 - 179.96$ & $0.15 - 2.01$ & $0.17 - 2.01$ & $0.52 - 2.07$ &$0.17 - 0.94$ & $0.18 - 2.01$ & $-0.20 - 0.04$  \\
 \hline
 $2~\sigma$ & $32.30 - 36.40$ & $47.37 - 50.71$ & $8.27 - 8.79$ & $152.02 - 253.93$ &$0.07 - 31.40 \cup 148.50 - 179.98$ & $0.08 - 82.54 \cup 125.10  - 179.99$ & $0.15 - 2.11$ & $0.18 - 2.11$ &$0.52 - 2.17$  & $0.17 - 1.07$ & $0.18 - 2.11$ & $-0.33 -  0.16$  \\
 \hline
 $3~\sigma$ &$31.40 - 37.40$& $41.21 - 51.32$ &$8.13 - 8.92$ & $128.06 - 261.10 \cup 267.60 - 271.30 \cup 276.90 - 358.95$ & $0.02 - 31.98 \cup 147.80 -179.98$ & $0.04- 73.93 \cup 79.54 - 82.88 \cup 92.62 - 179.98$ & $0.15 - 1.88$ & $0.17 - 1.88$ & $0.52 - 1.95$ & $0.17 - 1.26$ & $0.17 -  1.88$ & $-0.35 -  0.27$  \\
 \hline
 %%%%%%%%%%%%%%%%%%%%%%%%%%%%%%%%%%%%%%%%%%%%%%%%%%%%%%%%%%%%%%%%%%%%%%%%
 \multicolumn{13}{c}{\mbox{Inverted  Hierarchy}} \\
 \cline{1-13}
 $1~\sigma$ &$32.30 - 35.29$& $48.49  - 50.06$ &$8.44 - 8.70$ &$256.01 -266.10 \cup 274.60 - 309.99$ &$0.00 - 11.96 \cup 153.90 - 179.99$ & $14.16 - 168.70$& $0.57 - 2.32$ &$0.58 - 2.32$ & $0.29 - 2.27$ &$0.22 - 1.83$ & $0.57 - 2.32$ & $-0.34 -  -0.25$    \\
 \hline
  $2~\sigma$ &$32.30 -36.39$& $47.35 - 50.67 $ &$8.30- 8.83$ &$226.00- 264.60 \cup 274.90 -331.98$ &$0.00 - 30.54 \cup 149.10 -179.98$ & $13.57 - 169.24$& $0.53 - 2.29$ &$0.53 - 2.29$ & $0.19 - 2.24$ &$0.16 - 1.83$ & $0.52 - 2.29$ & $-0.35 - -0.15$    \\
 \hline
 $3~\sigma$ & $31.40 - 37.39$ & $41.16 - 51.25$ & $8.17 - 8.96$ & $200.00 - 263.50 \cup 271.10 - 274.60 \cup 279.80 -352.98 $ & $0.00 - 34.11 \cup 146.10 - 179.99$ & $14.82 - 167.74$ & $0.50 - 2.04$ & $0.51 -  2.04$ & $0.12 - 1.98$ & $0.13 - 1.81$ & $0.50 - 2.04$& $-0.36 -  -0.04$  \\
 \hline
 %%%%%%%%%%%%%%%%%%%%%%%%%%%%%%%%%%%%%%%%%%%%%%%%%%%%%%%%%%%%%%%%%%%%%%%%%%%%%%%%%%%%%%
 %%%%%%%%%%%%%%%%%%%%%%%%%%%%%%%%%%%%%%%%%%%%%%%%%%%%%%%%%%%%%%%%%%%%%%%%%%%%%%%%%
 \hline
 \multicolumn{13}{c}{\mbox{Pattern} $G_{2}\equiv M_{\nu~22} = M_{\nu~33} $} \\
\hline
\hline
 \mbox{quantity} & $\theta_{12}^{\circ}$ & $\theta_{23}^{\circ}$& $\theta_{13}^{\circ}$ & $\delta^{\circ}$ & $\rho^{\circ}$ & $\sigma^{\circ}$  & $m_{1}$ $(10^{-1} \text{eV})$ & $m_{2}$ $(10^{-1} \text{eV})$ & $m_{3}$ $(10^{-1} \text{eV})$ & $m_{ee}$ $(10^{-1} \text{eV})$ & $m_{e}$ $(10^{-1} \text{eV})$ & $J$ $(10^{-1})$\\
 \hline
 %%%%%%%%%%%%%%%%%%%%%%%%%%%%%%%%%%%%%%%%%%%%%%%%%%%%%%%%%%%%%%%%%
 \multicolumn{13}{c}{\mbox{Normal  Hierarchy}} \\
 \cline{1-13}
 $1~\sigma$ &$33.30 - 35.30$& $48.47 -50.04$ &$8.41 -8.66$ &$172.01 - 217.99$ &$0.00 - 38.63 \cup 165.90 - 179.90$ & $15.76 - 157.94$ & $0.27 - 2.08$ &$0.28 - 2.08$ & $0.57 - 2.14$ &$0.09 - 2.05$ & $0.28 - 2.08$ & $-0.21 - 0.04$  \\
 \hline
 $2~\sigma$ &$32.30 - 36.39$& $47.37 - 50.70$ & $8.27 - 8.77$ & $152.05 -254.95$ &$0.01 - 73.85 \cup 147.30 - 179.97$ & $21.59 - 164.13$ & $0.14 - 2.14$ & $0.17 - 2.15$ & $0.52 - 2.20$  &$0.04 - 1.94$ & $0.17 - 2.15$ & $-0.33 - 0.16$  \\
 \hline
 $3~\sigma$ &$31.40 - 37.40$ & $41.20 - 44.93 \cup 45.11 - 51.10$ & $8.13 - 8.92$ & $128.00 - 358.89$ & $0.02 -  179.99$ & $0.01 - 179.65$ & $2.50\times10^{-5} - 1.10$ &$0.08 - 1.10$ & $0.49 - 1.21$ & $3.44\times10^{-3} - 1.10$ & $0.08 -  1.10$ & $-0.36 - 0.28$  \\
 \hline
 %%%%%%%%%%%%%%%%%%%%%%%%%%%%%%%%%%%%%%%%%%%%%%%%%%%%%%%%%%%%%%%%%%%%%%%%
 \multicolumn{13}{c}{\mbox{Inverted  Hierarchy}} \\
 \cline{1-13}
 $1~\sigma$ & $33.30 - 35.30$ & $48.49 - 50.06$ &$8.44 - 8.70$ &$256.00 - 309.95$ &$0.00 - 21.86 \cup 25.04 - 41.16 \cup 53.31 - 179.98$ & $0.07 - 13.62 \cup 26.60 - 179.39$ & $0.49 - 1.95$ & $0.50 - 1.96$ & $6.05\times10^{-8} - 1.89$ &$0.40 - 1.90$ & $0.49 - 1.95$ & $-0.34 - -0.25$    \\
 \hline
  $2~\sigma$ &$32.30 - 36.39$& $47.35 - 50.67$ & $8.30 - 8.83$ &$226.05 - 325.97$ & $0.12 - 179.96$ & $0.37 - 179.85$ & $0.48 - 2.21$ &$0.49 - 2.22$ & $9.30\times10^{-5} - 2.16$ &$0.36 - 1.81$ & $0.48 - 2.21$ & $-0.35 - -0.18$    \\
 \hline
 $3~\sigma$ & $31.40 - 37.40$ & $41.16 - 44.96 \cup 45.03 - 51.25$ & $8.17 - 8.96$ & $205.29 - 326.90 \cup 331.10 - 334.30 \cup 340.80 - 352.54$ & $0.01  - 179.94$ & $0.15 - 179.70$ & $0.48 - 2.14$ & $0.49 - 2.14$ & $2.18\times10^{-5} - 2.08$ & $0.33 - 2.12$ & $0.48 - 2.14$ & $-0.36 -  -0.04$  \\
 \hline
 %%%%%%%%%%%%%%%%%%%%%%%%%%%%%%%%%%%%%%%%%%%%%%%%%%%%%%%%%%%%%%%%%%%%%%%%%%%%%%%%%%%%%%
 %%%%%%%%%%%%%%%%%%%%%%%%%%%%%%%%%%%%%%%%%%%%%%%%%%%%%%%%%%%%%%%%%%%%%%%%%%%%%%%%%
 \hline
\multicolumn{13}{c}{\mbox{Pattern} $G_{3}\equiv M_{\nu~12} = M_{\nu~13}$} \\
\hline
\hline
 \mbox{quantity} & $\theta_{12}^{\circ}$ & $\theta_{23}^{\circ}$& $\theta_{13}^{\circ}$ & $\delta^{\circ}$ & $\rho^{\circ}$ & $\sigma^{\circ}$ & $m_{1}$ $(10^{-1} \text{eV})$ & $m_{2}$ $(10^{-1} \text{eV})$ & $m_{3}$ $(10^{-1} \text{eV})$ & $m_{ee}$ $(10^{-1} \text{eV})$ & $m_{e}$ $(10^{-1} \text{eV})$ & $J$ $(10^{-1})$\\
 \hline
 %%%%%%%%%%%%%%%%%%%%%%%%%%%%%%%%%%%%%%%%%%%%%%%%%%%%%%%%%%%%%%%%%
 \multicolumn{13}{c}{\mbox{Normal  Hierarchy}} \\
 \cline{1-13}
 $1~\sigma$ &$33.30 - 35.30$& $48.47 - 50.05$ &$8.41 - 8.66$ & $186.40 - 188.20 \cup  190.00 - 217.97$ &$0.38 - 37.53 \cup 177.86 - 177.87$ & $0.38 - 36.53 \cup  177.91 -177.92 $ & $0.40 - 2.30$ &  $0.41 - 2.31$ &  $0.64 - 2.36$ & $0.41 - 2.31$ &  $0.41 - 2.31$ & $-0.21 -  0.04$  \\
 \hline
 $2~\sigma$ & $32.30 - 36.39$ & $47.37 - 50.70$ & $8.27 - 8.79$ & $152.09 - 177.80 \cup 186.20 - 255.00$ &$0.46 - 72.68 \cup 154.80 - 179.80$ & $0.45 - 71.44 \cup 155.50 - 179.80$ & $0.23 - 2.26$ & $0.24 - 2.26$ &  $0.55 - 2.31$ & $0.23 - 2.26$ &  $0.25 - 2.26$ & $-0.34 - 0.16$  \\
 \hline
 $3~\sigma$ & $31.40 - 37.39$ & $41.20 - 44.33 \cup 45.33 - 51.33$ &$8.13 -8.92$ & $128.01 - 358.83$ & $0.63 - 179.43$ & $0.66  - 179.38$ & $0.14 - 1.99$ & $0.17 - 1.99$ & $0.52 - 2.05$ & $0.14 - 1.99$ & $0.17 - 1.99$ & $-0.36 -  0.27$  \\
 \hline
 %%%%%%%%%%%%%%%%%%%%%%%%%%%%%%%%%%%%%%%%%%%%%%%%%%%%%%%%%%%%%%%%%%%%%%%%
 \multicolumn{13}{c}{\mbox{Inverted  Hierarchy}} \\
 \cline{1-13}
 $1~\sigma$ & $33.30 - 35.30$ & $48.49 - 50.06$ & $8.44- 8.70$ & $256.01 - 309.99$ & $0.04 - 151.20 \cup 161.70 - 179.90$ & $1.09 - 150.60 \cup 161.10 - 179.51$ & $0.49 - 1.75$ & $0.50 - 1.75$ & $ 2.84\times10^{-5} - 1.68$ & $0.48 - 1.75$ & $ 0.49 - 1.75$ & $-0.34 - -0.25$  \\
 \hline
 $2~\sigma$ & $32.30 - 36.39$ & $ 47.35  - 50.66$ & $8.30 - 8.83$ &  $ 226.77  - 331.99$ & $0.00 - 179.91$ & $0.00 - 179.98$ & $0.48 - 1.78$ & $0.49 - 1.79$ & $1.43\times10^{-5} - 1.72$ & $0.48 - 1.76$ & $0.48 - 1.78$ & $-0.35 - 0.15$  \\
 \hline
 $3~\sigma$ & $31.40 - 37.39$ & $41.16 - 44.52 \cup 45.43 - 51.25$ & $8.17 - 8.96$ &  $ 200.15 - 352.91$ & $ 0.01 - 179.90$ & $0.05 - 179.98$ & $0.48 - 2.32$ & $0.49 - 2.32$ & $1.77\times10^{-4} - 2.26$ & $0.47 - 2.32$ & $0.48 - 2.32$ & $-0.36 - 0.04$  \\
 \hline
 %%%%%%%%%%%%%%%%%%%%%%%%%%%%%%%%%%%%%%%%%%%%%%%%%%%%%%%%%%%%%%%%%%%%%%%%%%%%%%%%%%%%%%%%%%%%%%%%%
 %%%%%%%%%%%%%%%%%%%%%%%%%%%%%%%%%%%%%%%%%%%%%%%%%%%%%%%%%%%%%%%%%%%%%%%%%%%%%%%%%
 \hline
  \end{tabular}
 }}
 \end{center}
  \caption{\footnotesize  The 1-2-3 $\sigma$ predictions for the neutrino physical parameters ($\theta_{12}$, $\theta_{23}$, $\theta_{13}$, $\delta$, $\rho$, $\sigma$, $m_{1}$, $m_{2}$, $m_{3}$, $m_{ee}$, $m_{e}$ and $J$) with regard to E, F and G categories.}
  \label{numerical2}
 \end{table}
\end{landscape}
\restoregeometry

In addition to the fact that the $C_2$ texture of inverted type is very predictive, with $\theta_{23}$ in the second octant and $\d$ in the range $[325^o,353^o]$, all the patterns restrict well the allowed parameters space.  We pay a special attention to the correlation ($m_{ee}$ versus LNM), as it shows best this restriction. In fact we can look at any point of this correlation as a point satisfying, in addition to the two real constraints defining the one equality texture, seven experimental constraints (three mixing angles, one Dirac phase $\d$, two squared mass differences $\Delta^2 m_{\mbox{\tiny sol}}, \Delta^2 m_{\mbox{\tiny atm}}$ and one effective mass constraint, say $m_e$) which are enough to determine the 9-parameter $M_\n$ with vanishing nonphysical phases, and thus to determine the LNM and the other effective parameter $m_{ee}$. Had there been no incertitude on the experimental constraints, one would have obtained one point in the plan ($m_{ee}$-LNM), whereas the obtained plots do reflect these allowed ranges in the experimental constraints.

\begin{figure}[hbtp]
 \centering
\begin{minipage}[l]{0.5\textwidth}
\epsfxsize=24cm
\centerline{\epsfbox{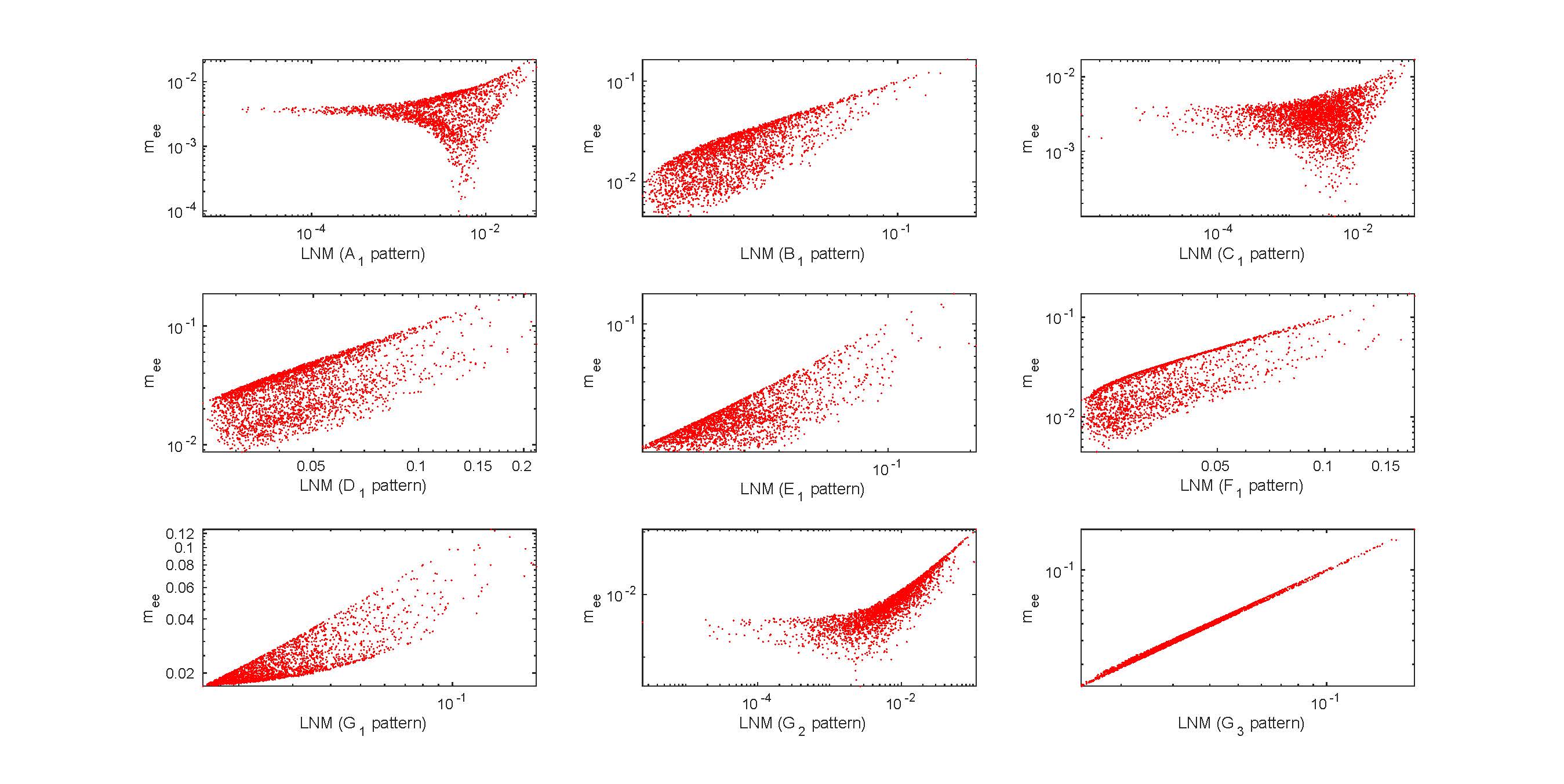}}
\end{minipage}%
\caption{ The correlation plots between LNM (lowest neutrino mass) and $m_{ee}$ for the nine independent patterns in the case of normal ordering. LNM and $m_{ee}$ are evaluated in eV.}
\label{LNM vs mee (N)}
\end{figure}

\begin{figure}[hbtp]
\centering
\begin{minipage}[l]{0.5\textwidth}
\epsfxsize=24cm
\centerline{\epsfbox{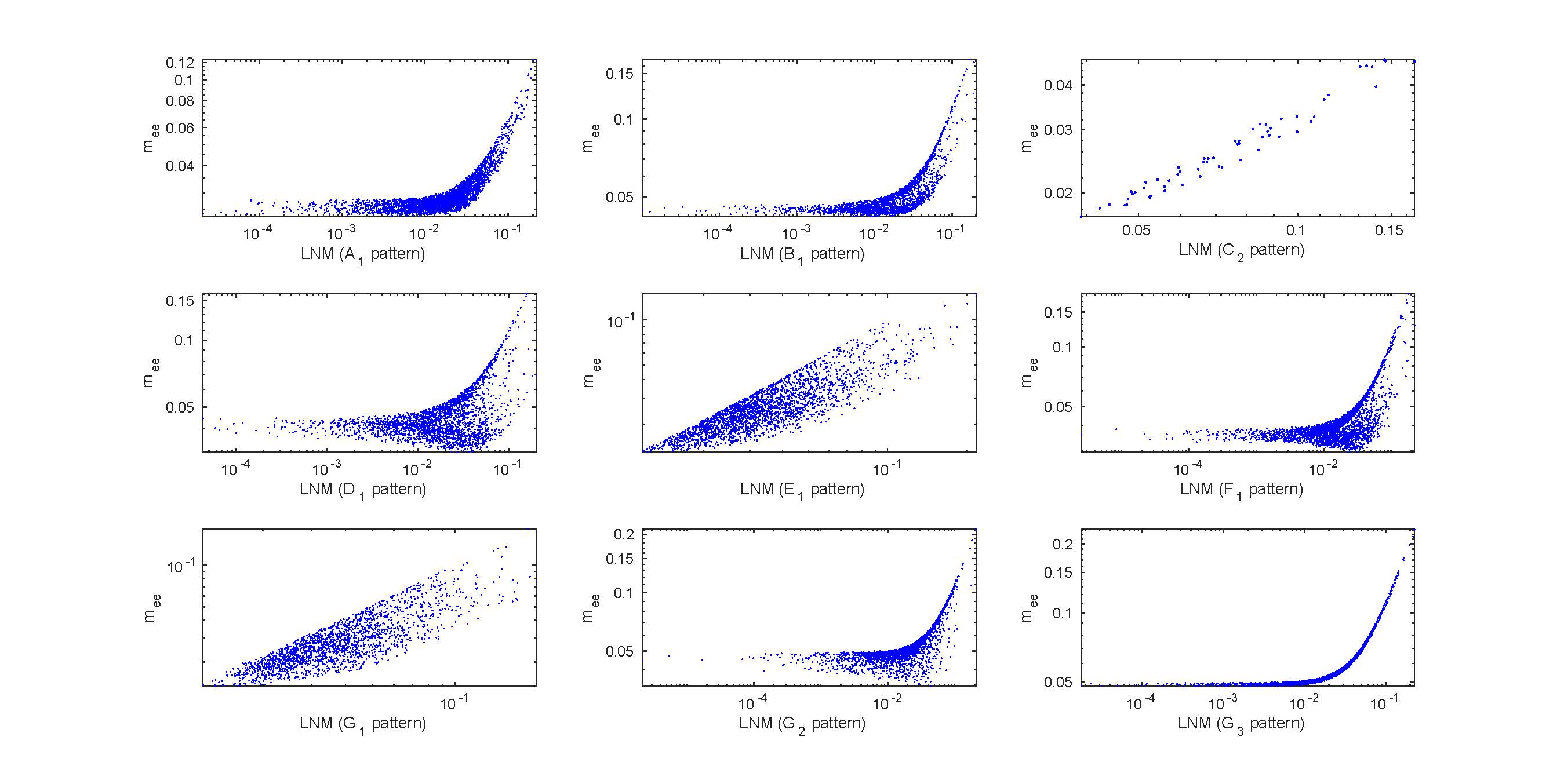}}
\end{minipage}%
\caption{ The correlation plots between LNM (lowest neutrino mass) and $m_{ee}$ for the nine independent patterns in the case of inverted ordering. LNM and $m_{ee}$ are evaluated in eV.}
\label{LNM vs mee (I)}
\end{figure}

Fig. (\ref{LNM vs mee (N)}) shows the $3\sigma$-level correlations ($m_{ee}$-$m_1$) for normal ordering on log-log scales in nine possible independent textures. The plots for the 6 remaining patterns are approximately the same as those in their interchange-symmetry related patterns, thus we opted not to present them. The plots show a direct relationship between the two parameters quite distinctive in the pattern $G_3$. However, for the nerve cell-like shape in pattern $A_1$, there is a singled out value for each parameter ($3$ m-eV for $m_{ee}$ and $3$-$9$ m-eV for $m_1$) when the other one is getting vanished. The inverted type nine correlations are shown in Fig. (\ref{LNM vs mee (I)}), where the corresponding LNM is now $m_3$, and where we omitted, due to near identicalness, what corresponds to the remaining 5 viable cases.  Again, the parameter space is restricted, especially for the $C_2$ and $G_3$ patterns, and we see a high density of allowed points around non-vanishing $m_3$ with decreasing $m_{ee}$ in the cases ($B_1, D_1, F_1, G_2$).

%\newpage
%\newpage
\newpage

\subsection{Pattern $A_1$: $M_{\nu~11}=M_{\nu~12}$}
The expressions for $A_{1}$, $A_{2}$ and $A_{3}$ coefficients for this pattern are
\begin{equation}
\begin{aligned}
A_{1}=&c_{12}^{2}c_{13}^{2}-c_{12}c_{13}(-c_{12}s_{23}s_{13}-s_{12}c_{23}e^{-i\delta}),\\
A_{2}=&s_{12}^{2}c_{13}^{2}-s_{12}c_{13}(-s_{12}s_{23}s_{13}+c_{12}c_{23}e^{-i\delta}),\\
A_{3}=&s_{13}^{2}-s_{13}s_{23}c_{13}.
\end{aligned}\label{A1coff}
\end{equation}
The leading order approximation in powers of $s_{13}$ for the mass ratios are given by
\begin{align}
\frac{m_1}{m_3}\approx&\frac{\big(\cos\theta_{23}\sin(2\sigma-\delta)-\tan\theta_{12}\sin2\sigma\big)}{\big(\cos2\theta_{12}\cos\theta_{23}\cos\delta-\frac{1}{2}\sin2\theta_{12}\sin^{2}\theta_{23}\big)\sin(2\sigma-2\rho)-\cos\theta_{23}\cos\delta\cos(2\sigma-2\rho)}\sin\theta_{13},\nonumber\\
\frac{m_2}{m_3}\approx&\frac{\big(\cos\theta_{23}\sin(2\rho-\delta)+\tan\theta_{12}\sin2\rho\big)}{\big(\cos2\theta_{12}\cos\theta_{23}\cos\delta-\frac{1}{2}\sin2\theta_{12}\sin^{2}\theta_{23}\big)\sin(2\sigma-2\rho)-\cos\theta_{23}\cos\delta\cos(2\sigma-2\rho)}\sin\theta_{13}\label{ratio2A1},
\end{align}
from which we get
\bea
\label{m2-m1}
1 < \frac{m_2}{m_1}  &\approx& \frac{\cos\theta_{23}\sin(2\rho-\delta)+\tan\theta_{12}\sin2\rho}{\cos\theta_{23}\sin(2\sigma-\delta)-\tan\theta_{12}\sin2\sigma}.
\eea
This imposes correlations amidst the mixing and phase angles, which, after fixing the mixing angles at their central values, can be checked easily to be met in the corresponding correlation plots amidst the phase angles.

For normal ordering, the representative point is taken as:
\begin{equation}
\begin{aligned}
(\theta_{12},\theta_{23},\theta_{13})=&(34.6452^{\circ},49.6598^{\circ},8.5176^{\circ}),\\
(\delta,\rho,\sigma)=&(198.4986^{\circ},105.8624^{\circ},13.6525^{\circ}),\\
(m_{1},m_{2},m_{3})=&(0.0076\textrm{ eV},0.0118\textrm{ eV},0.0513\textrm{ eV}),\\
(m_{ee},m_{e})=&(0.0010\textrm{ eV},0.0118\textrm{ eV}),
\end{aligned}
\end{equation}
the corresponding neutrino mass matrix (in eV) is
\begin{equation}
M_{\nu}=\left( \begin {array}{ccc} 0.0001 - 0.0010i&0.0001 - 0.0010i&0.0117 + 0.0013i\\ \noalign{\medskip}0.0001 - 0.0010i&0.0327 - 0.0002i&0.0219 + 0.0005i
\\ \noalign{\medskip}0.0117 + 0.0013i&0.0219 + 0.0005i&0.0228 - 0.0009i\end {array} \right).
\end{equation}
For inverted ordering, the representative point is taken as:
\begin{equation}
\begin{aligned}
(\theta_{12},\theta_{23},\theta_{13})=&(34.3293^{\circ},49.3765^{\circ},8.5673^{\circ}),\\
(\delta,\rho,\sigma)=&(235.3062^{\circ},178.3729^{\circ},113.5543^{\circ}),\\
(m_{1},m_{2},m_{3})=&(0.0573\textrm{ eV},0.0580\textrm{ eV},0.0298\textrm{ eV}),\\
(m_{ee},m_{e})=&(0.0307\textrm{ eV},0.0571\textrm{ eV}),
\end{aligned}
\end{equation}
the corresponding neutrino mass matrix (in eV) is
\begin{equation}
M_{\nu}=\left( \begin {array}{ccc} 0.0266 - 0.0154i&0.0266 - 0.0154i&-0.0302 + 0.0215i\\ \noalign{\medskip}0.0266 - 0.0154i&0.0005 + 0.0117i&0.0280 - 0.0100i
\\ \noalign{\medskip}-0.0302 + 0.0215i&0.0280 - 0.0100i&0.0042 + 0.0067i\end {array} \right).
\end{equation}

We see from Table (\ref{numerical1}) that the $A_{1}$ pattern is not viable at the 1-$\sigma$ level for inverted ordering. We find that the mixing angles ($\theta_{12}$,$\theta_{23}$,$\theta_{13}$) extend over their allowed experimental ranges for both normal and inverted hierarchies at all viable statistical levels. For inverted ordering, we find large forbidden regions for $\delta$ such as, $[247.25^{\circ},332.00^{\circ}] ([250.53^{\circ},353.00^{\circ}])$ at the 2(3)-$\sigma$ levels. For normal ordering, we find a strong restriction $[41.86^{\circ},122.87^{\circ}]$ on $\rho$ at the 1-$\sigma$ level, as well as a narrow forbidden gap $[164.51^{\circ},173.49^{\circ}]$ at the 2-$\sigma$ level. One also notes wide forbidden gaps $[33.38^{\circ},163.59^{\circ}]$($[87.02^{\circ},116^{\circ}]$) for the phase $\sigma$ at the 1(2)-$\sigma$ levels. We see that $m_{1}$ can reach a vanishing value in normal type at the 2-3 $\sigma$ levels, whereas $m_{3}$ approaches a zero value in inverted type at the same error levels. Therefore, the singular pattern is predicted in both normal and inverted types at the 2-3 $\sigma$ levels. %The allowed values of $J$ at all viable $\sigma$-levels for inverted ordering are negative, so the corresponding $\delta$ lies in the third or fourth quarters.

For normal ordering plots, there exists a narrow forbidden gap in all correlations including $\rho$ and $\sigma$ parameters. We find a moderate mass hierarchy where (0.16 $\leq$ $\frac{m_{2}}{m_{3}}$ $\leq$ 0.61) and a severe mass hierarchy where $m_{2}/m_{1}$ can reach $10^{3}$ indicating the possibility of a vanishing $m_{1}$.

For inverted ordering plots, we have a strong linear correlation between $\rho$ and $\sigma$ represented by two narrow ribbons. We find narrow disallowed regions for $\rho$ and $\sigma$ as in normal type. %We see from the correlation ($\theta_{23}, \delta$) that $\delta$ decreases when $\theta_{23}$ tends to increase.
We also find a quasi degeneracy characterized by $m_{1}\approx m_{2}$ as well as a very strong mass hierarchy where $m_{2}/m_{3}$ can reach $10^{3}$ indicating the possibility of a vanishing $m_{3}$. Due to a relatively narrow allowed range for $\delta$, the appearing portion of the sinusoidal curve corresponding to the correlation between (J,$\delta$) is very small.

 For both normal and inverted ordering, the correlations between ($\delta$,$m_{ee}$) and between ($\delta$,LNM) show that when $m_{ee}$ and LNM increase, the allowed parameter space becomes more restricted.
\begin{figure}[hbtp]
\centering
\begin{minipage}[l]{0.5\textwidth}
\epsfxsize=24.33cm
\centerline{\epsfbox{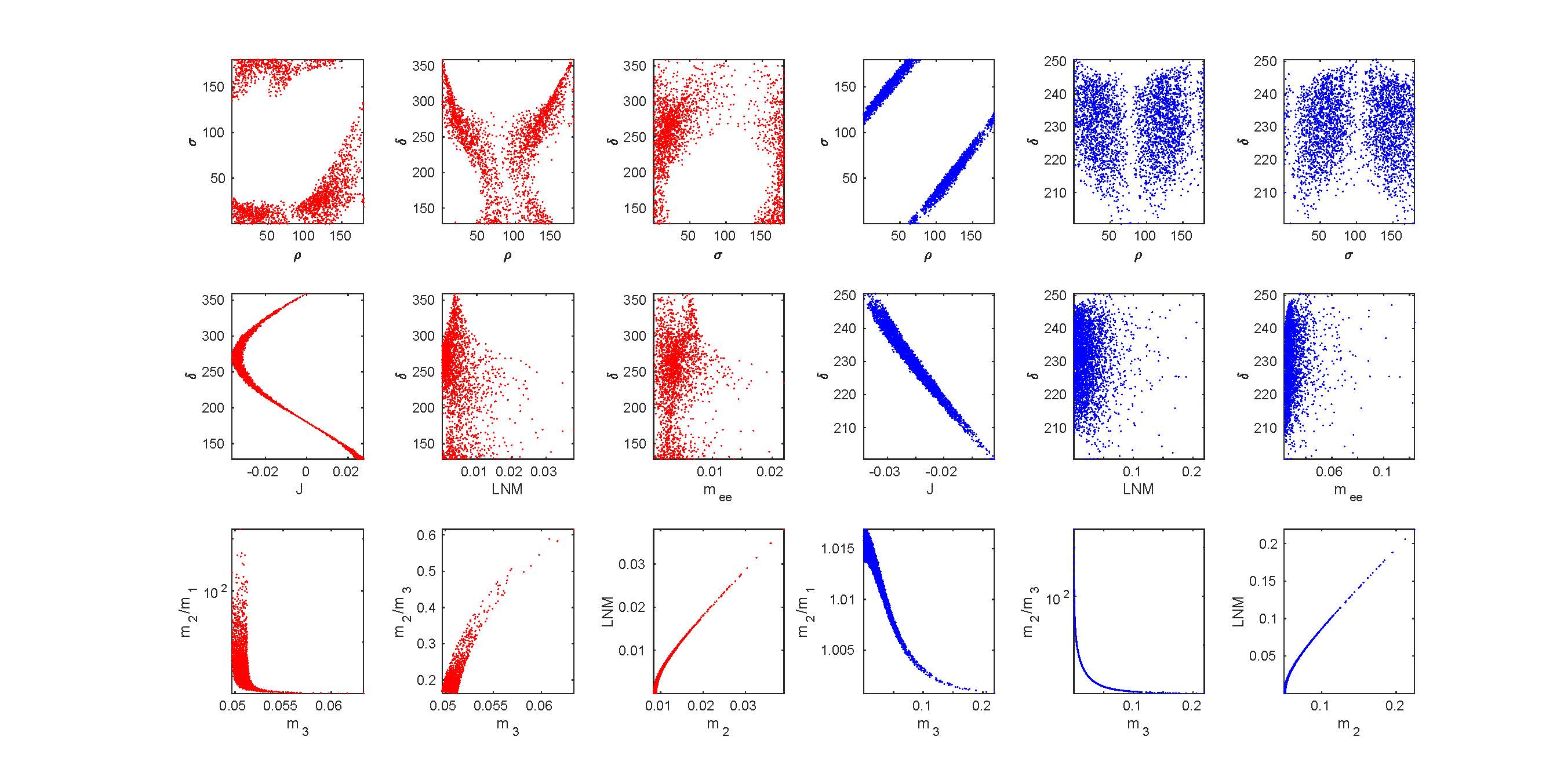}}
\end{minipage}%
\caption{ The correlation plots for $A_{1}$ pattern: the red (blue) plots represent the normal (inverted) ordering correlations. %The first row represents the correlations between the mixing angle $\theta_{23}$ and the CP-violating phases.
The first row represents the correlations amidst the CP-violating phases, the second one represents the correlations between $\delta$ and each of  J, LNM (the lowest neutrino mass), $m_{ee}$ parameters. The last one shows the degree of hierarchy. Angles (masses) are evaluated in degrees (eV).}
\label{A1fig}
\end{figure}
\newpage

\subsection{Pattern $A_2$: $M_{\nu~11}=M_{\nu~13}$}
The expressions for $A's$ coefficients and the approximate analytical formulas for the mass ratios can be obtained from Eqs. (\ref{A1coff},\ref{ratio2A1})by using the transformations in Eq. \ref{mutau}.

For normal ordering, the representative point is taken as:
\begin{equation}
\begin{aligned}
(\theta_{12},\theta_{23},\theta_{13})=&(34.1300^{\circ},49.6273^{\circ},8.5135^{\circ}),\\
(\delta,\rho,\sigma)=&(236.4442^{\circ},170.9022^{\circ},154.5267^{\circ}),\\
(m_{1},m_{2},m_{3})=&(0.0011\textrm{ eV},0.0088\textrm{ eV},0.0505\textrm{ eV}),\\
(m_{ee},m_{e})=&(0.0042\textrm{ eV},0.0090\textrm{ eV}),
\end{aligned}
\end{equation}
the corresponding neutrino mass matrix (in eV) is
\begin{equation}
M_{\nu}=\left( \begin {array}{ccc}  0.0035 - 0.0023i & 0.0062 + 0.0025i &  0.0035 - 0.0023i\\ \noalign{\medskip}0.0062 + 0.0025i  & 0.0260 - 0.0013i &  0.0274 + 0.0010i
\\ \noalign{\medskip}0.0035 - 0.0023i &  0.0274 + 0.0010i  & 0.0174 - 0.0007i
\end {array} \right).
\end{equation}
For inverted ordering, the representative point is taken as:
\begin{equation}
\begin{aligned}
(\theta_{12},\theta_{23},\theta_{13})=&(34.3729^{\circ},49.3046^{\circ},8.5548^{\circ}),\\
(\delta,\rho,\sigma)=&(299.3398^{\circ},32.1526^{\circ},93.7490^{\circ}),\\
(m_{1},m_{2},m_{3})=&(0.0615\textrm{ eV},0.0622\textrm{ eV},0.0359\textrm{ eV}),\\
(m_{ee},m_{e})=&(0.0344\textrm{ eV},0.0613\textrm{ eV}),
\end{aligned}
\end{equation}
the corresponding neutrino mass matrix (in eV) is
\begin{equation}
M_{\nu}=\left( \begin {array}{ccc} -0.0006 + 0.0344i &  0.0078 - 0.0364i & -0.0006 + 0.0344i\\ \noalign{\medskip}0.0078 - 0.0364i  & 0.0223 - 0.0070i &  0.0140 + 0.0165i
\\ \noalign{\medskip}-0.0006 + 0.0344i &  0.0140 + 0.0165i &  0.0198 - 0.0271i\end {array} \right).
\end{equation}
From Table (\ref{numerical1}), we find that the mixing angles $(\theta_{12},\theta_{23},\theta_{13}$) extend over their allowed experimental ranges at all $\sigma$ error levels for both orderings. For inverted ordering, we see that the Dirac phase $\delta$ is strongly restricted to the intervals: $[293.06^{\circ},309.99^{\circ}]$ at the 1-$\sigma$ level, $[290.70^{\circ},320.00^{\circ}]$ at the 2-$\sigma$ level and
$[287.81^{\circ},342.45^{\circ}]$ at the 3-$\sigma$ level. We also notice narrow disallowed regions for the phase $\rho$ at the 2-3-$\sigma$ levels in inverted type. There exist wide forbidden gaps for $\rho$ in normal type at the 1-2-$\sigma$ levels ($[23.44^{\circ} ,168.80^{\circ}]$ for 1-$\sigma$  and $[79.84^{\circ},90.10^{\circ}]\cup[90.83^{\circ},125.60^{\circ}]$ for 2-$\sigma$). Table (\ref{numerical1}) also shows that $m_{1}$ can reach zero value at the 2-3-$\sigma$ levels in normal type, whereas $m_{3}$ can reach a vanishing value at all $\sigma$ levels in inverted type. Therefore, the singular mass matrix is expected at the 2-3-$\sigma$ levels for normal ordering and all $\sigma$ levels for inverted ordering. %The allowed values of $J$ parameter are negative at all $\sigma$ levels for inverted ordering, so the corresponding $\delta$ lies in the third or fourth quarters.

For normal ordering plots, corresponding to $3\sigma$ level, we notice a narrow forbidden gap for the phase $\rho$. We also see a moderate mass hierarchy characterized by $0.16\leq\frac{m_2}{m_3}\leq0.85$ besides an acute mass hierarchy where $\frac{m_2}{m_1}$ can reach $10^4$ indicating the possibility of vanishing $m_1$.

As to the inverted ordering plots, we have a strong linear correlation between $\rho$ and $\sigma$ represented by two narrow ribbons. We find narrow disallowed regions for $\rho$ and $\sigma$ as in $A_{1}$ pattern. We also find a quasi degeneracy characterized by $m_{1}\approx m_{2}$ as well as a very strong mass hierarchy where $m_{2}/m_{3}$ can reach $10^{4}$ indicating the possibility of a vanishing $m_{3}$. As in $A_1$ case, and due to a relatively narrow allowed range for $\delta$, the appearing portion of the sinusoidal curve corresponding to the correlation between (J,$\delta$) is very small.

For both normal and inverted ordering, the correlations between ($\delta$,$m_{ee}$) and between ($\delta$,LNM) show that when $m_{ee}$ and LNM increase, the allowed parameter space becomes more restricted.

We note here that the plots of $A_2$ can approximately be deduced from those of $A_1$ case by doing the transformations of Eq. \ref{mutau}, and this can be justified because the experimental constraints at $3\s$ level for $\t_{23}$ do not break the interchange symmetry by much. This trend will be noted in all interchange-symmetry related patterns.

\begin{figure}[hbtp]
\centering
\begin{minipage}[l]{0.5\textwidth}
\epsfxsize=24cm
\centerline{\epsfbox{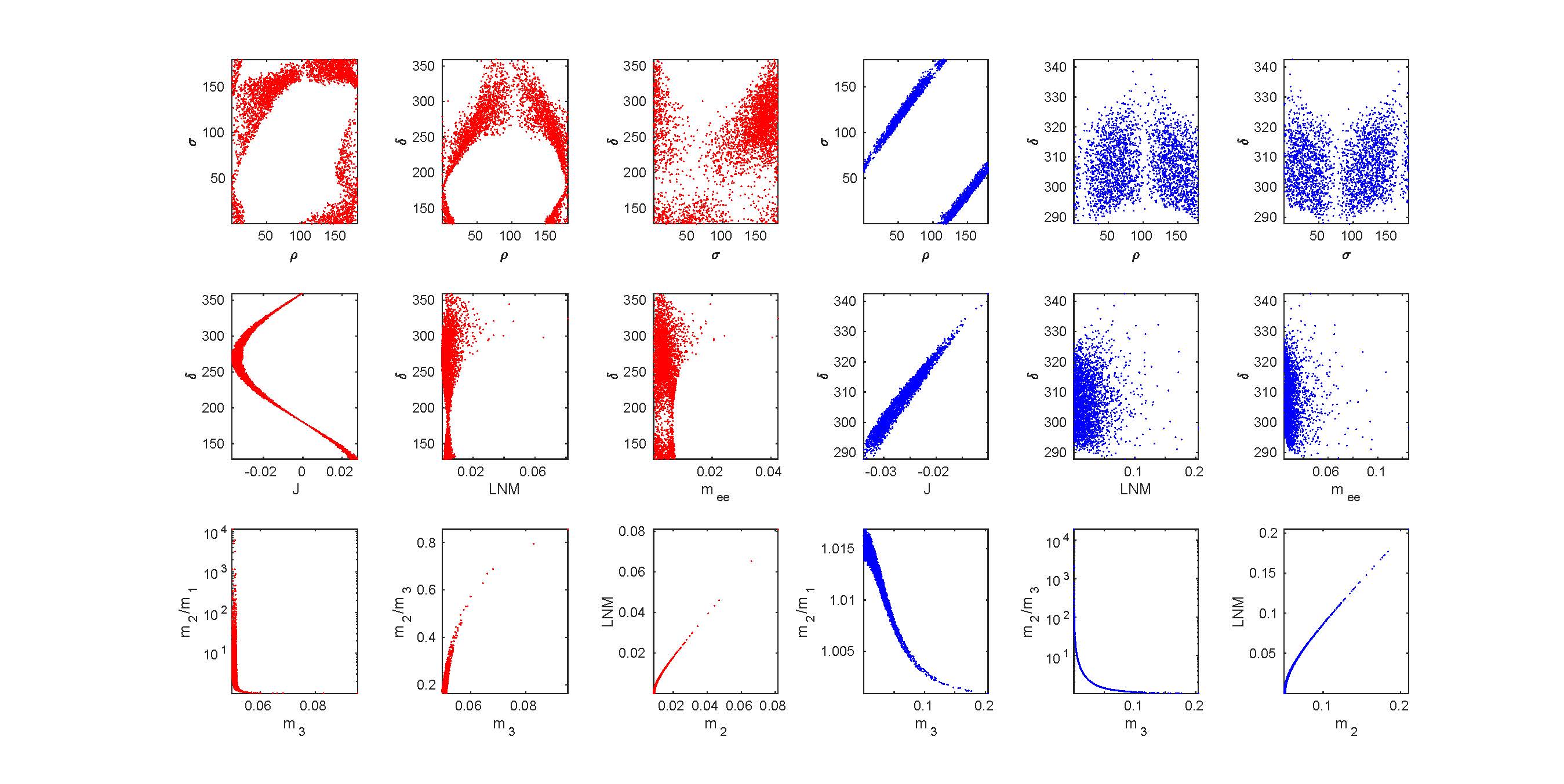}}
\end{minipage}%
\caption{ The correlation plots for $A_2$ pattern: the red (blue) plots represent the normal (inverted) ordering. The first row represents the correlations between the CP-violating phases, the second one represents the correlations between $\delta$ and each of J, LNM (the lowest neutrino mass), $m_{ee}$ parameters. The last row shows the degree of hierarchy. Angles (masses) are evaluated in degrees (eV).}
\label{A2fig}
\end{figure}
\newpage

\subsection{Pattern $B_1$: $M_{\nu~12}=M_{\nu~22}$}
The expressions for $A_{1}$, $A_{2}$ and $A_{3}$ coefficients for this pattern are
\begin{equation}
\begin{aligned}
A_{1}=&c_{12}c_{13}(-c_{12}s_{23}s_{13}-s_{12}c_{23}e^{-i\delta})-(-c_{12}s_{23}s_{13}-s_{12}c_{23}e^{-i\delta})^{2},\\
A_{2}=&s_{12}c_{13}(-s_{12}s_{23}s_{13}+c_{12}c_{23}e^{-i\delta})-(-s_{12}s_{23}s_{13}+c_{12}c_{23}e^{-i\delta})^{2},\\
A_{3}=&s_{13}s_{23}c_{13}-s_{23}^{2}c_{13}^{2}.
\end{aligned}\label{B1coff}
\end{equation}
The leading order approximation in $s_{13}$ for the mass ratios are given by
\begin{equation}
\hspace{-1em}\frac{m_1}{m_3}\approx\sin\theta_{23}\tan\theta_{23}\Bigg(\frac{\cot\theta_{12}\cos\theta_{23}\sin(2\sigma-2\delta)-\sin(2\sigma-\delta)}{\cos\theta_{23}\sin\delta\cos(2\rho-2\sigma)+\big(\cos2\theta_{12}\cos\theta_{23}\cos\delta-\sin\theta_{12}\cos\theta_{12}\sin^2\theta_{23}\big)\sin(2\rho-2\sigma)}\Bigg)
, \nonumber \end{equation}
\begin{equation}
\hspace{-1em}\frac{m_2}{m_3}\!\!\approx \!\! -\sin\theta_{23}\tan\theta_{23}\Bigg(\frac{\tan\theta_{12}\cos\theta_{23}\sin(2\rho-2\delta)+\sin(2\rho-\delta)}{\cos\theta_{23}\sin\delta\cos(2\rho-2\sigma)+\big(\cos2\theta_{12}\cos\theta_{23}\cos\delta-\sin\theta_{12}\cos\theta_{12}\sin^2\theta_{23}\big)\sin(2\rho-2\sigma)} \! \Bigg)\label{B1ratio2}
\!, \end{equation}
 whence
 \bea
 1 < \frac{m_2}{m_1}  &\approx& -\frac{\tan\theta_{12}\cos\theta_{23}\sin(2\rho-2\delta)+\sin(2\rho-\delta)}{\cot\theta_{12}\cos\theta_{23}\sin(2\sigma-2\delta)-\sin(2\sigma-\delta)}
 , \eea
so taking experimentally central, or best fit, values for the mixing angles leads to a condition on the phases which can be checked to be met in the corresponding correlation plots.

For normal ordering, the representative point is taken as:
\begin{equation}
\begin{aligned}
(\theta_{12},\theta_{23},\theta_{13})=&(34.3493^{\circ},49.0148^{\circ},8.5821^{\circ}),\\
(\delta,\rho,\sigma)=&(193.1429^{\circ},52.4608^{\circ},92.6505^{\circ}), \\
(m_{1},m_{2},m_{3})=&(0.0438\textrm{ eV},0.0446\textrm{ eV},0.0665\textrm{ eV}),\\
(m_{ee},m_{e})=&(0.0335\textrm{ eV},0.0447\textrm{ eV}),
\end{aligned}
\end{equation}
the corresponding neutrino mass matrix (in eV) is
\begin{equation}
M_{\nu}=\left( \begin {array}{ccc} -0.0199 + 0.0269i&0.0228 + 0.0083i&-0.0063 - 0.0158i\\ \noalign{\medskip}0.0228 + 0.0083i&0.0228 + 0.0083i&0.0450 - 0.0115i
\\\noalign{\medskip}-0.0063 - 0.0158i&0.0450 - 0.0115i
&0.0161 + 0.0169i\end {array} \right).
\end{equation}
For inverted ordering, the representative point is taken as:
\begin{equation}
\begin{aligned}
(\theta_{12},\theta_{23},\theta_{13})=&(34.3182^{\circ},49.1865^{\circ},8.5398^{\circ}),\\
(\delta,\rho,\sigma)=&(284.9976^{\circ},76.0464^{\circ},129.1494^{\circ}),\\
(m_{1},m_{2},m_{3})=&(0.1081\textrm{ eV},0.1084\textrm{ eV},0.0959\textrm{ eV}),\\
(m_{ee},m_{e})=&(0.0685\textrm{ eV},0.1080\textrm{ eV}),
\end{aligned}
\end{equation}
the corresponding neutrino mass matrix (in eV) is
\begin{equation}
M_{\nu}=\left( \begin {array}{ccc} -0.0685 + 0.0007i&0.0700 + 0.0091i&-0.0433 - 0.0107i\\ \noalign{\medskip}0.0700 + 0.0091i&0.0700 + 0.0091i&0.0140 - 0.0126i
\\ \noalign{\medskip}-0.0433 - 0.0107i&0.0140 - 0.0126i&0.0897 + 0.0171i\end {array} \right).
\end{equation}

We see from Table (\ref{numerical1}) that the allowed experimental ranges of the mixing angles ($\theta_{12}$,$\theta_{23}$,$\theta_{13}$) can be covered at all $\sigma$ levels for both normal and inverted ordering. For normal ordering, at the 1-$\sigma$-level, there exists a narrow forbidden gap $[91.84^{\circ},96.64^{\circ}]$ for $\rho$. We also find that the phase $\sigma$ is bound to the intervals: $[78.64^{\circ},124.38^{\circ}]$ at the 1-$\sigma$ level and $[63.79^{\circ},158.95^{\circ}]$ at the 2-$\sigma$ level. For inverted ordering, the parameter $\delta$ at 3-$\sigma$ level is restricted to lie in the interval $[224.88^{\circ},325.14^{\circ}]$. We note that $m_{1}$ does not reach zero at any $\sigma$ level in normal type, whereas $m_{3}$ can approach a vanishing value at all $\sigma$ levels in inverted type. Thus, the singular pattern is predicted in inverted type at all $\sigma$-levels. %The allowed values of $J$ at all $\sigma$ levels for inverted ordering are negative, so the corresponding $\delta$ lies in the third or fourth quarters.

For normal ordering plots, we see a mild mass hierarchy where $(0.33\leq\frac{m_{2}}{m_{3}}\leq0.96)$ as well as a quasi degenerate mass hierarchy where $(1 < \frac{m_{2}}{m_{1}}\leq1.15)$.

For inverted ordering plots, we see a quasi-linear correlation between $\rho$ and $\sigma$. We also see a quasi degenerate mass spectrum characterized  by $m_{1}\approx m_{2}$ in addition to an acute mass hierarchy where $\frac{m_{2}}{m_{3}}$ can reach $10^{3}$ indicating the possibility of a vanishing $m_{3}$.

For both normal and inverted ordering, the correlations between ($\delta$,$m_{ee}$) and between ($\delta$,LNM) show that when $m_{ee}$ and LNM increase, the allowed parameter space becomes more limited.
\begin{figure}[hbtp]
\centering
\begin{minipage}[l]{0.5\textwidth}
\epsfxsize=24cm
\centerline{\epsfbox{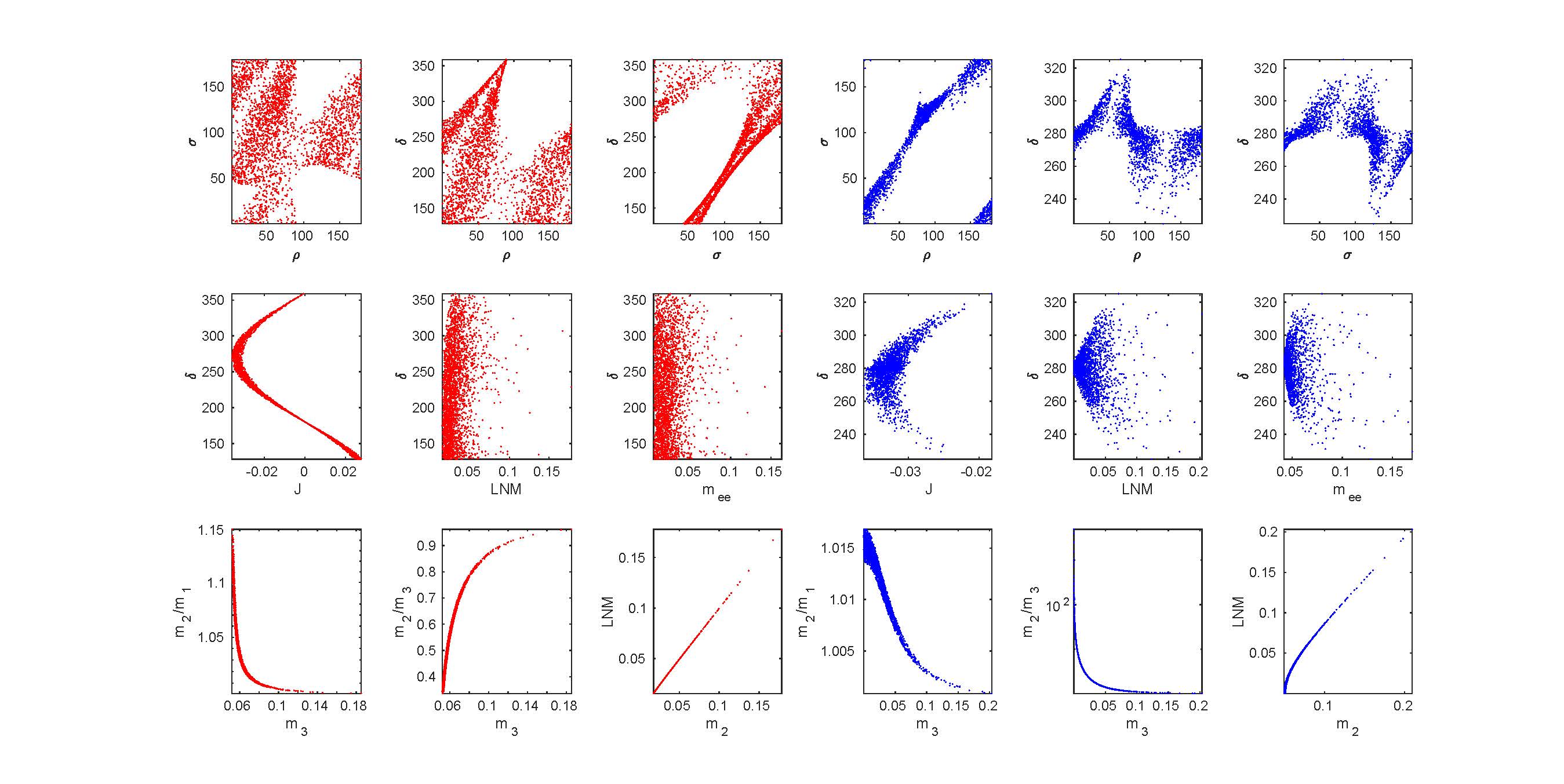}}
\end{minipage}%
\caption{ The correlation plots for $B_1$ pattern: the red (blue) plots represent the normal (inverted) ordering. The first row represents the correlations between the CP-violating phases, the second one represents the correlations between $\delta$ and each of J, LNM (the lowest neutrino mass), $m_{ee}$ parameters. The last row shows the degree of hierarchy. Angles (masses) are evaluated in degrees (eV).}
\label{B1fig}
\end{figure}
\newpage

\subsection{Pattern $B_2$: $M_{\nu~13}=M_{\nu~33}$}
The expressions for $A's$ coefficients and the approximate analytical formulas for the mass ratios can be obtained from Eqs. (\ref{B1coff},\ref{B1ratio2})by using the transformations in Eq. \ref{mutau}.

For normal ordering, the representative point is taken as:
\begin{equation}
\begin{aligned}
(\theta_{12},\theta_{23},\theta_{13})=&(34.4516^{\circ},49.4098^{\circ},8.5475^{\circ}),\\
(\delta,\rho,\sigma)=&(211.9931^{\circ},99.3031^{\circ},21.7817^{\circ}), \\
(m_{1},m_{2},m_{3})=&(0.0297\textrm{ eV},0.0309\textrm{ eV},0.0582\textrm{ eV}),\\
(m_{ee},m_{e})=&(0.0104\textrm{ eV},0.0310\textrm{ eV}),
\end{aligned}
\end{equation}
the corresponding neutrino mass matrix (in eV) is
\begin{equation}
M_{\nu}=\left( \begin {array}{ccc} -0.0104 + 0.0004i & -0.0099 + 0.0002i &  0.0274 - 0.0003i\\ \noalign{\medskip}-0.0099 + 0.0002i  & 0.0422 - 0.0003i &  0.0209 + 0.0003i
\\\noalign{\medskip}0.0274 - 0.0003i &  0.0209 + 0.0003i &  0.0274 - 0.0003i\end {array} \right).
\end{equation}
For inverted ordering, the representative point is taken as:
\begin{equation}
\begin{aligned}
(\theta_{12},\theta_{23},\theta_{13})=&(34.8159^{\circ},49.3943^{\circ},8.5900^{\circ}),\\
(\delta,\rho,\sigma)=&(262.2411^{\circ},62.7818^{\circ},32.3640^{\circ}),\\
(m_{1},m_{2},m_{3})=&(0.0493\textrm{ eV},0.0501\textrm{ eV},0.0076\textrm{ eV}),\\
(m_{ee},m_{e})=&(0.0426\textrm{ eV},0.0491\textrm{ eV}),
\end{aligned}
\end{equation}
the corresponding neutrino mass matrix (in eV) is
\begin{equation}
M_{\nu}=\left( \begin {array}{ccc} -0.0119 + 0.0409i & -0.0014 + 0.0101i &  0.0061 - 0.0212i\\ \noalign{\medskip}-0.0014 + 0.0101i &  0.0078 - 0.0212i &  0.0001 + 0.0224i
\\ \noalign{\medskip}0.0061 - 0.0212i &  0.0001 + 0.0224i &  0.0061 - 0.0212i\end {array} \right).
\end{equation}

We find from Table (\ref{numerical1}) that the allowed experimental ranges for the mixing angles $(\theta_{12},\theta_{23},\theta_{13})$ can be covered at all $\sigma$ levels with either hierarchy type. We see a narrow forbidden gap $[177.70^o,182.10^o]$ for the Dirac phase $\delta$ at the 3-$\sigma$ level in normal type. However, $\delta$ in the inverted type is bound to lie in the intervals $[256.00^{\circ},289.12^{\circ}] ([226.41^{\circ},291.28^{\circ}],[214.08^{\circ},311.63^{\circ}])$ at the 1(2,3)-$\sigma$ level. For normal ordering, the phase $\rho$ is strongly restricted to $[83.13^{\circ},123.54^{\circ}]$ at the 1-$\sigma$ level, and the range tends to widen into $[66.08^{\circ},165.04^{\circ}]$ at the 2-$\sigma$ level. One also notes that $m_{1}$ does not reach a vanishing value at all $\sigma$ levels in normal type. However,  $m_{3}$ approaches a zero value at all $\sigma$ levels. Thus, the singular pattern is predicted for inverted ordering at all statistical levels. %The allowed values of J in inverted type are negative, so the corresponding $\delta$ lies in the third or fourth quarters.

For normal ordering plots, we see a mild mass hierarchy where $(0.28\leq\frac{m_{2}}{m_{3}}\leq0.97)$ as well as a quasi degenerate mass hierarchy where $(1 < \frac{m_{2}}{m_{1}}\leq1.22)$.

For inverted ordering plots, we see a quasi-linear correlation between $\rho$ and $\sigma$ as in $B_{1}$ pattern. We also see a quasi degenerate mass spectrum characterized by $m_{1}\approx m_{2}$ in addition to an acute mass hierarchy where $\frac{m_{2}}{m_{3}}$ can reach $10^{3}$ indicating the possibility of a vanishing $m_{3}$.

For both normal and inverted ordering, the correlations between ($\delta$,$m_{ee}$) and between ($\delta$,LNM) show that when $m_{ee}$ and LNM increase, the allowed parameter space becomes more limited.
\begin{figure}[hbtp]
\centering
\begin{minipage}[l]{0.5\textwidth}
\epsfxsize=24cm
\centerline{\epsfbox{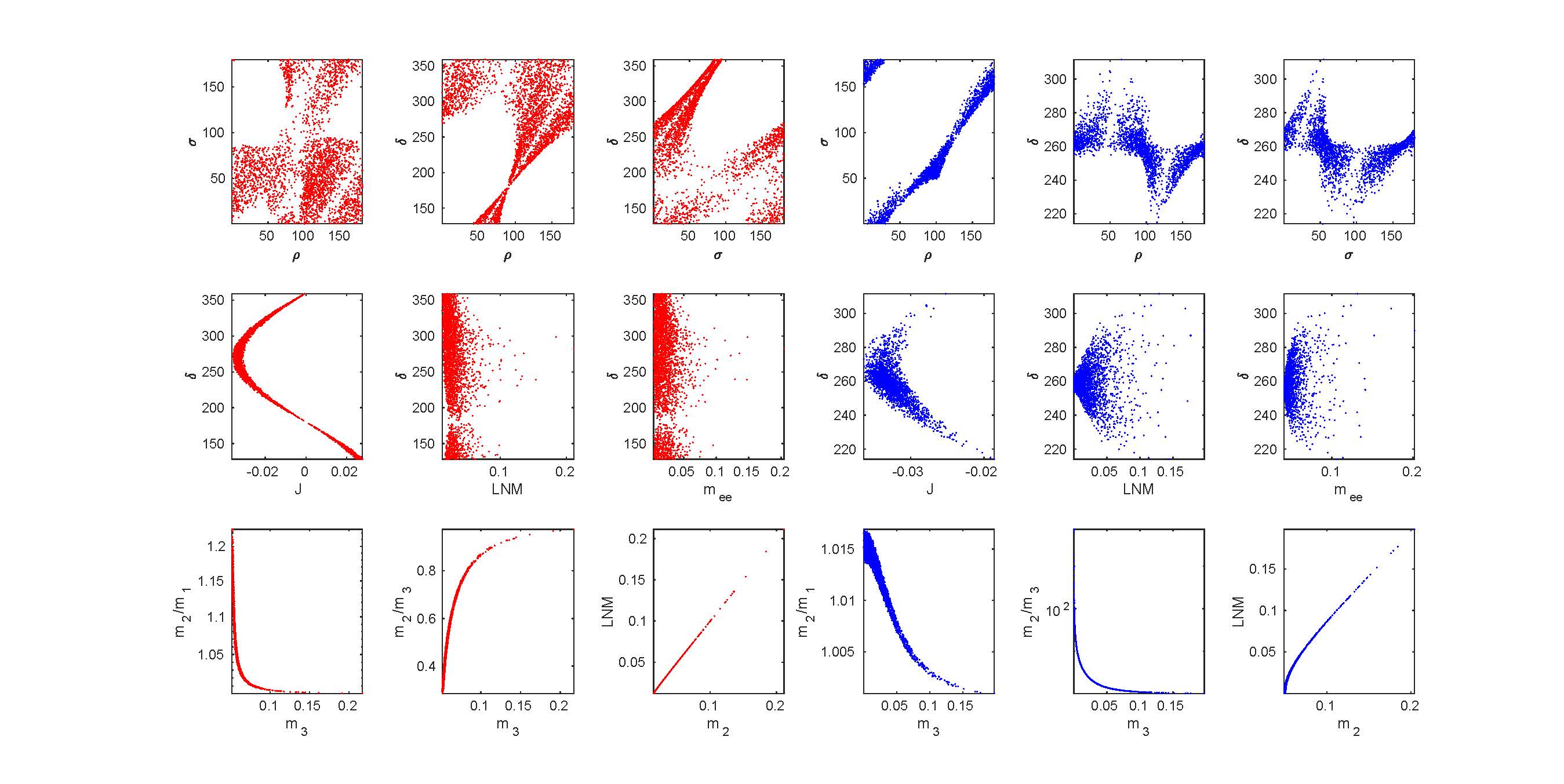}}
\end{minipage}%
\caption{ The correlation plots for $B_2$ pattern: the red (blue) plots represent the normal (inverted) ordering. The first row represents the correlations between the CP-violating phases, the second one represents the correlations between $\delta$ and each of J, LNM (the lowest neutrino mass), $m_{ee}$ parameters. The last row shows the degree of hierarchy. Angles (masses) are evaluated in degrees (eV).}
\label{B2fig}
\end{figure}
\newpage

\subsection{Pattern $C_1$: $M_{\nu~23}=M_{\nu~33}$}
The expressions for $A_{1}$, $A_{2}$ and $A_{3}$ coefficients for this pattern are
\begin{equation}
\begin{aligned}
A_{1}=&(-c_{12}s_{23}s_{13}-s_{12}c_{23}e^{-i\delta})(-c_{12}c_{23}s_{13}+s_{12}s_{23}e^{-i\delta})-(-c_{12}c_{23}s_{13}+s_{12}s_{23}e^{-i\delta})^{2},\\
A_{2}=&(-s_{12}s_{23}s_{13}+c_{12}c_{23}e^{-i\delta})^{2}(-s_{12}c_{23}s_{13}-c_{12}s_{23}e^{-i\delta})-(-s_{12}c_{23}s_{13}-c_{12}s_{23}e^{-i\delta})^{2},\\
A_{3}=&s_{23}c_{13}^{2}c_{23}-c_{23}^{2}c_{13}^{2}.
\end{aligned}\label{C1coff}
\end{equation}
The leading order approximation in powers of $s_{13}$ for the mass ratios and $m_{ee}$ are given by
\begin{align}
\frac{m_1}{m_3}\approx&\frac{1}{2}\frac{\tan2\theta_{23}(\sin2\theta_{23}-1)\sin(2\sigma-2\delta)}{\sin^2\theta_{12}\sin^2\theta_{23}\sin(2\sigma-2\rho)}, \nonumber\\
\frac{m_2}{m_3}\approx&-\frac{1}{2}\frac{\tan2\theta_{23}(\sin2\theta_{23}-1)\sin(2\rho-2\delta)}{\cos^2\theta_{12}\sin^2\theta_{23}\sin(2\sigma-2\rho)}, \nonumber\\
m_{ee}\approx&m_{3}\Bigg|\frac{4\cos^{4}\theta_{12}(\sin2\theta_{23}-1)e^{2i\sigma}\big(\tan^4\theta_{12}\sin(2\delta-2\rho)-\sin(2\delta-2\sigma)e^{2i(\rho-\sigma)}\big)}{\sin^22\theta_{12}\cos2\theta_{23}\tan\theta_{23}\sin(2\sigma-2\rho)}\Bigg|,\label{Cratios}
\end{align}
so we get
\bea
1 < \frac{m_2}{m_1} &\approx& - \frac{\tan^2 \t_{12} \sin(2\rho-2\delta) }{\sin(2\sigma-2\delta)}
.\eea
Fixing the mixing angles around their central values leads to an inequality involving the phases which can be checked to be met in the corresponding correlation plots.

For normal ordering, the representative point is taken as:
\begin{equation}
\begin{aligned}
(\theta_{12},\theta_{23},\theta_{13})=&(35.3523^{\circ},50.9121^{\circ},8.5430^{\circ}),\\
(\delta,\rho,\sigma)=&(187.7299^{\circ},104.5948^{\circ},10.5286^{\circ}),\\
(m_{1},m_{2},m_{3})=&(0.0057\textrm{ eV},0.0103\textrm{ eV},0.0516\textrm{ eV}),\\
(m_{ee},m_{e})=&(0.0012\textrm{ eV},0.0107\textrm{ eV}),
\end{aligned}
\end{equation}
the corresponding neutrino mass matrix (in eV) is
\begin{equation}
M_{\nu}=\left( \begin {array}{ccc} 0.0011 - 0.0006i&0.0014 - 0.0012i&0.0103 + 0.0017i\\ \noalign{\medskip}0.0014 - 0.0012i&0.0334 + 0.0004i&0.0220 - 0.0002i
\\\noalign{\medskip} 0.0103 + 0.0017i&0.0220 - 0.0002i
&0.0220 - 0.0002i\end {array} \right).
\end{equation}

From Table (\ref{numerical1}), we find that this texture can accommodate the experimental data for only normal ordering at the 2-3-$\sigma$ levels. The mixing angles $\theta_{12}(\theta_{23})$ are tightly restricted to the intervals $[35.39^{\circ},36.40^{\circ}]([50.06^{\circ},50.70^{\circ}])$ at the 2-$\sigma$ level. However, at the 3-$\sigma$ level, we see that $\theta_{12}$ is bound to the interval $[34.49^{\circ}, 37.40^{\circ}]$, in addition to a wide forbidden gap $[42.16^{\circ},49.52^{\circ}]$ for $\theta_{23}$. We also find that the Dirac phase $\delta$ is bound to the intervals $[152.02^{\circ}, 227.99^{\circ}]$ at the 2-$\sigma$ level and $[128.05^{\circ}, 257.64^{\circ}]$ at the 3-$\sigma$ level. There exists a strong restriction on $\rho$ at the 2-$\sigma$ level, whereas a wide disallowed region $[48.35^{\circ},145.00^{\circ}]$ is noticed for the phase $\sigma$. Table (\ref{numerical1}) also reveals that $m_{1}$ can not reach a vanishing value at every viable $\sigma$-levels. Therefore, the singular pattern is not expected for this texture.

\begin{figure}[hbtp]
\centering
\begin{minipage}[l]{0.5\textwidth}
\epsfxsize=24cm
\centerline{\epsfbox{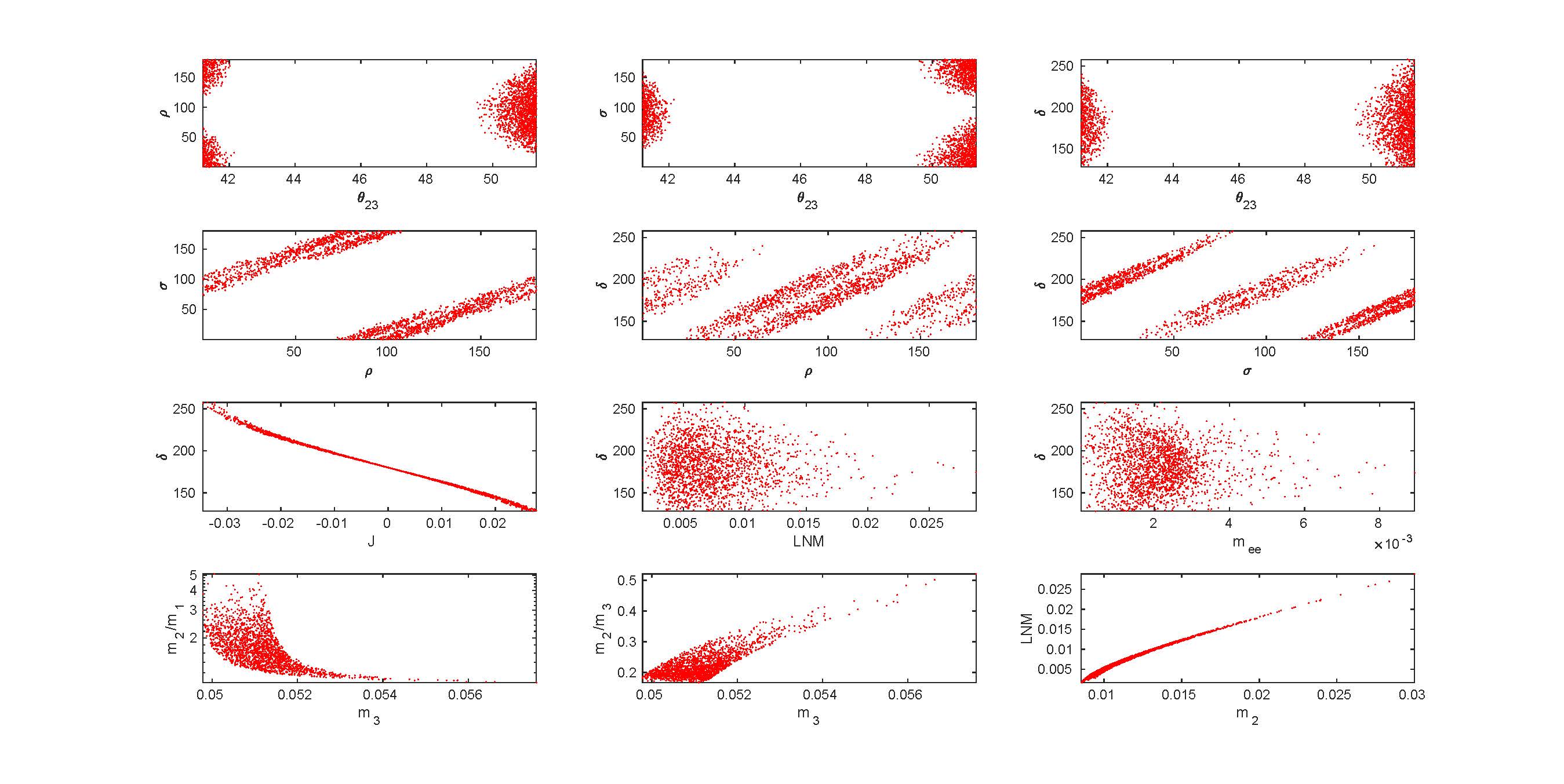}}
\end{minipage}%
\caption{The correlation plots for $C_{1}$ pattern: the red (blue) plots represent the normal (inverted) ordering correlations. The first row represents the correlations between the mixing angle $\theta_{23}$ and the CP-violating phases.
The second row represents the correlations amidst the CP-violating phases, the third one represents the correlations between $\delta$ and each of  J, LNM (the lowest neutrino mass), $m_{ee}$ parameters. The last one shows the degree of hierarchy. Angles (masses) are evaluated in degrees (eV).}
\label{C1fig}
\end{figure}

From Fig. \ref{C1fig}, we see a wide forbidden gap for $\theta_{23}$ as well as quasi linear relations represented by ribbons in the correlations of ($\sigma$,$\delta$) and of ($\rho$,$\sigma$), and so approximately also for ($\rho$,$\delta$) where the quasi linear ribbons become larger and contain forbidden bands. The correlations between ($\delta$,$m_{ee}$) and between ($\delta$,LNM) show that when $m_{ee}$ and LNM increase, the allowed parameter space becomes more restricted. Due to a relatively narrow allowed range for $\delta$, the appearing portion of the sinusoidal curve corresponding to the correlation between (J,$\delta$) is very small. We find a mild mass hierarchy where $(0.16\leq\frac{m_{2}}{m_{3}}\leq0.52)$.
\newpage

\subsection{Pattern $C_2$: $M_{\nu~23}=M_{\nu~22}$}
The expressions for $A's$ coefficients as well as the approximate analytical formulas for the mass ratios and $m_{ee}$ can be obtained from Eqs. (\ref{C1coff},\ref{Cratios})by using the transformations in Eq. \ref{mutau}.

For normal ordering, the representative point is taken as:
\begin{equation}
\begin{aligned}
(\theta_{12},\theta_{23},\theta_{13})=&(34.6096^{\circ},49.4957^{\circ},8.4636^{\circ}),\\
(\delta,\rho,\sigma)=&(311.0930^{\circ},125.4990^{\circ},36.1443^{\circ}),\\
(m_{1},m_{2},m_{3})=&(0.0028\textrm{ eV},0.0090\textrm{ eV},0.0508\textrm{ eV}),\\
(m_{ee},m_{e})=&(0.0016\textrm{ eV},0.0093\textrm{ eV}),
\end{aligned}
\end{equation}
the corresponding neutrino mass matrix (in eV) is
\begin{equation}
M_{\nu}=\left( \begin {array}{ccc} 0.0014 + 0.0009i &  0.0038 + 0.0029i &  0.0069 - 0.0037i\\ \noalign{\medskip}0.0038 + 0.0029i  & 0.0270 - 0.0003i &  0.0270 - 0.0003i
\\\noalign{\medskip} 0.0069 - 0.0037i & 0.0270 - 0.0003i &  0.0176 + 0.0012i\end {array} \right).
\end{equation}
For inverted ordering, the representative point is taken as:
\begin{equation}
\begin{aligned}
(\theta_{12},\theta_{23},\theta_{13})=&(36.5151^{\circ},50.6039^{\circ},8.6868^{\circ}),\\
(\delta,\rho,\sigma)=&(347.0386^{\circ},159.9094^{\circ},69.0954^{\circ}),\\
(m_{1},m_{2},m_{3})=&(0.1543\textrm{ eV},0.1545\textrm{ eV},0.1461\textrm{ eV}),\\
(m_{ee},m_{e})=&(0.0466\textrm{ eV},0.1542\textrm{ eV}),
\end{aligned}
\end{equation}
the corresponding neutrino mass matrix (in eV) is
\begin{equation}
M_{\nu}=\left( \begin {array}{ccc} 0.0379 - 0.0272i & -0.0690 + 0.0467i &  0.1100 - 0.0504i\\ \noalign{\medskip}-0.0690 + 0.0467i &  0.0877 - 0.0051i &  0.0877 - 0.0051i
\\\noalign{\medskip} 0.1100 - 0.0504i &  0.0877 - 0.0051i &  0.0128 + 0.0183i\end {array} \right).
\end{equation}

From Table (\ref{numerical1}), we see that the $C_2$ pattern is not viable at the 1-$\sigma$ level for normal ordering as well as 1-2-$\sigma$ levels for inverted ordering. We also find that $\theta_{23}$ is strongly restricted to the ranges $[49.98^{\circ},50.71^{\circ}]([48.15^{\circ},51.33^{\circ}])$ in normal type at the 2(3)-$\sigma$ levels. However, there exists a tight allowed range $[36.42^o,37.39^o]([50.14^o,51.24^o])$ for $\theta_{12}$ ($\theta_{23}$) at the 3-$\sigma$ level in inverted type. For the Dirac phase $\delta$, one notes a narrow forbidden gap $[181.41^{\circ},189.39^{\circ}]$ at the 2-$\sigma$ level in normal type, whereas it is strongly bounded to lie in the interval $[325.11^{\circ},352.72^{\circ}]$ in inverted type. For inverted ordering, a large forbidden gap $[13.86^{\circ},136.29^{\circ}]$ exists for $\rho$, in addition to a tight restriction imposed on the phase $\sigma$. Table \ref{numerical1} also reveals that $m_{1}$ approaches a vanishing value at the 2-3-$\sigma$ levels in normal type. Therefore, the singular mass matrix is predicted at the 2-3-$\sigma$ levels in the case of normal ordering. %The allowed values of J parameter are negative in inverted type, so the corresponding $\delta$ lies in the third or fourth quarters.

For normal ordering plots, we see a strong linear relation for the correlation $(\sigma,\delta)$, besides narrow forbidden ribbons for the correlations $(\rho,\sigma)$ and $(\rho,\delta)$. We also see large disallowed regions for the plots including $\theta_{23}$, but, unlike the $C_1$ case, it can not be in the first octant. The correlations between ($\delta$,$m_{ee}$) and between ($\delta$,LNM) show that when $m_{ee}$ and LNM increase, the allowed parameter space becomes more restricted. We find a moderate mass hierarchy characterized by $0.16\leq\frac{m_2}{m_3}\leq0.77$ as well as a severe mass hierarchy where $\frac{m_2}{m_1}$ can reach $10^3$ indicating the possibility of vanishing $m_1$.

As to the inverted ordering plots, we notice that the allowed parameter space is very limited. We find a wide forbidden gap for $\rho$, and that $\theta_{23}$ is again restricted to fall in the second octant. A quasi degeneracy characterized by $m_1\approx m_2\approx m_3$ is also found.

\begin{figure}[hbtp]
\centering
\begin{minipage}[l]{0.5\textwidth}
\epsfxsize=24cm
\centerline{\epsfbox{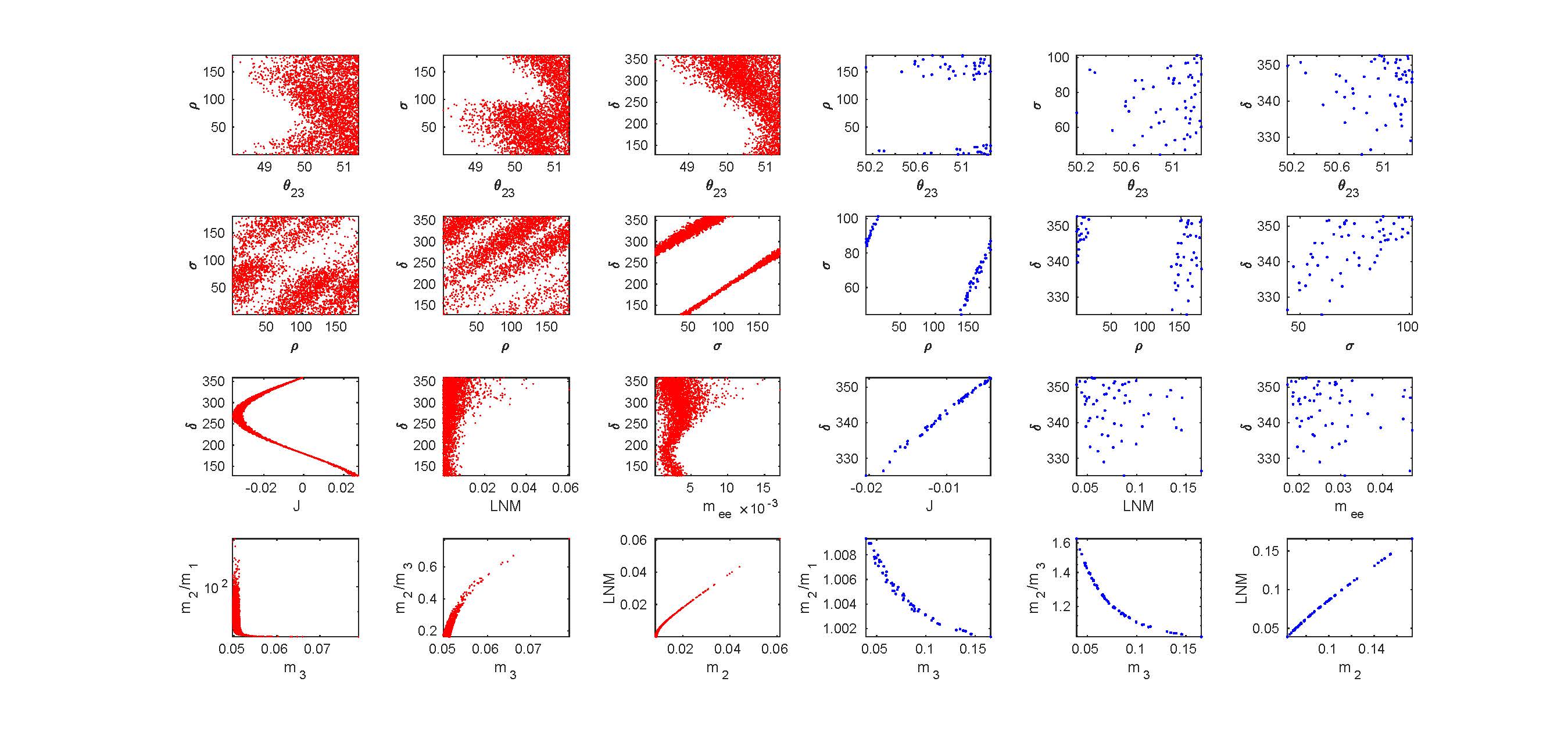}}
\end{minipage}%
\caption{The correlation plots for $C_{2}$ pattern: the red (blue) plots represent the normal (inverted) ordering correlations. The first row represents the correlations between the mixing angle $\theta_{23}$ and the CP-violating phases.
The second row represents the correlations amidst the CP-violating phases, the third one represents the correlations between $\delta$ and each of  J, LNM (the lowest neutrino mass), $m_{ee}$ parameters. The last one shows the degree of hierarchy. Angles (masses) are evaluated in degrees (eV).}
\label{C2fig}
\end{figure}

Another remark here concerns the consistency of the T-symmetry (Eq. \ref{mutau}) with the numerical results of Tables (\ref{TableLisi:as},\ref{numerical1},\ref{numerical2}), taking into consideration that the experimental data of Table (\ref{TableLisi:as}) are not invariant by T-symmetry. Restricting to the two dimensions $(\theta_{23}, \delta)$ under study, let us denote by $(M,M')$ the two sets of admissible points corresponding to two T-symmetry-related textures, and by $(E)$ the experimentally allowed region. Let us denote by $(\Pi_{1(2)})$ the projection onto the $\theta_{23} (\delta)$-axis, and by $(T=T_1 \times T_2)$ the T-symmetry transformation:($T_1: \theta_{23} \rightarrow \pi/2 - \theta_{23}, T_2: \delta \rightarrow \delta \pm \pi $) . The consistency condition gives:
\bea \label{ccondition} T(M) \cap E \subseteq M'\eea
which leads to the `necessary' (non-sufficient) condition \bea \label{ccondition_proj} \Pi_{k} \left( T(M) \cap E \right)  \subseteq \Pi_{k} \left( M' \right), k=1,2 \eea
From ($\Pi_{k}  \left( T(M) \cap E \right) \subseteq \Pi_{k}(T(M))  \cap   \Pi_{k}(E)  $) and ($\Pi_k T = T_k \Pi_k$), one can test the stronger `sufficient' (non-necessary) condition:
\bea \label{ccondition_test}  T_k(\Pi_{k}(M))  \cap   \Pi_{k}(E)  \subseteq \Pi_{k} \left( M' \right), k=1,2 \eea

   For illustration, we can take here the two textures $C_1$ and $C_2$ related by T-symmetry. Suppose an interval $I \equiv \Pi_1(C) (J \equiv \Pi_2(C))$ is covered by the values of $\theta_{23} (\delta)$ of the admissible points of the texture $C$, then denoting by $TI \equiv T_1(I) (TJ\equiv T_2(J))$  the T-symmetry transform of $I(J)$, we see from the condition (\ref{ccondition_test}) that if the first (second) projection of the admissible points of $C'$  contains $TI \cap \theta_{23}^{exp} (TJ \cap \delta^{exp})$, where $\theta_{23}^{exp} (\delta^{exp})$ represents the experimental data allowed for $\theta_{23} (\delta)$, then the `necessary' consistency condition (\ref{ccondition_proj}) is met.  Table (\ref{C1C2cosistency}) illustrates testing the condition (\ref{ccondition_test} for the two textures ($C_1, C_2$).

\newpage
\begin{table}[h]
\centering
\scalebox{0.8}{
\begin{tabular}{ccccc}
\toprule
Parameter/Hierarchy &Texture & $1 \sigma$ & $2 \sigma$ & $3 \sigma$ \\
\toprule
\midrule
\multirow{7}{*}
{$\theta_{23}^{NH}$ ($^{\circ}$)}   & $\theta_{23}^{exp,NH}$ & [48.47,50.05]& [47.37,50.71] & [41.20,51.33] \\
&$I_{C_1}^{NH}$ & $\phi$  & $[50.06,50.07]$ & $[41.20,42.15]\cup [49.53,51.33]$   \\
&$TI_{C_1}^{NH}$ & $\phi$  & $[39.3,39.94]$ & $[47.85,48.8] \cup [38.67,40.47]$   \\
& $TI_{C_1}^{NH}\cap \theta_{23}^{exp,NH}$ & $\phi  \subseteq  I_{C_2}^{NH}$  & $\phi \subseteq  I_{C_2}^{NH}$ &  $ [47.85,48.8] \precapprox  I_{C_2}^{NH}$   \\
& $I^{NH}_{C_2}$& $\phi$  & $[49.98,50.00] \cup [50.09,50.71]$ & [48.15,51.33]   \\
&$TI^{NH}_{C_2}$ & $\phi$  & $[39.29,39.91] \cup [40.0,40.02]$   & [38.67,41.85]   \\
& $ TI^{NH}_{C_2}\cap \theta_{23}^{exp,NH}$ & $\phi \subseteq  I_{C_1}^{NH}$  & $\phi \subseteq  I_{C_1}^{NH}$   & $[41.20,41.85] \subseteq  I_{C_1}^{NH}$   \\
\midrule
\multirow{7}{*}
 {$\theta_{23}^{IH}$ ($^{\circ}$)} & $\theta_{23}^{exp,IH}$ & [48.49,50.06]& [47.35,50.67] & [41.16,51.25] \\
&$I_{C_1}^{IH}$ & $\phi$  & $\phi$ & $\phi$   \\
&$TI_{C_1}^{IH}$ & $\phi$  & $\phi$ & $\phi$    \\
& $TI_{C_1}^{IH}\cap \theta_{23}^{exp,IH}$ & $\phi  \subseteq  I_{C_2}^{IH}$  & $\phi \subseteq  I_{C_2}^{IH}$ &  $ \phi \subseteq  I_{C_2}^{IH}$   \\
& $I^{IH}_{C_2}$& $\phi$  & $\phi$ & [50.14,51.24]   \\
&$TI^{IH}_{C_2}$ & $\phi$  & $\phi$   & [38.76,39.86]   \\
& $ TI^{IH}_{C_2}\cap \theta_{23}^{exp,IH}$ & $\phi \subseteq  I_{C_1}^{IH}$  & $\phi \subseteq  I_{C_1}^{IH}$   &  $ \phi \subseteq  I_{C_1}^{IH}$   \\
\midrule
\midrule
\multirow{7}{*}
{$\delta^{NH}$ ($^{\circ}$)}  & $\delta^{exp,NH}$ & [172.00,218.00]& [152.00,255.00] & [128.00,359.00] \\
&$J_{C_1}^{NH}$ & $\phi$  & $[152.02,227.99]$ & $[128.05,257.64]$   \\
&$TJ_{C_1}^{NH}$ & $\phi$  & $[0,47.99] \cup [332.02,360]$ & $[0,77.64] \cup [308.05,360.00]$   \\
& $TJ_{C_1}^{NH}\cap \delta^{exp,NH}$ & $\phi  \subseteq  J_{C_2}^{NH}$  & $\phi \subseteq  J_{C_2}^{NH}$ &  $ [308.05,359.00] \subseteq  J_{C_2}^{NH}$   \\
& $J^{NH}_{C_2}$& $\phi$  & $[152.02,181.40] \cup [189.40,254.99]$ & [128.05,359.00]   \\
&$TJ^{NH}_{C_2}$ & $\phi$  & $[0,1.4] \cup [332.02,360] \cup [9.4,74.99]$  & $[0.0,179.0] \cup [308.05,360]$   \\
& $ TJ^{NH}_{C_2}\cap \delta^{exp,NH}$ & $\phi \subseteq  J_{C_1}^{NH}$  & $\phi \subseteq  J_{C_1}^{NH}$   &  $ [128,179]\cup [308.05,359] \nsubseteq  J_{C_1}^{NH}$   \\
\midrule
 \multirow{7}{*}
 {$\delta^{IH}$ ($^{\circ}$)} & $\delta^{exp,IH}$ & [256.00,310.00]& [226.00,332.00] & [200.00,353.00] \\
 &$J_{C_1}^{IH}$ & $\phi$  & $\phi$ & $\phi$   \\
&$TJ_{C_1}^{IH}$ & $\phi$  & $\phi$ & $\phi$   \\
& $TJ_{C_1}^{IH}\cap \delta^{exp,IH}$ & $\phi  \subseteq  J_{C_2}^{IH}$  & $\phi \subseteq  J_{C_2}^{IH}$ &  $\phi \subseteq  J_{C_2}^{IH}$   \\
& $J^{IH}_{C_2}$& $\phi$  & $\phi$ & [325.11,352.72]   \\
&$TJ^{IH}_{C_2}$ & $\phi$  & $\phi$   & [145.11,172.72]   \\
& $ TJ^{IH}_{C_2}\cap \delta^{exp,IH}$ & $\phi \subseteq  J_{C_1}^{IH}$  & $\phi \subseteq  J_{C_1}^{IH}$   &  $ \phi \subseteq  J_{C_1}^{IH}$   \\
\bottomrule
\end{tabular}}
\caption{\footnotesize Consistency of the $T$-symmetry ($\theta_{23} \rightarrow \pi/2 - \theta_{23}, \delta \rightarrow \delta \pm \pi $) with the numerical results of T-related textures $C_1, C_2$.
 $TI (TJ)$  represents the T-symmetry transform of $I(J)$, the range of $\theta_{23} (\delta)$ covered by admissible points in one of the textures. The intersection of $TI (TJ)$ with the experimental ranges, which do not respect T-symmetry, is contrasted to the range  $I(J)$ corresponding to the other T-related texture.}
\label{C1C2cosistency}
\end{table}

We see from Table {\ref{C1C2cosistency}}, that at $1$-$2\sigma$ levels, the consistency condition (\ref{ccondition_proj}) is met, as the stronger condition (\ref{ccondition_test}) is satisfied. In the $3\sigma$ level, the strong condition (\ref{ccondition_test}) is met for the IH type, whereas it is not satisfied for the NH type. Whereas ($TI_{C_1}^{NH} \cap \theta_{23}^{exp,NH} = [47.85,48.8]$) is `almost' included in ($I^{NH}_{C_2} = [48.15,51.33]$), where the disappearance of $[47.85,48.15]$ is actually due to numerics originating from throwing random values in the parameter space which may not cover it wholly, however the noninclusion of ($[128,179] \cap [308.05,359] $), pertaining to ($ TJ^{NH}_{C_2}\cap \delta^{exp,NH}$), in ($J_{C_1}^{NH}=[128.05,257.64] $) can be interpreted by the correlation plot ($\delta, \theta_{23}$) in the left part of Fig. (\ref{C2fig}) in the form of a triangle with `small' $\delta$-values implying `large' $\theta_{23}$-values. As a matter of fact, the failing to appear, in $J_{C_1}^{NH}$, part $[308.05,359]$ of $TJ^{NH}_{C_2}$ corresponds to ($[128,179]$) in $J_{C_2}^{NH}$, which, according to the plot ($\delta, \theta_{23}$) for NH type in Fig. (\ref{C2fig}), is correlated with ($\theta_{23} > 50.4$) which in turn is T-symmetry-related to the experimentally excluded range ($\theta_{23} < 39.6$).

\newpage

\subsection{Pattern $D_1$: $M_{\nu~13}=M_{\nu~22}$}
The expressions for $A_{1}$, $A_{2}$ and $A_{3}$ coefficients for this pattern are
\begin{equation}
\begin{aligned}
A_{1}=&c_{12}c_{13}(-c_{12}c_{23}s_{13}+s_{12}s_{23}e^{-i\delta})-(-c_{12}s_{23}s_{13}-s_{12}c_{23}e^{-i\delta})^{2},\\
A_{2}=&s_{12}c_{13}(-s_{12}c_{23}s_{13}-c_{12}s_{23}e^{-i\delta})-(-s_{12}s_{23}s_{13}+c_{12}c_{23}e^{-i\delta})^{2},\\
A_{3}=&s_{13}c_{23}c_{13}-s_{23}^{2}c_{13}^{2}.
\end{aligned}\label{D1coff}
\end{equation}
The leading order approximation in powers of $s_{13}$ for the mass ratios are given by
\begin{align}
\frac{m_1}{m_3}\approx&-\frac{-\tan^2\theta_{23}\big[\sin(2\sigma-\delta)+\cot\theta_{12}\cot\theta_{23}\cos\theta_{23}\sin(2\sigma-2\delta)\big]}{\sin\delta\cos(2\rho-2\sigma)+\big[\cos2\theta_{12}\cos\delta+\frac{\cos\theta_{12}\sin\theta_{12}\tan\theta_{23}}{\cos\theta_{23}}(1-\cos^2\theta_{23}\cot^2\theta_{23})\big]},\nonumber \\
\frac{m_2}{m_3}\approx&-\frac{-\tan^2\theta_{23}\big[\sin(2\rho-\delta)-\tan\theta_{12}\cot\theta_{23}\cos\theta_{23}\sin(2\rho-2\delta)\big]}{\sin\delta\cos(2\rho-2\sigma)+\big[\cos2\theta_{12}\cos\delta+\frac{\cos\theta_{12}\sin\theta_{12}\tan\theta_{23}}{\cos\theta_{23}}(1-\cos^2\theta_{23}\cot^2\theta_{23})\big]}\label{D1ratio}
, \end{align}
from which
\bea
1 < \frac{m_2}{m_1} &\approx& \frac{\sin(2\rho-\delta)-\tan\theta_{12}\cot\theta_{23}\cos\theta_{23}\sin(2\rho-2\delta)}{\sin(2\sigma-\delta)+\cot\theta_{12}\cot\theta_{23}\cos\theta_{23}\sin(2\sigma-2\delta)}
, \eea
so again fixing the mixing angles around their experimentally central values leads to an inequality involving the phases which can be checked to be met in the corresponding correlation plots.

For normal ordering, the representative point is taken as:
\begin{equation}
\begin{aligned}
(\theta_{12},\theta_{23},\theta_{13})=&(34.7345^{\circ},47.7461^{\circ},8.5536^{\circ}),\\
(\delta,\rho,\sigma)=&(200.2544^{\circ},107.0959^{\circ},101.3076^{\circ}),\\
(m_{1},m_{2},m_{3})=&(0.0757\textrm{ eV},0.0762\textrm{ eV},0.0909\textrm{ eV}),\\
(m_{ee},m_{e})=&(0.0722\textrm{ eV},0.0763\textrm{ eV}),
\end{aligned}
\end{equation}
the corresponding neutrino mass matrix (in eV) is
\begin{equation}
M_{\nu}=\left( \begin {array}{ccc} -0.0617 - 0.0374i&0.0178 - 0.0005i&0.0145 + 0.0090i\\ \noalign{\medskip}0.0178 - 0.0005i&0.0145 + 0.0090i&0.0801 - 0.0097i
\\ \noalign{\medskip}0.0145 + 0.0090i&0.0801 - 0.0097i&-0.0005 + 0.0087i\end {array} \right).
\end{equation}
For inverted ordering, the representative point is taken as:
\begin{equation}
\begin{aligned}
(\theta_{12},\theta_{23},\theta_{13})=&(34.6393^{\circ},49.7237^{\circ},8.6990^{\circ}),\\
(\delta,\rho,\sigma)=&(283.8818^{\circ},35.28160^{\circ},23.1221^{\circ}),\\
(m_{1},m_{2},m_{3})=&(0.0572\textrm{ eV},0.0579\textrm{ eV},0.0288\textrm{ eV}),\\
(m_{ee},m_{e})=&(0.0553\textrm{ eV},0.0570\textrm{ eV}),
\end{aligned}
\end{equation}
the corresponding neutrino mass matrix (in eV) is
\begin{equation}
M_{\nu}=\left( \begin {array}{ccc} 0.0259 + 0.0489i&0.0054 - 0.0005i & -0.0057 - 0.0110i\\ \noalign{\medskip}0.0054 - 0.0005i&  -0.0057 - 0.0110i &  0.0394 + 0.0131i
\\ \noalign{\medskip}-0.0057 - 0.0110i &  0.0394 + 0.0131i & -0.0163 - 0.0128i\end {array} \right).
\end{equation}
\newpage
We see from Table (\ref{numerical1}) that the allowed experimental ranges for the mixing angles ($\theta_{12}$,$\theta_{23}$,$\theta_{13}$) can be covered for both normal and inverted orderings at all $\sigma$-levels. For normal ordering, we find that $\rho$ is strongly restricted to fall in the interval $[82.42^{\circ},119.94^{\circ}] ([66.45^{\circ},152.53^{\circ}])$ at the 1(2)-$\sigma$ level. We also see narrow forbidden gaps for the phase $\sigma$ at the 1-2-$\sigma$ levels. For inverted ordering, we see a wide disallowed region $[316.78^{\circ},353^{\circ}]$ for $\delta$ at the 3-$\sigma$ level. The narrow forbidden gaps are also noticed for the phases $\rho$ and $\sigma$ at the 1-$\sigma$ level.  Table (\ref{numerical1}) also shows that $m_{1}$ does not reach zero value at all $\sigma$-levels, whereas $m_{3}$ can reach a vanishing value at all $\sigma$-levels. Thus, the singular mass matrix is predicted in inverted type at all $\sigma$-levels. %The allowed values of J at all $\sigma$-levels for inverted ordering are negative, so the corresponding $\delta$ lies in the third or fourth quarters.

For normal ordering plots, we find a mild mass hierarchy where $(0.45\leq\frac{m_{2}}{m_{3}}\leq0.97)$ together with a quasi degeneracy characterized by $m_{1}\approx m_{2}$.

As to the inverted plots, we get an acute mass hierarchy where $\frac{m_{2}}{m_{3}}$ can reach $10^{3}$ besides a quasi degeneracy where $(1 < \frac{m_{2}}{m_{1}}\leq1.01)$.

For both normal and inverted ordering, the correlations between ($\delta$,$m_{ee}$) and between ($\delta$,LNM) show that when $m_{ee}$ and LNM increase, the allowed parameter space becomes more restricted.
\begin{figure}[hbtp]
\centering
\begin{minipage}[l]{0.5\textwidth}
\epsfxsize=24.33cm
\centerline{\epsfbox{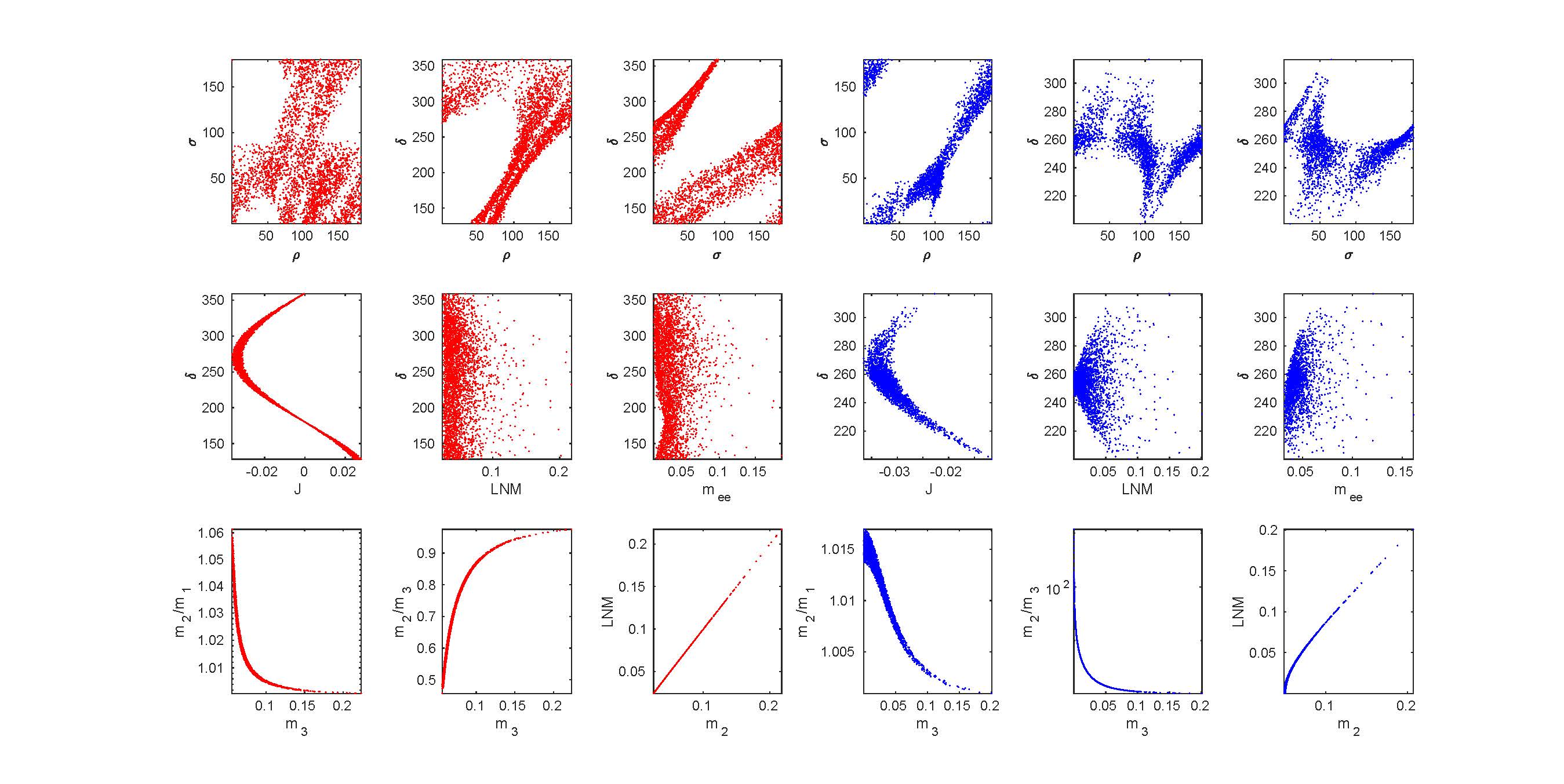}}
\end{minipage}%
\caption{The correlation plots for $D_1$ pattern, the red (blue) plots represent the normal (inverted) ordering. The first row represents the correlations between the CP-violating phases, the second one represent the correlations between $\delta$ and each of J, LNM (the lowest neutrino mass) and $m_{ee}$ parameters. The last one shows the degree of hierarchy. Angles (masses) are evaluated in degrees (eV).}
\label{D1fig}
\end{figure}
\newpage

\subsection{Pattern $D_2$: $M_{\nu~12}=M_{\nu~33}$}
The expressions for $A's$ coefficients and the approximate analytical formulas for the mass ratios can be obtained from Eqs. (\ref{D1coff},\ref{D1ratio})by using the transformations in Eq. \ref{mutau}.

For normal ordering, the representative point is taken as:
\begin{equation}
\begin{aligned}
(\theta_{12},\theta_{23},\theta_{13})=&(34.1052^{\circ},49.4172^{\circ},8.5161^{\circ}),\\
(\delta,\rho,\sigma)=&(213.1118^{\circ},175.2291^{\circ},114.2297^{\circ}),\\
(m_{1},m_{2},m_{3})=&(0.0242\textrm{ eV},0.0257\textrm{ eV},0.0565\textrm{ eV}),\\
(m_{ee},m_{e})=&(0.0148\textrm{ eV},0.0258\textrm{ eV}),
\end{aligned}
\end{equation}
the corresponding neutrino mass matrix (in eV) is
\begin{equation}
M_{\nu}=\left( \begin {array}{ccc} 0.0120 - 0.0086i &  0.0178 - 0.0019i & -0.0105 + 0.0042i\\ \noalign{\medskip}0.0178 - 0.0019i  & 0.0228 - 0.0003i &  0.0352 + 0.0008i
\\ \noalign{\medskip}-0.0105 + 0.0042i &  0.0352 + 0.0008i &  0.0178 - 0.0019i
\end {array} \right).
\end{equation}
For inverted ordering, the representative point is taken as:
\begin{equation}
\begin{aligned}
(\theta_{12},\theta_{23},\theta_{13})=&(34.3488^{\circ},49.4372^{\circ},8.5251^{\circ}),\\
(\delta,\rho,\sigma)=&(282.4787^{\circ},140.0014^{\circ},1.9822^{\circ}),\\
(m_{1},m_{2},m_{3})=&(0.0506\textrm{ eV},0.0514\textrm{ eV},0.0127\textrm{ eV}),\\
(m_{ee},m_{e})=&(0.0390\textrm{ eV},0.0504\textrm{ eV}),
\end{aligned}
\end{equation}
the corresponding neutrino mass matrix (in eV) is
\begin{equation}
M_{\nu}=\left( \begin {array}{ccc} 0.0221 - 0.0321i & -0.0139 + 0.0196i &  0.0141 - 0.0154i\\ \noalign{\medskip}-0.0139 + 0.0196i & -0.0017 + 0.0079i &  0.0200 - 0.0137i
\\ \noalign{\medskip}0.0141 - 0.0154i &  0.0200 - 0.0137i & -0.0139 + 0.0196i\end {array} \right).
\end{equation}

We see from Table (\ref{numerical1}) that the allowed experimental ranges for the mixing angles ($\theta_{12},\theta_{23},\theta_{13}$) can be covered at all $\sigma$ levels with either hierarchy type. We find that the Dirac phase $\delta$ is restricted at the 2-3-$\sigma$ levels in the case of inverted ordering. For normal ordering, we notice that the phase $\sigma$ is bounded to the intervals: $[82.32^{\circ},122.38^{\circ}]$ at the 1-$\sigma$ level and $[66.96^{\circ},161.38^{\circ}]$ at the 2-$\sigma$ level. Table (\ref{numerical1}) also shows that $m_{1}$ does not approach a vanishing value in normal type. However, $m_{3}$ reaches a zero value in inverted type at all $\sigma$ error levels. Thus, the singular mass matrix is predicted for inverted ordering at all $\sigma$ levels. %The allowed values of J are negative in inverted type at all $\sigma$ levels, so the corresponding $\delta$ lies in the third or fourth quarters.

For normal ordering plots, we find a mild mass hierarchy where $(0.37\leq\frac{m_{2}}{m_{3}}\leq0.97)$ together with a quasi degeneracy characterized by $m_{1}\approx m_{2}$.

As to the inverted plots, we get an acute mass hierarchy where $\frac{m_{2}}{m_{3}}$ can reach $10^{4}$ besides a quasi degeneracy where $(1 < \frac{m_{2}}{m_{1}}\leq1.015)$.

For both normal and inverted ordering, the correlations between ($\delta$,$m_{ee}$) and between ($\delta$,LNM) show that when $m_{ee}$ and LNM increase, the allowed parameter space becomes more restricted.
\begin{figure}[hbtp]
\centering
\begin{minipage}[l]{0.5\textwidth}
\epsfxsize=24cm
\centerline{\epsfbox{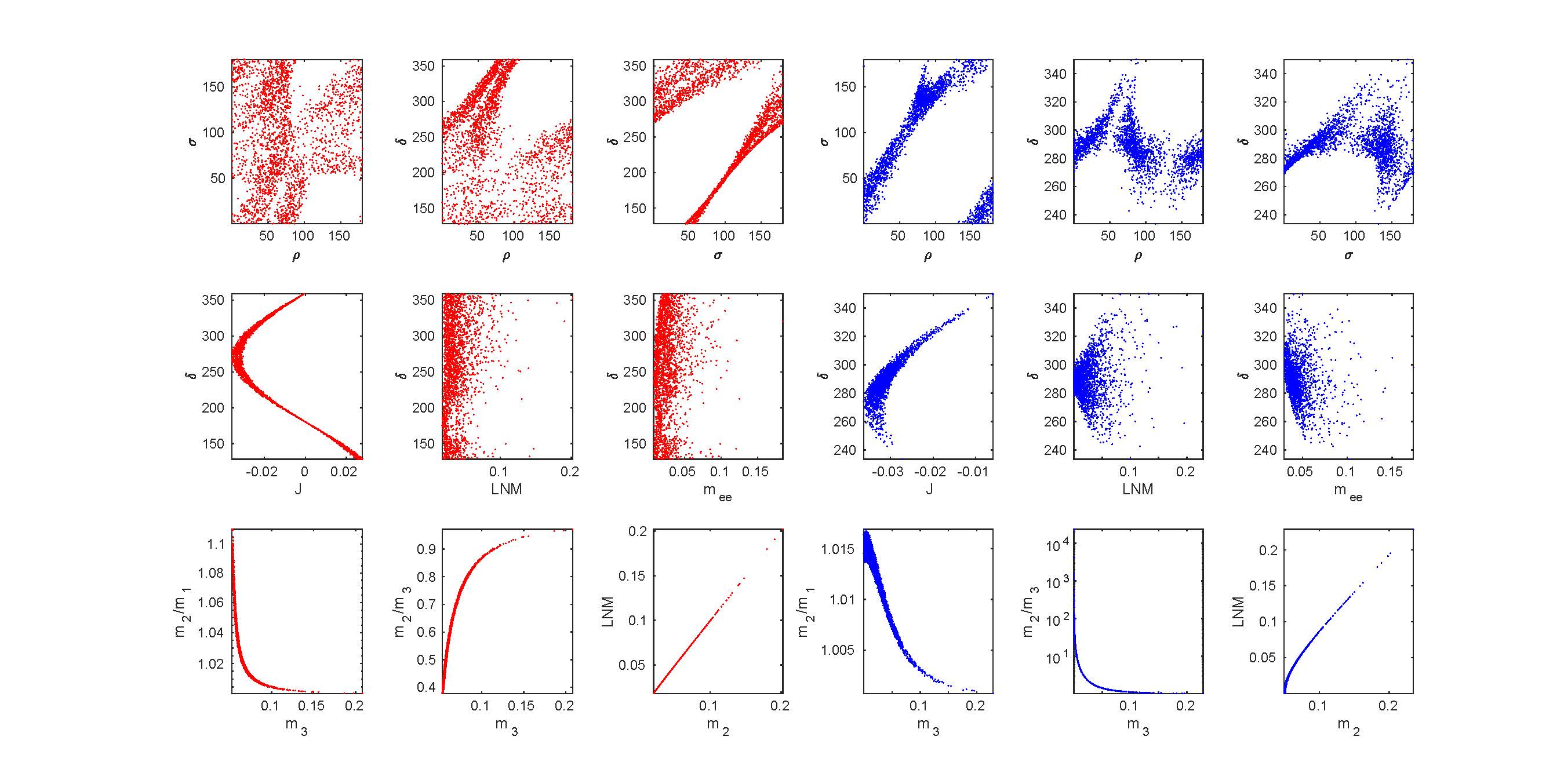}}
\end{minipage}%
\caption{The correlation plots for $D_2$ pattern, the red (blue) plots represent the normal (inverted) ordering. The first row represents the correlations between the CP-violating phases, the second one represent the correlations between $\delta$ and each of J, LNM (the lowest neutrino mass) and $m_{ee}$ parameters. The last one shows the degree of hierarchy. Angles (masses) are evaluated in degrees (eV).}
\label{D1fig}
\end{figure}
\newpage

\subsection{Pattern $E_1$: $M_{\nu~11}=M_{\nu~22}$}
The expressions for $A_{1}$, $A_{2}$ and $A_{3}$ coefficients for this pattern are
\begin{equation}
\begin{aligned}
A_{1}=&c_{12}^{2}c_{13}^{2}-(-c_{12}s_{23}s_{13}-s_{12}c_{23}e^{-i\delta})^{2},\\
A_{2}=&s_{12}^{2}c_{13}^{2}-(-s_{12}s_{23}s_{13}+c_{12}c_{23}e^{-i\delta})^{2},\\
A_{3}=&s_{13}^{2}-s_{23}^{2}c_{13}^{2}.
\end{aligned}\label{E1coff}
\end{equation}
The leading order approximation in powers of $s_{13}$ for the mass ratios are given by
\begin{align}
\hspace{-6em}\frac{m_1}{m_3}\approx&\frac{\sin^2\theta_{23}\big[\sin^2\theta_{12}\sin2\sigma-\cos^2\theta_{12}\cos^2\theta_{23}\sin(2\sigma-2\delta)\big]}{\cos2\theta_{12}\cos^2\theta_{23}\sin2\delta\cos(2\rho-2\sigma)-\big[\cos^2\theta_{12}\sin^2\theta_{12}(1+\cos^4\theta_{23})+\frac{1}{2}(\sin^22\theta_{12}-2)\cos^2\theta_{23}\cos2\delta\big]\sin(2\rho-2\sigma)},\nonumber\\
\hspace{-6em}\frac{m_2}{m_3}\approx&\frac{\sin^2\theta_{23}\big[-\cos^2\theta_{12}\sin2\rho+\sin^2\theta_{12}\cos^2\theta_{23}\sin(2\rho-2\delta)\big]}{\cos2\theta_{12}\cos^2\theta_{23}\sin2\delta\cos(2\rho-2\sigma)-\big[\cos^2\theta_{12}\sin^2\theta_{12}(1+\cos^4\theta_{23})+\frac{1}{2}(\sin^22\theta_{12}-2)\cos^2\theta_{23}\cos2\delta\big]\sin(2\rho-2\sigma)},\label{E1ratio}
\end{align}
from which we have
\bea
1 < \frac{m_2}{m_1} &\approx& \frac{-\cos^2\theta_{12}\sin2\rho+\sin^2\theta_{12}\cos^2\theta_{23}\sin(2\rho-2\delta)}{\sin^2\theta_{12}\sin2\sigma-\cos^2\theta_{12}\cos^2\theta_{23}\sin(2\sigma-2\delta)}
.\eea
Again, one can check that this inequality is satisfied for the phase angles fixing the mixing angles at their central experimental values.

For normal ordering, the representative point is taken as:
\begin{equation}
\begin{aligned}
(\theta_{12},\theta_{23},\theta_{13})=&(34.3021^{\circ},49.3949^{\circ},8.4496^{\circ}),\\
(\delta,\rho,\sigma)=&(213.8835^{\circ},3.2180^{\circ},133.5755^{\circ}),\\
(m_{1},m_{2},m_{3})=&(0.0351\textrm{ eV},0.0361\textrm{ eV},0.0618\textrm{ eV}),\\
(m_{ee},m_{e})=&(0.0255\textrm{ eV},0.0362\textrm{ eV}),
\end{aligned}
\end{equation}
the corresponding neutrino mass matrix (in eV) is
\begin{equation}
M_{\nu}=\left( \begin {array}{ccc} 0.0240 - 0.0086i &  0.0201 + 0.0048i & -0.0148 - 0.0036i\\ \noalign{\medskip}0.0201 + 0.0048i&   0.0240 - 0.0086i &  0.0395 + 0.0089i
\\ \noalign{\medskip}-0.0148 - 0.0036i &  0.0395 + 0.0089i  & 0.0191 - 0.0096i\end {array} \right).
\end{equation}
For inverted ordering, the representative point is taken as:
\begin{equation}
\begin{aligned}
(\theta_{12},\theta_{23},\theta_{13})=&(34.7416^{\circ},49.4454^{\circ},8.5335^{\circ}),\\
(\delta,\rho,\sigma)=&(284.2698^{\circ},23.0607^{\circ},118.3590^{\circ}),\\
(m_{1},m_{2},m_{3})=&(0.0556\textrm{ eV},0.0563\textrm{ eV},0.0243 \textrm{ eV}),\\
(m_{ee},m_{e})=&(0.0199 \textrm{ eV},0.0554\textrm{ eV}),
\end{aligned}
\end{equation}
the corresponding neutrino mass matrix (in eV) is
\begin{equation}
M_{\nu}=\left( \begin {array}{ccc} 0.0162 + 0.0115i  & 0.0212 - 0.0281i & -0.0229 + 0.0301i\\ \noalign{\medskip}0.0212 - 0.0281i  & 0.0162 + 0.0115i &  0.0046 - 0.0070i
\\ \noalign{\medskip}-0.0229 + 0.0301i &  0.0046 - 0.0070i &  0.0242 + 0.0012i\end {array} \right).
\end{equation}
\newpage
We see from Table (\ref{numerical2}), the mixing angles ($\theta_{12}$,$\theta_{23}$,$\theta_{13}$) extend over their allowed experimental ranges for both normal and inverted ordering at all $\sigma$-levels. We notice a persistent wide forbidden gap  for the phase $\rho$ with each hierarchy type at all $\sigma$-levels. For inverted ordering, we find that $\sigma$ is restricted at all $\sigma$-levels ( $[36.32^{\circ},150.39^{\circ}]$, $[26.78^{\circ},158.93^{\circ}]$ and $[16.70^{\circ},164.72^{\circ}]$ at the 1,2 and 3-$\sigma$ levels respectively). Table   (\ref{numerical2}) also reveals that neither $m_{1}$ in normal type nor $m_{3}$ in inverted type can reach a vanishing value. Thus, the singular pattern is not predicted at any $\sigma$-level. %The allowed values of J at all $\sigma$-levels for inverted ordering are negative, so the corresponding $\delta$ lies in the third or fourth quarters.

For normal ordering plots, a large forbidden gap exists for $\rho$. We see a moderate mass hierarchy where $(0.29\leq\frac{m_{2}}{m_{3}}\leq0.97)$ and a quasi degenerate mass spectrum characterized by $(1 < \frac{m_{2}}{m_{1}}\leq1.22)$.

As to inverted ordering plots, we see a wide forbidden gap for the phase $\rho$ as in normal type. We also find that $\sigma$ is bound to the interval $[16^{\circ},164^{\circ}]$. One notes a quasi degenerate mass spectrum characterized by $(1\leq\frac{m_{2}}{m_{1}}\leq1.015)$ and a mild mass hierarchy where $(1.02\leq\frac{m_{2}}{m_{3}}\leq4.41)$.

For both normal and inverted ordering, the correlations between ($\delta$,$m_{ee}$) and between ($\delta$,LNM) show that when $m_{ee}$ and LNM increase, the allowed parameter space becomes more restricted.
\begin{figure}[hbtp]
\centering
\begin{minipage}[l]{0.5\textwidth}
\epsfxsize=24.33cm
\centerline{\epsfbox{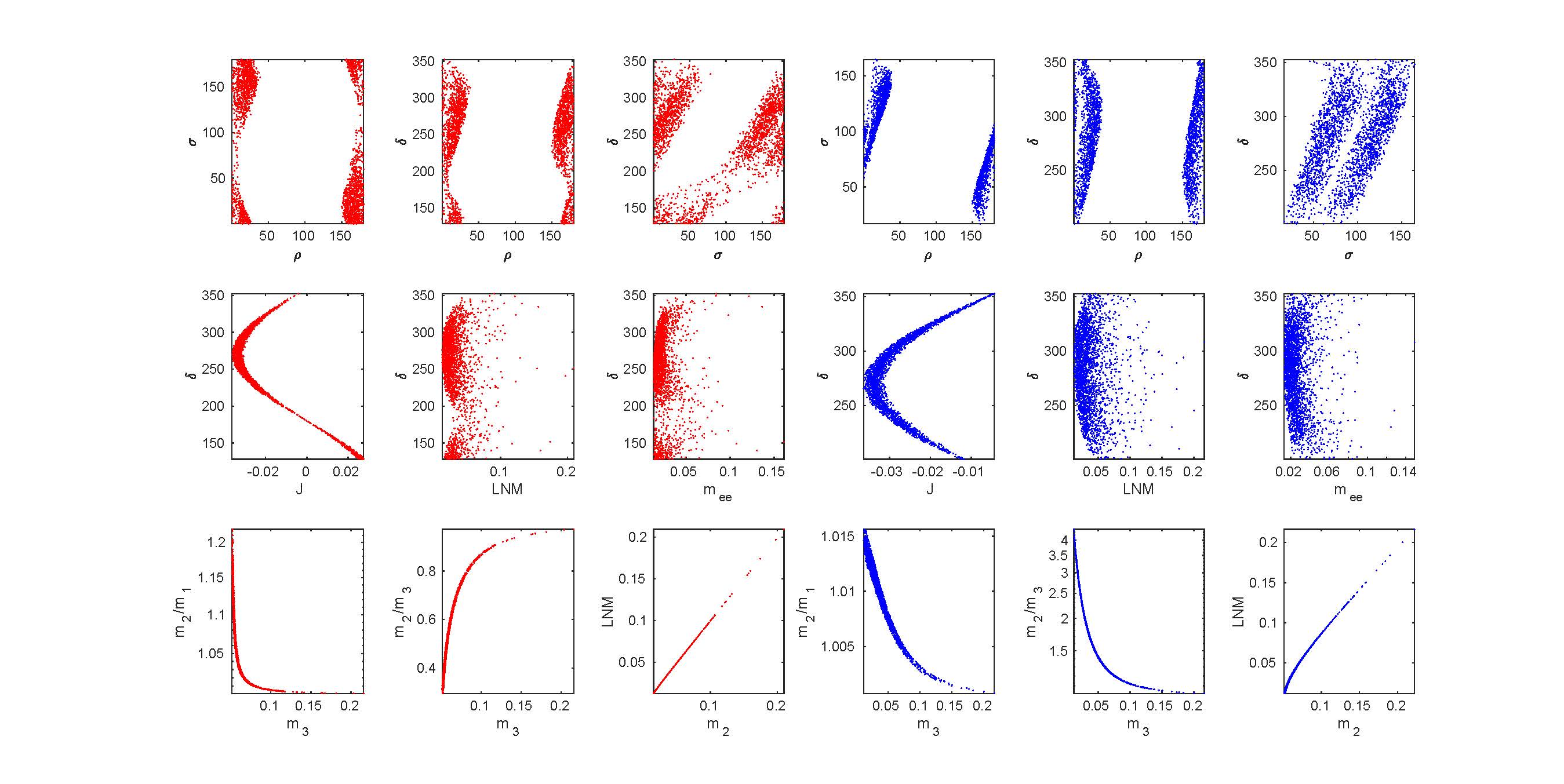}}
\end{minipage}%
\caption{ The correlation plots for $E_1$ pattern, the red (blue) plots represents the normal (inverted) ordering. The first row represents the correlations between the CP-violating phases, the second one represent the correlations between $\delta$ and each of J, LNM (the lowest neutrino mass) and $m_{ee}$ parameters. The last one shows the degree of hierarchy. Angles (masses) are evaluated in degrees (eV).}
\label{E1fig}
\end{figure}
\newpage

\subsection{Pattern $E_2$: $M_{\nu~11}=M_{\nu~33}$}
The expressions for $A's$ coefficients and the approximate analytical formulas for the mass ratios can be obtained from Eqs. (\ref{E1coff},\ref{E1ratio})by using the transformations in Eq. \ref{mutau}.

For normal ordering, the representative point is taken as:
\begin{equation}
\begin{aligned}
(\theta_{12},\theta_{23},\theta_{13})=&(34.2867^{\circ},49.4424^{\circ},8.4681^{\circ}),\\
(\delta,\rho,\sigma)=&(248.3912^{\circ},146.1343^{\circ},27.1265^{\circ}),\\
(m_{1},m_{2},m_{3})=&(0.0372\textrm{ eV},0.0382\textrm{ eV},0.0632\textrm{ eV}),\\
(m_{ee},m_{e})=&(0.0222\textrm{ eV},0.0383\textrm{ eV}),
\end{aligned}
\end{equation}
the corresponding neutrino mass matrix (in eV) is
\begin{equation}
M_{\nu}=\left( \begin {array}{ccc} 0.0177 - 0.0133i & -0.0140 - 0.0034i &  0.0267 + 0.0070i\\ \noalign{\medskip}-0.0140 - 0.0034i & 0.0371 - 0.0079i &  0.0337 + 0.0100i
\\ \noalign{\medskip}0.0267 + 0.0070i &  0.0337 + 0.0100i &  0.0177 - 0.0133i\end {array} \right).
\end{equation}
For inverted ordering, the representative point is taken as:
\begin{equation}
\begin{aligned}
(\theta_{12},\theta_{23},\theta_{13})=&(34.4883^{\circ},49.1484^{\circ},8.5615^{\circ}),\\
(\delta,\rho,\sigma)=&(285.2591^{\circ},162.4256^{\circ},77.9786^{\circ}),\\
(m_{1},m_{2},m_{3})=&(0.0569\textrm{ eV},0.0575\textrm{ eV},0.0281 \textrm{ eV}),\\
(m_{ee},m_{e})=&(0.0208 \textrm{ eV},0.0566\textrm{ eV}),
\end{aligned}
\end{equation}
the corresponding neutrino mass matrix (in eV) is
\begin{equation}
M_{\nu}=\left( \begin {array}{ccc} 0.0150 - 0.0144i & -0.0227 - 0.0227i &  0.0293 + 0.0296i\\ \noalign{\medskip}-0.0227 - 0.0227i &  0.0279 - 0.0011i &  0.0055 + 0.0066i
\\ \noalign{\medskip}0.0293 + 0.0296i &  0.0055 + 0.0066i &  0.0150 - 0.0144i\end {array} \right).
\end{equation}

We see from Table (\ref{numerical2}) that the mixing angles ($\theta_{12}$,$\theta_{23}$,$\theta_{13}$) extend over their allowed experimental ranges for both normal and inverted hierarchies at all statistical levels. There exists a persistent forbidden gap for $\rho$ at all $\sigma$ levels with either hierarchy type. One also notes disallowed regions for the phases $\delta$ and $\sigma$ in the case of normal ordering at all $\sigma$ levels. For inverted ordering, we see that the phase $\sigma$ is restricted at all $\sigma$ levels, %and it's allowed range tends to be wider at the 3-$\sigma$ level.
as was the case of $E_1$ pattern. Moreover, a narrow forbidden gap $[346.14^{\circ},353^{\circ}]$ is found for $\delta$ at the 3-$\sigma$ level. Neither $m_{1}$ in normal type nor $m_{3}$ in inverted type approach a vanishing value. Therefore, the singular mass matrix is not expected for this pattern. %The allowed values of J are negative at all $\sigma$ levels in inverted type, so the corresponding $\delta$ lies in the third or fourth quarters.

For normal ordering plots, a large forbidden gap exists for $\rho$ as in the pattern $E_1$, whereas narrow forbidden gaps are noticed for $\delta$ and $\sigma$. We see a moderate mass hierarchy where $(0.27\leq\frac{m_{2}}{m_{3}}\leq0.97)$ and a quasi degenerate mass spectrum characterized by $(1 < \frac{m_{2}}{m_{1}}\leq1.33)$.

As to inverted ordering plots, we see a wide forbidden gap for the phase $\rho$ as in normal type. We also find that $\sigma$ is bound to the interval $[16^{\circ},167^{\circ}]$. One notes a quasi degenerate mass spectrum characterized by $(1 < \frac{m_{2}}{m_{1}}\leq1.015)$ and a mild mass hierarchy where $(1.03\leq\frac{m_{2}}{m_{3}}\leq4.40)$.

For both normal and inverted ordering, the correlations between ($\delta$,$m_{ee}$) and between ($\delta$,LNM) show that when $m_{ee}$ and LNM increase, the allowed parameter space becomes more restricted.
\begin{figure}[hbtp]
\centering
\begin{minipage}[l]{0.5\textwidth}
\epsfxsize=24cm
\centerline{\epsfbox{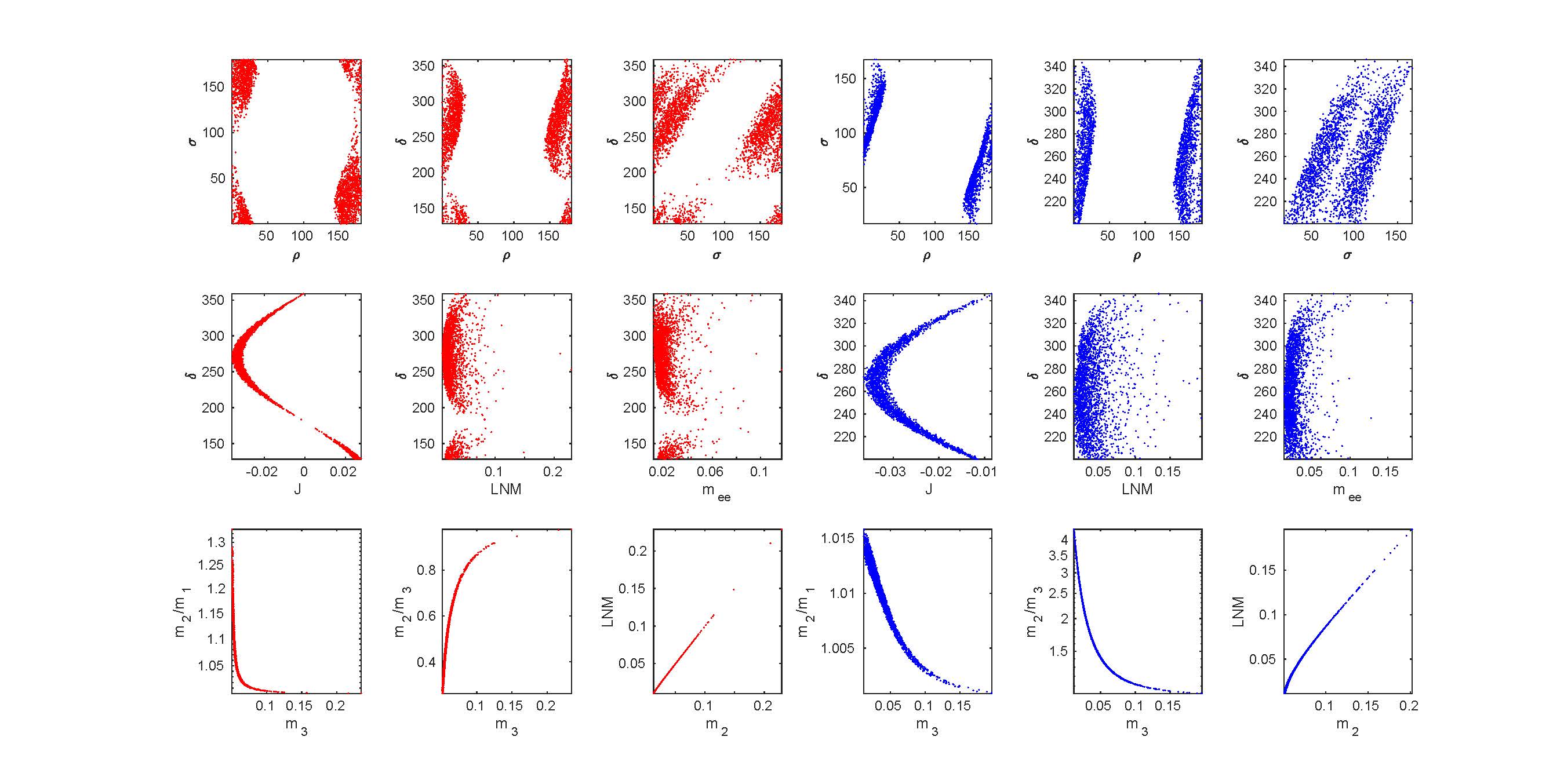}}
\end{minipage}%
\caption{ The correlation plots for $E_2$ pattern, the red (blue) plots represents the normal (inverted) ordering. The first row represents the correlations between the CP-violating phases, the second one represent the correlations between $\delta$ and each of J, LNM (the lowest neutrino mass) and $m_{ee}$ parameters. The last one shows the degree of hierarchy. Angles (masses) are evaluated in degrees (eV).}
\label{E2fig}
\end{figure}
\newpage

\subsection{Pattern $F_1$: $M_{\nu~12}=M_{\nu~23}$}
The expressions for $A_{1}$, $A_{2}$ and $A_{3}$ coefficients for this pattern are
\begin{equation}
\begin{aligned}
A_{1}=&c_{12}c_{13}(-c_{12}s_{23}s_{13}-s_{12}c_{23}e^{-i\delta})-(-c_{12}s_{23}s_{13}-s_{12}c_{23}e^{-i\delta})(-c_{12}c_{23}s_{13}+s_{12}s_{23}e^{-i\delta}),\\
A_{2}=&s_{12}c_{13}(-s_{12}s_{23}s_{13}+c_{12}c_{23}e^{-i\delta})-(-s_{12}s_{23}s_{13}+c_{12}c_{23}e^{-i\delta})(-s_{12}c_{23}s_{13}-c_{12}s_{23}e^{-i\delta}),\\
A_{3}=&s_{13}s_{23}c_{13}-s_{23}c_{13}^{2}c_{23}.
\end{aligned}\label{F1coff}
\end{equation}
The leading order approximation in $s_{13}$ for the mass ratios are given by
\begin{align}
\frac{m_1}{m_3}\approx&\frac{\sin(2\sigma-\delta)+\cot\theta_{12}\sin\theta_{23}\sin(2\sigma-2\delta)}{\sin\delta\cos(2\rho-2\sigma)+\big(\cos2\theta_{12}\cos\delta+\sin\theta_{12}\cos\theta_{12}\cos\theta_{23}\cot\theta_{23}\big)\sin(2\rho-2\sigma)}, \nonumber\\
\frac{m_2}{m_3}\approx&\frac{\sin(2\rho-\delta)-\tan\theta_{12}\sin\theta_{23}\sin(2\rho-2\delta)}{\sin\delta\cos(2\rho-2\sigma)+\big(\cos2\theta_{12}\cos\delta+\sin\theta_{12}\cos\theta_{12}\cos\theta_{23}\cot\theta_{23}\big)\sin(2\rho-2\sigma)},\label{F1ratio}
\end{align}
whence
\bea
1 < \frac{m_2}{m_1} &\approx& \frac{\sin(2\rho-\delta)-\tan\theta_{12}\sin\theta_{23}\sin(2\rho-2\delta)}{\sin(2\sigma-\delta)+\cot\theta_{12}\sin\theta_{23}\sin(2\sigma-2\delta)},
\eea
and one can check on the correlation plots, after having fixed the mixing angles at their central experimental values, that the phase angles satisfy this inequality.

For normal ordering, the representative point is taken as:
\begin{equation}
\begin{aligned}
(\theta_{12},\theta_{23},\theta_{13})=&(34.0311^{\circ},49.8152^{\circ},8.4294^{\circ}),\\
(\delta,\rho,\sigma)=&(225.9590^{\circ},52.7640^{\circ},
53.2878^{\circ}),\\
(m_{1},m_{2},m_{3})=&(0.0514\textrm{ eV},0.0521\textrm{ eV},0.0716\textrm{ eV}),\\
(m_{ee},m_{e})=&(0.0501\textrm{ eV},0.0521\textrm{ eV}),
\end{aligned}
\end{equation}
the corresponding neutrino mass matrix (in eV) is
\begin{equation}
M_{\nu}=\left( \begin {array}{ccc} -0.0123 + 0.0486i &  0.0096 - 0.0058i &  0.0079 - 0.0043i\\ \noalign{\medskip} 0.0096 - 0.0058i &  0.0616 + 0.0060i &  0.0096 - 0.0058i
\\ \noalign{\medskip}0.0079 - 0.0043i &  0.0096 - 0.0058i  & 0.0584 + 0.0079i
\end {array} \right).
\end{equation}
For inverted ordering, the representative point is taken as:
\begin{equation}
\begin{aligned}
(\theta_{12},\theta_{23},\theta_{13})=&(34.2844^{\circ},49.8543^{\circ},8.4406^{\circ}),\\
(\delta,\rho,\sigma)=&(269.4737^{\circ},128.1922^{\circ},88.8523^{\circ}),\\
(m_{1},m_{2},m_{3})=&(0.0536\textrm{ eV},0.0543\textrm{ eV},0.02120\textrm{ eV}),\\
(m_{ee},m_{e})=&(0.0422\textrm{ eV},0.0533\textrm{ eV}),
\end{aligned}
\end{equation}
the corresponding neutrino mass matrix (in eV) is
\begin{equation}
M_{\nu}=\left( \begin {array}{ccc} -0.0248 - 0.0341i & -0.0108 - 0.0086i &  0.0234 + 0.0181i\\ \noalign{\medskip}-0.0108 - 0.0086i &  0.0324 + 0.0090i & -0.0108 - 0.0086i
\\ \noalign{\medskip}0.0234 + 0.0181i & -0.0108 - 0.0086i &  0.0286 + 0.0061i
\end {array} \right).
\end{equation}
\newpage
We see from Table (\ref{numerical2}) that the mixing angles $(\theta_{12},\theta_{23},\theta_{23})$ extend over their allowed experimental ranges with each hierarchy type at all statistical levels. For normal ordering, there exists a persistent forbidden gap for the Dirac phase $\delta$ at all $\sigma$ levels. We also find, for normal type, wide disallowed regions for $\rho$ besides restrictions on the phase $\sigma$ at the 1-2-$\sigma$ levels. One notes also that there exists a wide forbidden gap $[26.04^{\circ},92.05^{\circ}]$ for $\rho$ together with a strong restriction onthe phase $\sigma$ at the 1-$\sigma$ level in inverted type. We see narrow forbidden regions for $\delta$ at the 2-3-$\sigma$ levels for inverted ordering. Table (\ref{numerical2}) also reveals that $m_{1}$ does not reach zero value at any $\sigma$-level, whereas $m_{3}$ can reach a vanishing value at the 2-3-$\sigma$ levels. Thus, the singular pattern is predicted for inverted ordering at the 2-3-$\sigma$ levels. %The allowed values of J at all $\sigma$ levels for inverted ordering are negative, so the corresponding $\delta$ lies in the third or fourth quarters.

For normal ordering plots, we see forbidden gaps for the phases $\delta$ and $\sigma$. We also find a mild mass hierarchy where $(0.41\leq\frac{m_{2}}{m_{3}}\leq0.96)$ and a quasi degeneracy characterized by $m_{1}\approx m_{2}$.

As to inverted ordering plots, we find disallowed regions for $\delta$. We also see a quasi degeneracy characterized by $m_{1}\approx m_{2}$ as well as an acute mass hierarchy where $\frac{m_{2}}{m_{3}}$ can reach $10^{4}$ indicating the possibility of a vanishing $m_{3}$.

For both normal and inverted ordering, the correlations between ($\delta$,$m_{ee}$) and between ($\delta$,LNM) show that when $m_{ee}$ and LNM increase, the allowed parameter space becomes more restricted.
\begin{figure}[hbtp]
\centering
\begin{minipage}[l]{0.5\textwidth}
\epsfxsize=24.33cm
\centerline{\epsfbox{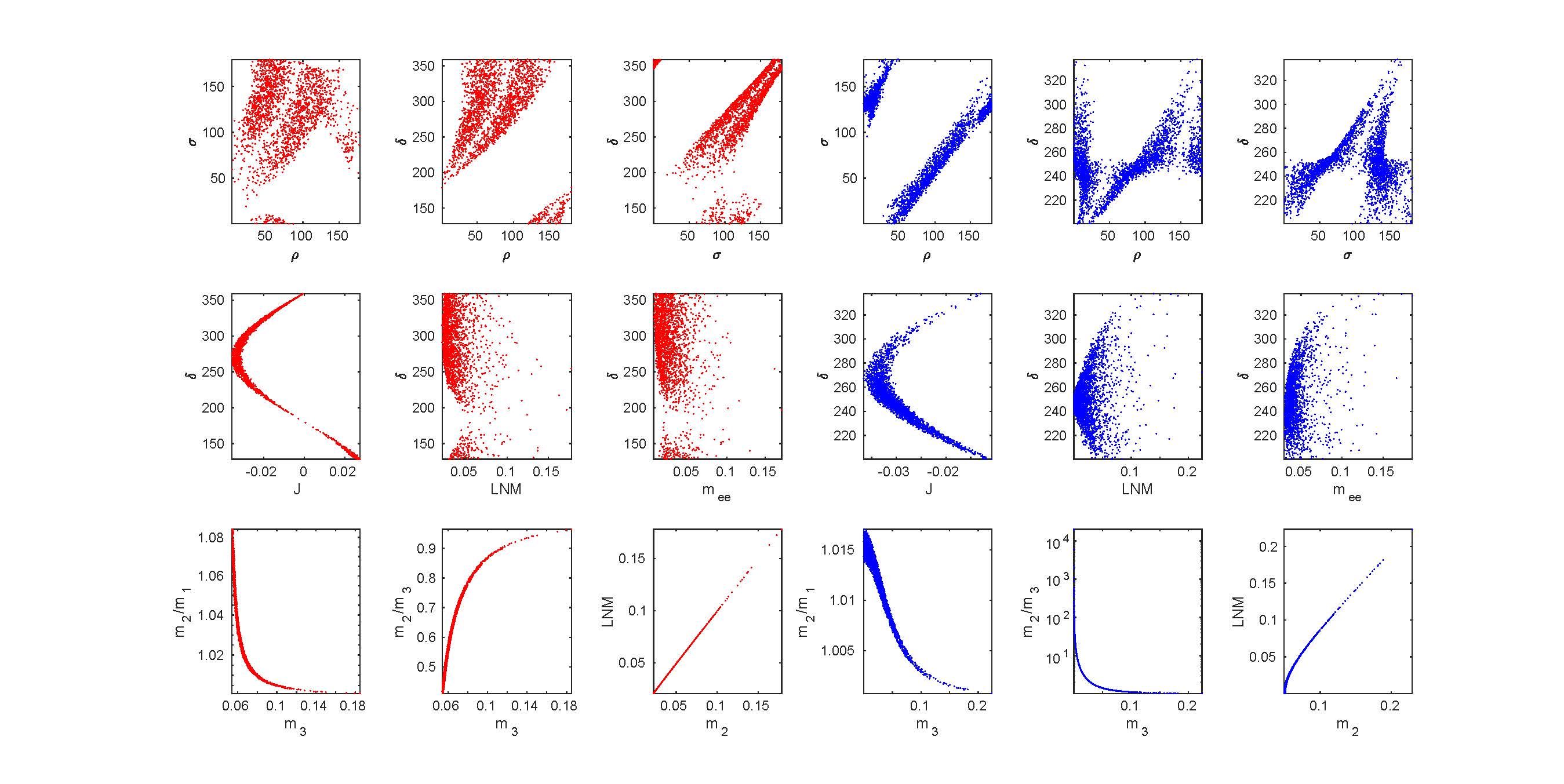}}
\end{minipage}%
\caption{ The correlation plots for $F_1$ pattern, the red (blue) plots represent the normal (inverted) ordering. The first row represents the correlations between the CP-violating phases, the second one represents the correlations between $\delta$ and each of J, LNM (the lowest neutrino mass) and $m_{ee}$ parameters. The last one shows the degree of hierarchy. Angles (masses) are evaluated in degrees (eV).}
\label{F1fig}
\end{figure}
\newpage

\subsection{Pattern $F_2$: $M_{\nu~13}=M_{\nu~23}$}
The expressions for $A's$ coefficients and the approximate analytical formulas for the mass ratios can be obtained from Eqs. (\ref{F1coff},\ref{F1ratio})by using the transformations in Eq. \ref{mutau}.

For normal ordering, the representative point is taken as:
\begin{equation}
\begin{aligned}
(\theta_{12},\theta_{23},\theta_{13})=&(34.5291^{\circ},49.4592^{\circ},8.4391^{\circ}),\\
(\delta,\rho,\sigma)=&(199.7745^{\circ},66.9448^{\circ},
24.8158^{\circ}),\\
(m_{1},m_{2},m_{3})=&(0.0268\textrm{ eV},0.0281\textrm{ eV},0.0577\textrm{ eV}),\\
(m_{ee},m_{e})=&(0.0203\textrm{ eV},0.0282\textrm{ eV}),
\end{aligned}
\end{equation}
the corresponding neutrino mass matrix (in eV) is
\begin{equation}
M_{\nu}=\left( \begin {array}{ccc} -0.0054 + 0.0196i & -0.0035 + 0.0009i &  0.0185 - 0.0056i\\ \noalign{\medskip} -0.0035 + 0.0009i &  0.0426 + 0.0046i &  0.0185 - 0.0056i
\\ \noalign{\medskip} 0.0185 - 0.0056i &  0.0185 - 0.0056i &  0.0319 + 0.0078i
\end {array} \right).
\end{equation}
For inverted ordering, the representative point is taken as:
\begin{equation}
\begin{aligned}
(\theta_{12},\theta_{23},\theta_{13})=&(34.5483^{\circ},49.5221^{\circ},8.5678^{\circ}),\\
(\delta,\rho,\sigma)=&(287.6146^{\circ},173.6865^{\circ},38.8534^{\circ}),\\
(m_{1},m_{2},m_{3})=&(0.0505\textrm{ eV},0.0513\textrm{ eV},0.0086\textrm{ eV}),\\
(m_{ee},m_{e})=&(0.0373\textrm{ eV},0.0502\textrm{ eV}),
\end{aligned}
\end{equation}
the corresponding neutrino mass matrix (in eV) is
\begin{equation}
M_{\nu}=\left( \begin {array}{ccc}  0.0363 + 0.0084i & -0.0241 - 0.0064i &  0.0218 + 0.0055i\\ \noalign{\medskip}-0.0241 - 0.0064i & -0.0053 - 0.0035i &  0.0218 + 0.0055i
\\ \noalign{\medskip} 0.0218 + 0.0055i &  0.0218 + 0.0055i & -0.0221 - 0.0078i
\end {array} \right).
\end{equation}

We see from Table (\ref{numerical2}) that the allowed experimental ranges for the mixing angles ($\theta_{12},\theta_{23},\theta_{13}$) can be covered at all statistical levels for both mass orderings. For normal ordering, we find wide forbidden gaps for the phase $\sigma$ such as, $[43.81^{\circ},161.79^{\circ}]$ at the 1-$\sigma$ level and $[79.31^{\circ},145.89^{\circ}]$ at the 2-$\sigma$ level. We also notice that the phase $\rho$ is restricted at all $\sigma$ levels in the normal type. For inverted ordering, we see narrow disallowed regions for $\delta$ at the 2-3-$\sigma$ levels. Table (\ref{numerical2}) also reveals that $m_{3}$ approaches a vanishing value at all $\sigma$ levels for inverted ordering. Thus, the singular mass matrix is predicted at all $\sigma$ levels in an inverted type. %The allowed values of the J parameter are negative in inverted type at all $\sigma$ levels, so the corresponding $\delta$ lies in the third or fourth quarters.

For normal ordering plots, we see a quasi-linear relation for ($\sigma,\delta$) correlation. We also find a moderate mass hierarchy where $(0.40\leq\frac{m_{2}}{m_{3}}\leq0.98)$ and a quasi degeneracy characterized by $m_{1}\approx m_{2}$.

As to the inverted plots, we find disallowed regions for the correlations including $\delta$. We also see a quasi degeneracy characterized by $m_{1}\approx m_{2}$ as well as an acute mass hierarchy where $\frac{m_{2}}{m_{3}}$ can reach $10^{4}$ indicating the possibility of a vanishing $m_{3}$.

For both normal and inverted ordering, the correlations between ($\delta$,$m_{ee}$) and between ($\delta$,LNM) show that when $m_{ee}$ and LNM increase, the allowed parameter space becomes more restricted.
\begin{figure}[hbtp]
\centering
\begin{minipage}[l]{0.5\textwidth}
\epsfxsize=24cm
\centerline{\epsfbox{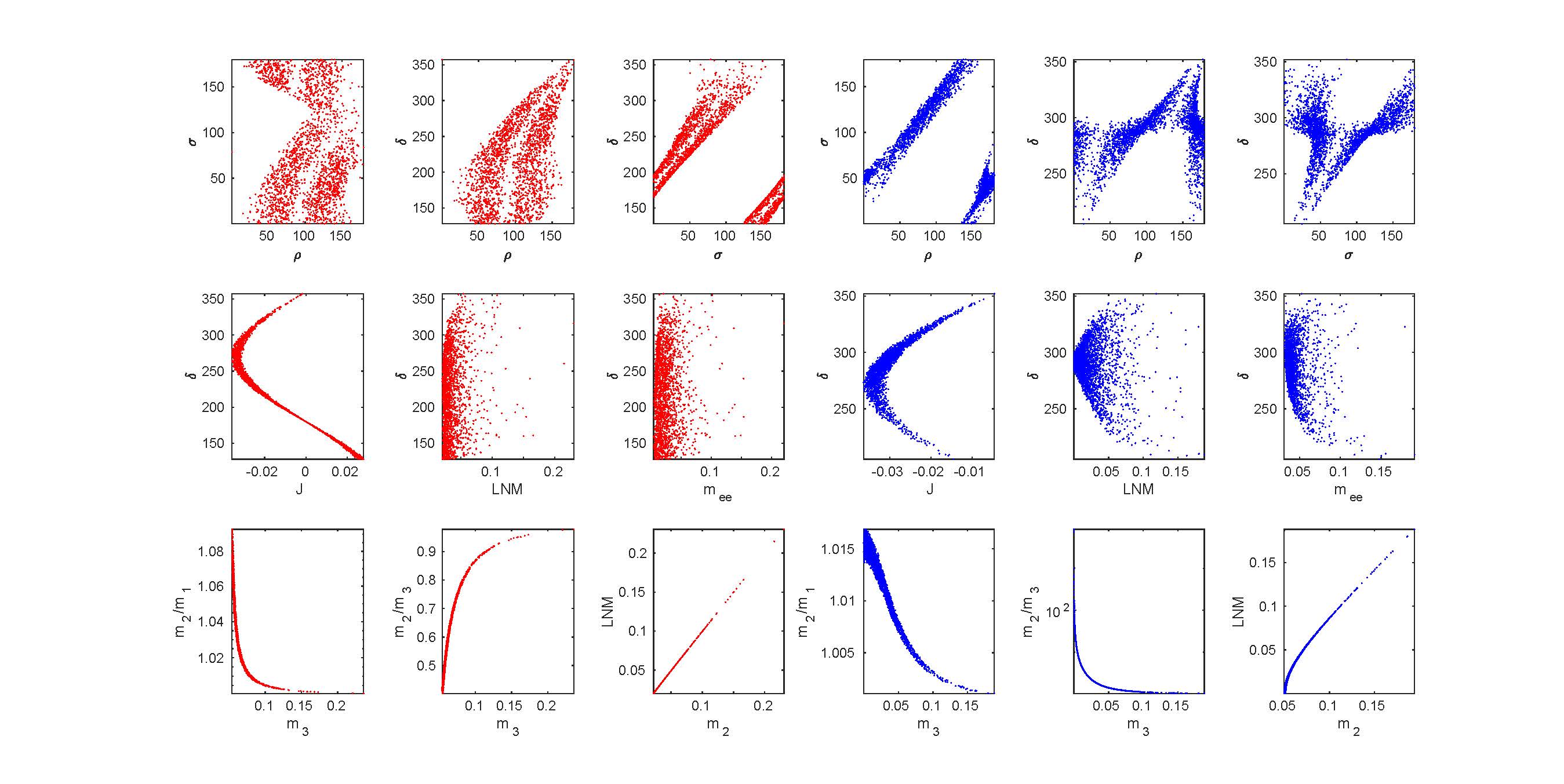}}
\end{minipage}%
\caption{ The correlation plots for $F_2$ pattern, the red (blue) plots represent the normal (inverted) ordering. The first row represents the correlations between the CP-violating phases, the second one represents the correlations between $\delta$ and each of J, LNM (the lowest neutrino mass) and $m_{ee}$ parameters. The last one shows the degree of hierarchy. Angles (masses) are evaluated in degrees (eV).}
\label{F2fig}
\end{figure}
\newpage

\subsection{Pattern $G_{1}$: $M_{\nu~11}=M_{\nu~23}$}
The expressions for $A_{1}$, $A_{2}$ and $A_{3}$ coefficients for this pattern are
\begin{equation}
\begin{aligned}
A_{1}=&c_{12}^{2}c_{13}^{2}-(-c_{12}s_{23}s_{13}-s_{12}c_{23}e^{-i\delta})(-c_{12}c_{23}s_{13}+s_{12}s_{23}e^{-i\delta}),\\
A_{2}=&s_{12}^{2}c_{13}^{2}-(-s_{12}s_{23}s_{13}+c_{12}c_{23}e^{-i\delta})(-s_{12}c_{23}s_{13}-c_{12}s_{23}e^{-i\delta}),\\
A_{3}=&s_{13}^{2}-s_{23}c_{13}^{2}c_{23}.
\end{aligned}
\end{equation}
The leading order approximation in $s_{13}$ for the mass ratios are given by
\begin{align}
\hspace{-1em}\frac{m_1}{m_3}\approx&\frac{-\sin^2\theta_{12}\big[\sin2\sigma+\cot^2\theta_{12}\cot\theta_{23}\sin^2\theta_{23}\sin(2\sigma-2\delta)\big]}{\cos2\theta_{12}\sin2\delta\cos(2\rho-2\sigma)+\big[\cos^2\theta_{12}\sin^2\theta_{12}\big(\frac{1+\cos^2\theta_{23}\sin^2\theta_{23}}{\cos\theta_{23}\sin\theta_{23}}\big)-\frac{1}{2}(\sin^22\theta_{12}-2)\cos2\delta\big]\sin(2\rho-2\sigma)\big]},\nonumber\\
\hspace{-1em}\frac{m_2}{m_3}\approx&\frac{\cos^2\theta_{12}\big[\sin2\rho+\tan^2\theta_{12}\sin\theta_{23}\cos\theta_{23}\sin(2\sigma-2\delta)\big]}{\cos2\theta_{12}\sin2\delta\cos(2\rho-2\sigma)+\big[\cos^2\theta_{12}\sin^2\theta_{12}\big(\frac{1+\cos^2\theta_{23}\sin^2\theta_{23}}{\cos\theta_{23}\sin\theta_{23}}\big)-\frac{1}{2}(\sin^22\theta_{12}-2)\cos2\delta\big]\sin(2\rho-2\sigma)\big]},
\end{align}
whence
\bea
1 < \frac{m_2}{m_1} &\approx&  \frac{ -\cot^2 \t_{12} \big[\sin2\rho+\tan^2\theta_{12}\sin\theta_{23}\cos\theta_{23}\sin(2\sigma-2\delta)\big] }{\sin2\sigma+\cot^2\theta_{12}\cot\theta_{23}\sin^2\theta_{23}\sin(2\sigma-2\delta)}
, \eea
and we can check on the correlation plots that, after having fixed the mixing angles at their central values, that the phase angles satisfy the above inequality.

For normal ordering, the representative point is taken as follows.
\begin{equation}
\begin{aligned}
(\theta_{12},\theta_{23},\theta_{13})=&(34.2586^{\circ},49.3215^{\circ},8.4114^{\circ}),\\
(\delta,\rho,\sigma)=&(205.0568^{\circ},21.3766^{\circ},172.1662^{\circ}),\\
(m_{1},m_{2},m_{3})=&(0.0230\textrm{ eV},0.0246\textrm{ eV},0.0549\textrm{ eV}),\\
(m_{ee},m_{e})=&(0.0215\textrm{ eV},0.0246\textrm{ eV}),
\end{aligned}
\end{equation}
The corresponding neutrino mass matrix (in eV) is
\begin{equation}
M_{\nu}=\left( \begin {array}{ccc} 0.0198 + 0.0084i &  0.0049 + 0.0060i &  0.0023 - 0.0088i\\ \noalign{\medskip}0.0049 + 0.0060i  & 0.0369 - 0.0084i &  0.0198 + 0.0084i
\\ \noalign{\medskip}0.0023 - 0.0088i  & 0.0198 + 0.0084i  & 0.0313 - 0.0077i
\end {array} \right).
\end{equation}
For inverted ordering, the representative point is taken as follows.
\begin{equation}
\begin{aligned}
(\theta_{12},\theta_{23},\theta_{13})=&(34.6418^{\circ},49.0016^{\circ},8.4594^{\circ}),\\
(\delta,\rho,\sigma)=&(326.4538^{\circ},6.3879^{\circ},90.1508^{\circ}),\\
(m_{1},m_{2},m_{3})=&(0.0578\textrm{ eV},0.0585\textrm{ eV},0.0293\textrm{ eV}),\\
(m_{ee},m_{e})=&(0.0212\textrm{ eV},0.0575\textrm{ eV}),
\end{aligned}
\end{equation}
The corresponding neutrino mass matrix (in eV) is
\begin{equation}
M_{\nu}=\left( \begin {array}{ccc} 0.0195 + 0.0084i & -0.0257 - 0.0235i &  0.0318 + 0.0251i\\ \noalign{\medskip}-0.0257 - 0.0235i &  0.0175 - 0.0026i &  0.0195 + 0.0084i
\\ \noalign{\medskip}0.0318 + 0.0251i &  0.0195 + 0.0084i & -0.0003 - 0.0153i
\end {array} \right).
\end{equation}

From Table (\ref{numerical2}), we see that the mixing angles ($\theta_{12}$,$\theta_{23}$,$\theta_{13}$) extend over their allowed experimental ranges at all $\sigma$-levels for both normal and inverted ordering. There exists a large forbidden gap for $\rho$ at all $\sigma$-levels with each hierarchy type. For normal ordering, we find large disallowed regions $[60.62^{\circ},134^{\circ}]([82.55^{\circ},125.09^{\circ}])$ for the phase $\sigma$ at the 1(2)-$\sigma$ levels. For inverted ordering, the phase $\sigma$ is restricted at all $\sigma$-levels to be nearly in the interval $[14^{\circ},169^{\circ}]$. We also note persistent forbidden gaps for $\delta$ in inverted type. %The allowed values of $J$ at all error levels for inverted ordering are negative, so the corresponding $\delta$ lies in the third or fourth quarters.
Neither $m_{1}$ in normal type nor $m_{3}$ in inverted type can reach a vanishing value. Thus, the singular pattern is not predicted at any $\sigma$-level.

For normal ordering plots, a large forbidden gap exists for $\rho$, whereas narrow gaps are noticed for $\sigma$ and $\delta$. We find a moderate mass hierarchy where $(0.33\leq\frac{m_{2}}{m_{3}}\leq0.97)$ together with a quasi degenerate mass spectrum characterized by $(1 < \frac{m_{2}}{m_{1}}\leq1.15)$.

For inverted ordering plots, as in normal type, a large forbidden gap exists for $\rho$. We also see a tight forbidden gap in all correlations including $\delta$. We get a mild mass hierarchy characterized by $(1.03\leq\frac{m_{2}}{m_{3}}\leq4.21)$ and a quasi degeneracy where  $(1 < \frac{m_{2}}{m_{1}}\leq1.015)$ .

For both normal and inverted ordering, the correlations between ($\delta$,$m_{ee}$) and between ($\delta$,LNM) show that when $m_{ee}$ and LNM increase, the allowed parameter space becomes more restricted.
\begin{figure}[hbtp]
\centering
\begin{minipage}[l]{0.5\textwidth}
\epsfxsize=24.33cm
\centerline{\epsfbox{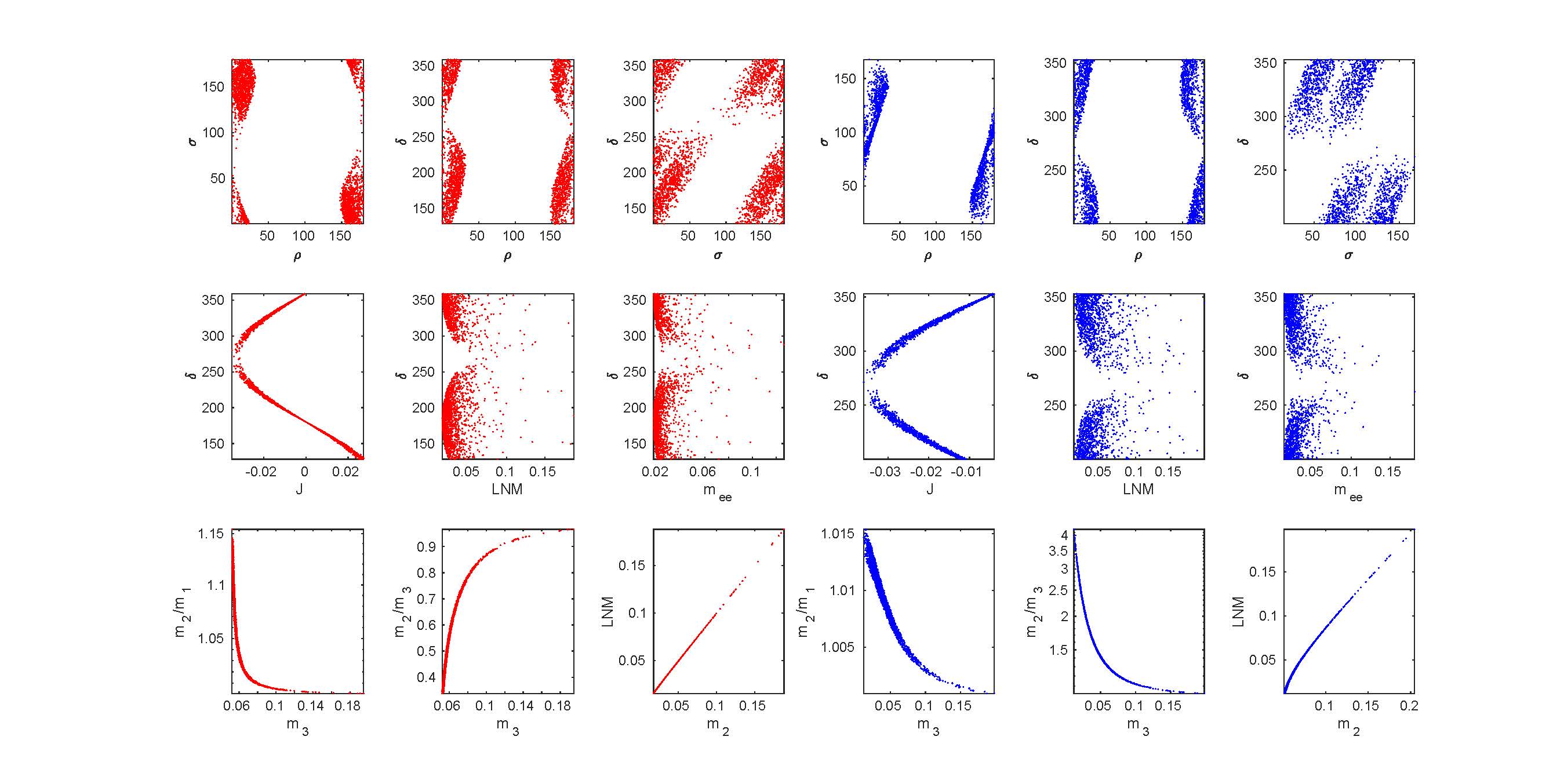}}
\end{minipage}%
\caption{ The correlation plots for $G_{1}$ pattern, the red (blue) plots represent the normal ordering. The first row represents the correlations between the CP-violating phases, the second one represent the correlations between $\delta$ and each of J, LNM (the lowest neutrino mass) and $m_{ee}$ parameters. The last one shows the degree of hierarchy. Angles (masses) are evaluated in degrees (eV).}
\label{G1fig}
\end{figure}
\newpage

\subsection{Pattern $G_{2}$: $M_{\nu~22}=M_{\nu~33}$}
The expressions for $A_{1}$, $A_{2}$ and $A_{3}$ coefficients for this pattern are
\begin{equation}
\begin{aligned}
A_{1}=&(-c_{12}s_{23}s_{13}-s_{12}c_{23}e^{-i\delta})^{2}-(-c_{12}c_{23}s_{13}+s_{12}s_{23}e^{-i\delta})^{2},\\
A_{2}=&(-s_{12}s_{23}s_{13}+c_{12}c_{23}e^{-i\delta})^{2}-(-s_{12}c_{23}s_{13}-c_{12}s_{23}e^{-i\delta})^{2},\\
A_{3}=&s_{23}^{2}c_{13}^{2}-c_{23}^{2}c_{13}^{2}.
\end{aligned}
\end{equation}
The leading order approximation in $s_{13}$ for the mass ratios and $m_{ee}$ are given by
\begin{align}
\frac{m_1}{m_3}\approx&\frac{\sin(2\sigma-2\delta)}{\sin^2\theta_{12}\sin(2\sigma-2\rho)},\nonumber\\
\frac{m_2}{m_3}\approx&\frac{-\sin(2\rho-2\delta)}{\cos^2\theta_{12}\sin(2\sigma-2\rho)},\nonumber \\
m_{ee}\approx&m_{3}\Bigg|\frac{\tan^2\theta_{12}e^{2i\sigma}}{\sin(2\sigma-2\rho)}\bigg(\sin(2\rho-2\delta)-\cot^4\theta_{12}\sin(2\sigma-2\delta)e^{2i(\rho-\sigma)}\bigg)\Bigg|,
\end{align}
so
\bea
1 < \frac{m_2}{m_1} &\approx& \frac{-\tan^2 \t_{12} \sin(2\rho-2\delta) }{\sin(2\sigma-2\delta)}.
\eea
After fixing the mixing angles at their experimentally central vlaues, one can check that this inequality is met on the phase angles correlation plots.

For normal ordering, the representative point is taken as follows.
\begin{equation}
\begin{aligned}
(\theta_{12},\theta_{23},\theta_{13})=&(34.6200^{\circ},45.3514^{\circ},8.5418^{\circ}),\\
(\delta,\rho,\sigma)=&(207.9270^{\circ},86.8034^{\circ},
92.7703^{\circ}),\\
(m_{1},m_{2},m_{3})=&(0.0060\textrm{ eV},0.0106\textrm{ eV},0.0504\textrm{ eV}),\\
(m_{ee},m_{e})=&(0.0062\textrm{ eV},0.0107\textrm{ eV}),
\end{aligned}
\end{equation}
The corresponding neutrino mass matrix (in eV) is
\begin{equation}
M_{\nu}=\left( \begin {array}{ccc} -0.0062 + 0.0001i &  0.0076 - 0.0002i &  0.0044 + 0.0002i\\ \noalign{\medskip}0.0076 - 0.0002i  & 0.0218 + 0.0036i &  0.0273 - 0.0036i
\\ \noalign{\medskip} 0.0044 + 0.0002i &  0.0273 - 0.0036i  & 0.0218 + 0.0036i\end {array} \right).
\end{equation}
For inverted ordering, the representative point is taken as follows.
\begin{equation}
\begin{aligned}
(\theta_{12},\theta_{23},\theta_{13})=&(34.5244^{\circ},49.3872^{\circ},8.4267^{\circ}),\\
(\delta,\rho,\sigma)=&(289.8576^{\circ},138.3062^{\circ},118.9371^{\circ}),\\
(m_{1},m_{2},m_{3})=&(0.0576\textrm{ eV},0.0583\textrm{ eV},0.0306\textrm{ eV}),\\
(m_{ee},m_{e})=&(0.0537\textrm{ eV},0.0574\textrm{ eV}),
\end{aligned}
\end{equation}
The corresponding neutrino mass matrix (in eV) is
\begin{equation}
M_{\nu}=\left( \begin {array}{ccc} -0.0047 - 0.0535i & -0.0021 - 0.0038i &  0.0105 + 0.0166i\\ \noalign{\medskip}-0.0021 - 0.0038i &  0.0387 + 0.0133i & -0.0091 - 0.0147i
\\ \noalign{\medskip}0.0105 + 0.0166i & -0.0091 - 0.0147i  & 0.0387 + 0.0133i\end {array} \right).
\end{equation}
We see from Table (\ref{numerical2}) that the mixing angles ($\theta_{12}$,$\theta_{13}$) extend over their allowed experimental ranges with each hierarchy type at all $\sigma$-levels. However, we notice a tight forbidden gap for $\theta_{23}$ around $45^{\circ}$ at the 3-$\sigma$ level for both orderings. The allowed experimental ranges for the Dirac phase $\delta$ can be covered in normal type at all $\sigma$-levels, whereas narrow forbidden gaps exist in inverted type at the 2-3-$\sigma$ levels. For normal ordering, we find that the phase $\sigma$ is bound to be in the intervals $[15.76^{\circ},157.94^{\circ}]$ at the 1-$\sigma$ level and $[21.59^{\circ},164.13^{\circ}]$ at the 2-$\sigma$ level. We also see wide forbidden gaps for $\rho$ at the 1-2-$\sigma$ levels in normal type. As to the inverted ordering, one notes narrow disallowed regions for the phases $\rho$ and $\sigma$ at the 1-$\sigma$ level. Table  (\ref{numerical2}) also shows that $m_{1}$ can reach zero value at the 3-$\sigma$ level while $m_{3}$ can reach a vanishing value at all $\sigma$-levels. Thus, we predict a singular case with each hierarchy type, at the 3-$\sigma$-level in normal type and at all $\sigma$-levels in inverted type. %The allowed values of J at all $\sigma$ levels for inverted ordering  are negative, so the corresponding $\delta$ lies in the third or fourth quarters.

In a similar way to the patterns of the category $C$, we include correlation plots involving $\theta_{23}$. For normal ordering plots,  we see clearly a tight forbidden gap around $45^{\circ}$ for $\theta_{23}$. We also see  forbidden bands in the correlation plots between the CP-violating phases $\rho, \delta$ and $\sigma$. We notice a moderate mass hierarchy where $(0.16\leq\frac{m_{2}}{m_{3}}\leq0.91)$ besides a severe hierarchy where $\frac{m_2}{m_1}$ reaches $10^3$ indicating the possibility of vanishing $m_1$.

As to inverted ordering plots, there is also a tight forbidden gap around $45^{\circ}$ for $\theta_{23}$, albeit not so clear as in the normal case. We find an approximate linear correlation between $(\rho,\sigma)$. We also find a quasi degeneracy characterized by $m_{1}\approx m_{2}$ as well as an acute hierarchy where $\frac{m_{2}}{m_{3}}$ can reach $10^{4}$ indicating the possibility of a vanishing $m_{3}$.

For both normal and inverted ordering, the correlations between ($\delta$,$m_{ee}$) and between ($\delta$,LNM) show that when $m_{ee}$ and LNM increase, the allowed parameter space becomes more limited.
\begin{figure}[hbtp]
\centering
\begin{minipage}[l]{0.5\textwidth}
\epsfxsize=24cm
\centerline{\epsfbox{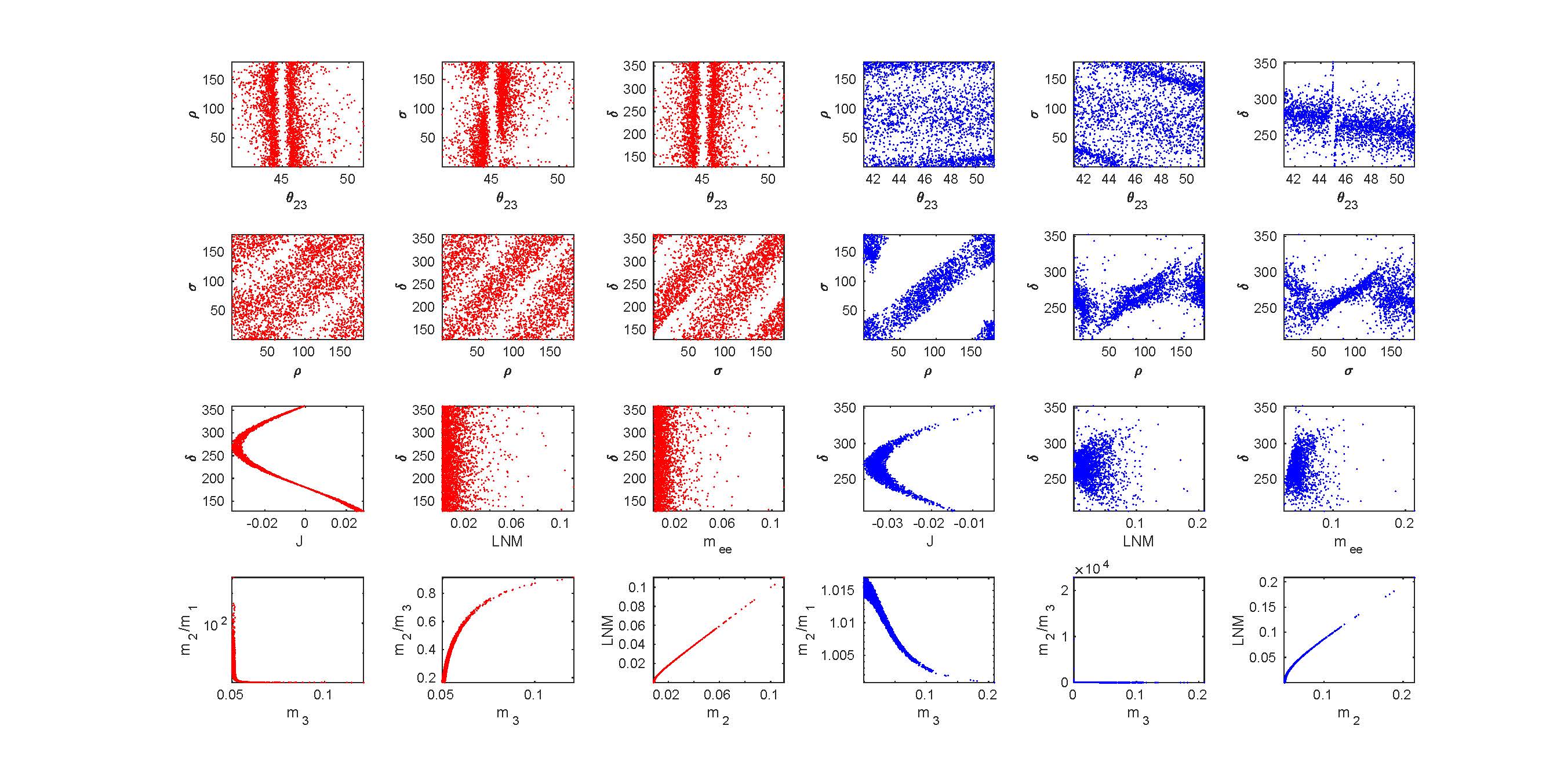}}
\end{minipage}%
\caption{The correlation plots for $G_{2}$ pattern, the red (blue) plots represent the normal (inverted) ordering. The first row represents the correlations with the CP-violating phases involving $\theta_{23}$. The second row represents the correlations between the CP-violating phases, the third one represent the correlations between $\delta$ and each of  J, LNM (the lowest neutrino mass) and $m_{ee}$ parameters. The last one shows the degree of hierarchy. Angles (masses) are evaluated in degrees (eV).}
\label{fig13}
\end{figure}
\newpage

\subsection{Pattern $G_{3}$: $M_{\nu~12}=M_{\nu~13}$}
The expressions for $A_{1}$, $A_{2}$ and $A_{3}$ coefficients for this pattern are
\begin{equation}
\begin{aligned}
A_{1}=&c_{12}c_{13}(-c_{12}s_{23}s_{13}-s_{12}c_{23}e^{-i\delta})-c_{12}c_{13}(-c_{12}c_{23}s_{13}+s_{12}s_{23}e^{-i\delta}),\\
A_{2}=&s_{12}c_{13}(-s_{12}s_{23}s_{13}+c_{12}c_{23}e^{-i\delta})-s_{12}c_{13}(-s_{12}c_{23}s_{13}-c_{12}s_{23}e^{-i\delta}),\\
A_{3}=&s_{13}s_{23}c_{13}-s_{13}c_{23}c_{13}.
\end{aligned}
\end{equation}
The leading order approximation in powers of $s_{13}$ for the mass ratios are given by
\begin{equation}
\begin{aligned}
\frac{m_{1}}{m_{3}}\approx&\frac{2(\sin 2\theta_{23}-1)\sin(2\sigma-\delta)}{\cos 2\theta_{23}\sin 2\theta_{12}\sin(2\sigma-2\rho)}\sin\theta_{13},\\
\frac{m_{2}}{m_{3}}\approx&-\frac{2\cos2\theta_{23}\sin(2\rho-\delta)}{\sin 2\theta_{12}(\sin 2\theta_{23}+1)\sin(2\sigma-2\rho)}\sin\theta_{13}.
\end{aligned}
\end{equation}
Thus, we obtain
\begin{equation}
1 < \frac{m_2}{m_1}\approx\frac{\sin(2\rho-\delta)}{\sin(2\sigma-\delta)}.
\end{equation}
One can check on the phases correlation plots that this inequality is met. Moreover,  the leading order approximations in powers of $s_{13}$ for $R_{\nu}$ and $m_{ee}$ are given by
\begin{align}
R_{\nu}\approx&\frac{8\cos^{2}2\theta_{23}\big[\sin^{2}(2\sigma-\delta)+\sin^{2}(2\rho-\delta)\big]\sin^{2}\theta_{13}}{\sin^{2}2\theta_{12}(\sin2\theta_{23}+1)^2\sin^{2}(2\sigma-2\rho)}\label{RG3},\\
m_{ee}\approx&m_{3}\Bigg|\frac{\tan\theta_{12}\cos2\theta_{23}e^{2i\sigma}}{(\sin2\theta_{23}+1)\sin(2\rho-2\sigma)}\bigg(\sin(2\rho-\delta)+\frac{e^{2i(\rho-\sigma)}\sin(2\sigma-\delta)}{\tan^{2}\theta_{12}}\bigg)\Bigg|\sin\theta_{13}.
\end{align}
%We see from Eq. \ref{RG3} that $\theta_{23}\neq\frac{\pi}{4}$. However, taking the central values of the mixing angles we see that the numerator of $R_\nu = O(10^{-2})$ is small of order of $\cos^2 2\t_{23} \sin^2 \t_{13} \approx \cos^2 99^o \sin^2 8.5^o =O(10^{-4})$, so we deduce that the denominator should be of order $O(10^{-2})$, which means ($\r - \s$) is very small, but not vanishing, justifying the first bisectrix in the corresponding correlation plot, and we can check also that the correlations ($\d-\r$) and ($\d-\s$) are similar for both types of hierarchy.
We see from Eq. \ref{RG3} that $\theta_{23}\neq\frac{\pi}{4}$. However, taking the central values of the mixing angles we see that the numerator of $R_\nu$, where $R_\nu$ is of order $O(10^{-2})$, is small of order of $\cos^2 2\t_{23} \sin^2 \t_{13} \approx \cos^2 99^o \sin^2 8.5^o =O(10^{-4})$, so we deduce that the denominator should be of order $O(10^{-2})$, which means ($\r - \s$) is very small, but not vanishing, justifying the first bisectrix in the corresponding correlation plot, and we can check also that the correlations ($\d-\r$) and ($\d-\s$) are similar for both types of hierarchy.

For normal ordering, the representative point is taken as follows.
\begin{equation}
\begin{aligned}
(\theta_{12},\theta_{23},\theta_{13})=&(34.4170^{\circ},49.4488^{\circ},8.5185^{\circ}),\\
(\delta,\rho,\sigma)=&(218.0135^{\circ},30.9730^{\circ},
30.1741^{\circ}),\\
(m_{1},m_{2},m_{3})=&(0.0603\textrm{ eV},0.0609\textrm{ eV},0.0786\textrm{ eV}),\\
(m_{ee},m_{e})=&(0.0600\textrm{ eV},0.0609\textrm{ eV}),
\end{aligned}
\end{equation}
The corresponding neutrino mass matrix (in eV) is
\begin{equation}
M_{\nu}=\left( \begin {array}{ccc} 0.0300 + 0.0519i &  0.0052 - 0.0055i &  0.0052 - 0.0055i\\ \noalign{\medskip} 0.0052 - 0.0055i &  0.0696 - 0.0061i &  0.0093 + 0.0084i
\\ \noalign{\medskip}0.0052 - 0.0055i &  0.0093 + 0.0084i  & 0.0665 - 0.0086i\end {array} \right).
\end{equation}
For inverted ordering, the representative point is taken as follows.
\begin{equation}
\begin{aligned}
(\theta_{12},\theta_{23},\theta_{13})=&(34.8938^{\circ},49.2853^{\circ},8.4769^{\circ}),\\
(\delta,\rho,\sigma)=&(274.9649^{\circ},25.4637^{\circ},25.0458^{\circ}),\\
(m_{1},m_{2},m_{3})=&(0.0586\textrm{ eV},0.0592\textrm{ eV},0.0315\textrm{ eV}),\\
(m_{ee},m_{e})=&(0.0579\textrm{ eV},0.0583\textrm{ eV}),
\end{aligned}
\end{equation}
The corresponding neutrino mass matrix (in eV) is
\begin{equation}
M_{\nu}=\left( \begin {array}{ccc} 0.0371 + 0.0445i & -0.0006 - 0.0047i & -0.0006 - 0.0047i\\ \noalign{\medskip}-0.0006 - 0.0047i & -0.0010 - 0.0158i &  0.0378 + 0.0194i
\\ \noalign{\medskip}-0.0006 - 0.0047i &  0.0378 + 0.0194i & -0.0123 - 0.0215i\end {array} \right).
\end{equation}

We see from Table (\ref{numerical2}) that the allowed experimental ranges for the mixing angles $\theta_{12}$ and $\theta_{13}$ can be covered for both normal and inverted ordering at all $\sigma$-levels. However, we find a very tight forbidden gap $[44.33^{\circ},45.33^{\circ}]$ for the mixing angle $\theta_{23}$ with each hierarchy type at the 3-$\sigma$-level. For the phases $\rho$ and $\sigma$, we find wide forbidden gaps at the 1-2-$\sigma$ levels for normal type. However, the narrow disallowed regions are noticed only at the 1-$\sigma$ level in the inverted type. Table  (\ref{numerical2}) also reveals that $m_{1}$ does not approach zero value at any $\sigma$-level, whereas $m_{3}$ can reach a vanishing value at all $\sigma$-levels. Therefore, the singular pattern is predicted in inverted type at all $\sigma$-levels.% The allowed values of J at all $\sigma$ levels for inverted ordering are negative, so the corresponding $\delta$ lies in the third or fourth quarters.

As in the case of $G_2$ pattern, correlations involving $\theta_{23}$ do show a gap around $45^o$, so we include them. For normal ordering plots, a strong linear correlation exists between $(\rho,\sigma)$. We see a moderate mass hierarchy where $(0.32\leq\frac{m_{2}}{m_{3}}\leq0.97)$ together with a quasi degeneracy characterized by $m_{1}\approx m_{2}$.

For inverted ordering plots, we find a strong linear correlation between $(\rho,\sigma)$ as in normal type and a severe hierarchy where $\frac{m_{2}}{m_{3}}$ can reach $10^{3}$ indicating the possibility of a vanishing $m_{3}$.

For both normal and inverted ordering, the correlations between ($\delta$,$m_{ee}$) and between ($\delta$,LNM) show that when $m_{ee}$ and LNM increase, the allowed parameter space becomes more restricted. We also see a tight forbidden gap around $45^{\circ}$ for $\theta_{23}$.
\begin{figure}[hbtp]
\centering
\begin{minipage}[l]{0.5\textwidth}
\epsfxsize=24.33cm
\centerline{\epsfbox{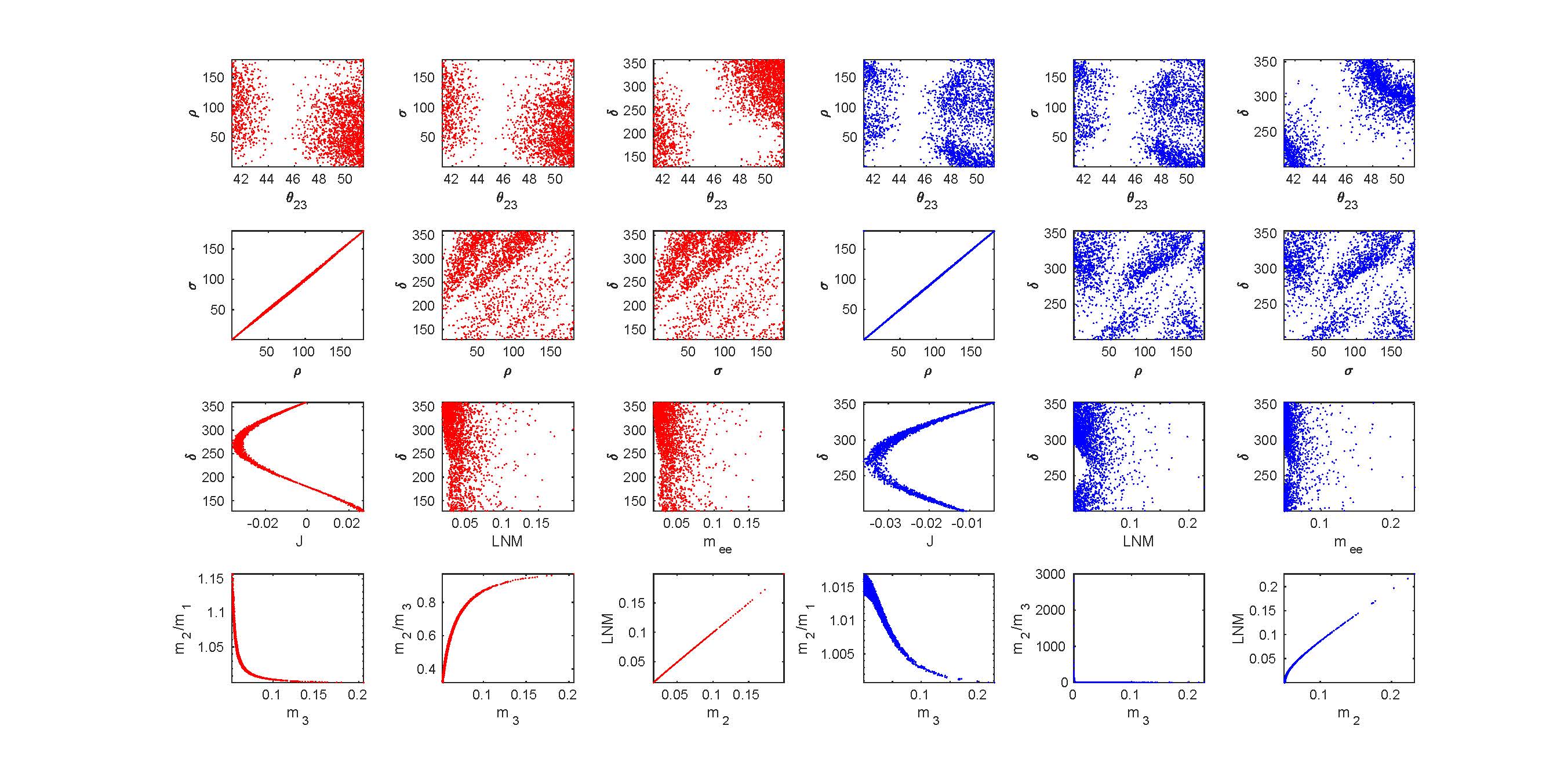}}
\end{minipage}%
\caption{The correlation plots for $G_{3}$ pattern, the red (blue) plots represent the normal (inverted) ordering.The first row represents the correlations with the CP-violating phases involving $\theta_{23}$. The second row represents the correlations between the CP-violating phases, the third one represent the correlations between $\delta$ and each of  J, LNM (the lowest neutrino mass) and $m_{ee}$ parameters. The last one shows the degree of hierarchy. Angles (masses) are evaluated in degrees (eV).}
\label{G3fig}
\end{figure}
\newpage

\section{Symmetry realization}
In this section, we introduce symmetry realizations for four, out of the stated nine phenomenologically allowed textures, in different seesaw scenarios. As a matter of fact, there are various realization methods which have been used to realize the Equality textures in $M_{\nu}$. Neutrino mass matrices with two equalities have been realized through a non-Abelian $A_{4}$ flavor symmetry in \cite{Dev_2013}, whereas the authors of \cite{Liu_2013} used $S_{3}$ symmetry to realize one equality texture in $M_{\nu}$. In \cite{Han_2017}, the Abelian $S_{\mu-\tau}\times Z_8$ symmetry with Froggatt-Nielsen mechanism including non-renormalizable 5-dim operators has been introduced to realize equalities in $M_{\nu}$ with either $m_{1}$ or $m_{3}$ equal to zero. In this study, we adopt realization methods depending on Abelian discrete flavor symmetries by extending the standard model with some additional matter fields. We present symmetry realizations for the patterns $E_1$, $F_1$, $G_2$ and $G_3$, and the obtained matrices would be parameterized by a sufficient number of free parameters (Yukawa couplings and VEVs) enough to satisfy the corresponding mathematical condition of equality between two entries with vanishing nonphysical phases. We use $Z_2\times Z_6$, $Z_2\times Z_{2}^{'}$ and $Z_2\times Z_2^{'}\times U(1)^3$ within type-I, type-II and mixed type (I+II) seesaw scenarios respectively. For the purpose of illustration, we take the pattern $E_1$  where $(M_{\nu~11}=M_{\nu~22})$ as a concrete example whereas the realizations for the remaining three patterns are summarized in the tables.

An `indirect' method to find realization models starting from zero-texture realizations was stated in detail in \cite{Lashin_2017,ismael}. Since the one-zero textures realizations are known \cite{Lashin_2012}, we briefly explain in the last subsection how we can enforce this for the one-equality texture. Whereas the method is applied for one particular pattern,  we state in one table the unitary matrices relating the other remaining 5 equality-textures which can be related to zero-textures. Quite interestingly, we find that the 6 patterns out of 15 which can be related to zero-textures in this way are exactly the 4 patterns, with another two patterns related by $\mu$-$\tau$-interchange symmetry, for which we could find ``direct'' realizations based on abelian groups. Moreover, they are the patterns which belong to the two classes $(Y_1 \cup Y_2)$ to which the `perturbed' $\mu$-$\tau$-symmetry patterns, defined by ($M_{\n 12}=M_{\n 13}$ or $M_{\n 22}=M_{\n 33}$) belong.

\subsection{Type-I seesaw}
The standard model (SM) particle content is extended with three-right handed neutrinos $\nu_{Ri}(i=1,2,3)$, four SM Higgs doublets $\phi_{i}(i=1,2,3,4)$ and a scalar singlet $\chi$. The general gauge-invariant Lagrangian is given by
\begin{equation}
\mathcal{L}=Y_{lij}^{k}\bar{D}_{Li}\phi_{k}l_{Rj}+Y_{\nu ij}^{k}\bar{D}_{Li}\tilde{\phi_{k}}\nu_{Rj}+X_{ij}\chi\nu_{Ri}\nu_{Rj}+h.c\label{type-I Lag}
\end{equation}
where $D_{Li}=(\nu_{Li},l_{Li})^{T}$ are three left-handed lepton doublets, $l_{Rj}$ are three right-handed charged leptons and $\tilde{\phi_{k}}=i\sigma_{2}\phi_{k}^{*}$. The effective Majorana mass matrix of light neutrino $M_{\nu}$ is generated through a type-I seesaw formula $M_{\nu}=M_{D}M_{R}^{-1}M_{D}^{T}$. We impose $Z_{2}\times Z_{6}$ flavor symmetry on the Lagrangian in Eq.(\ref{type-I Lag}), and assume the leptons transform under $Z_{6}$ as
\begin{equation}
\begin{aligned}
D_{L1}\rightarrow& D_{L1},~~~~D_{L2}\rightarrow D_{L2},~~~~D_{L3}\rightarrow\omega D_{L3},\\
\nu_{R1}\rightarrow&\nu_{R1},~~~~~~\nu_{R2}\rightarrow\nu_{R2},~~~~~\nu_{R3}\rightarrow\omega\nu_{R3},\\
l_{R1}\rightarrow&\omega^{3}l_{R1},~~~~l_{R2}\rightarrow\omega^{3}l_{R2},~~~l_{R3}\rightarrow\omega^{4}l_{R3},
\end{aligned}
\end{equation}
while the scalar multiplets transform as
\begin{equation}
\begin{aligned}
\phi_{1}\rightarrow&\phi_{1},~\phi_{2}\rightarrow\omega\phi_{2},~\phi_{3}\rightarrow\omega^{5}\phi_{3},~\phi_{4}\rightarrow\omega^{3}\phi_{4},~\phi_{5}\rightarrow\omega^{3}\phi_{5},~\chi\rightarrow\omega^{4}\chi,
\end{aligned}
\end{equation}
where $\omega=e^{\pi i/3}$. The bilinear $\nu_{Ri}\nu_{Rj}$ transform under $Z_{6}$ as
\begin{equation}
\nu_{Ri}\nu_{Rj}\cong\left( \begin {array}{ccc} 1&1&\omega\\ \noalign{\medskip}1&1&\omega
\\ \noalign{\medskip}\omega&\omega&\omega^{2}\end {array} \right).
\end{equation}
The elements (1,1), (1,2), (2,1) and (2,2) are thus directly existent in the Lagrangian because they are $Z_{6}$ invariant. A real scalar singlet ($\chi$) generates a non-vanishing matrix element (3,3) when acquiring a vacuum expectation value (VEV) at the seesaw scale. Under the action of $Z_{6}$, $\bar{D}_{Li}\nu_{Rj}$ and $\bar{D}_{Li}l_{Rj}$ transform as
\begin{equation}
\bar{D}_{Li}\nu_{Rj}\cong\left( \begin {array}{ccc} 1&1&\omega\\ \noalign{\medskip}1&1&\omega
\\ \noalign{\medskip}\omega^{5}&\omega^{5}&1\end {array} \right),~~~\bar{D}_{Li}l_{Rj}\cong\left( \begin {array}{ccc} \omega^{3}&\omega^{3}&\omega^{4}\\ \noalign{\medskip}\omega^{3}&\omega^{3}&\omega^{4}
\\ \noalign{\medskip}\omega^{2}&\omega^{2}&\omega^{3}\end {array} \right).
\end{equation}
The SM Higgs doublet $\phi_{1}$ generates (1,1), (1,2), (2,1), (2,2) and (3,3) matrix elements in the Dirac neutrino mass matrix $M_{D}$ while $\phi_{2}$ and $\phi_{3}$ generate (1,3), (2,3) and (3,1), (3,2) non-vanishing elements. The scalar doublets $\phi_{4}$ and $\phi_{5}$ are responsible for generating  (1,1), (1,2), (2,1), (2,2) and (3,3) matrix elements in the charged lepton mass matrix $M_{l}$.

 We set up that under the action of $Z_{2}$, the leptons transform as
\begin{equation}
\begin{aligned}
D_{L1}\rightarrow& D_{L2},~~~~D_{L2}\rightarrow D_{L1},~~~~D_{L3}\rightarrow D_{L3},\\
\nu_{R1}\rightarrow&\nu_{R2},~~~~~~\nu_{R2}\rightarrow\nu_{R1},~~~~~\nu_{R3}\rightarrow\nu_{R3},\\
l_{R1}\rightarrow&l_{R1},~~~~~~~l_{R2}\rightarrow l_{R2},~~~~~~l_{R3}\rightarrow l_{R3},
\end{aligned}\label{trans1}
\end{equation}
while the scalar multiplets transform as:
\begin{equation}
\begin{aligned}
\phi_{1}\rightarrow&\phi_{1},~\phi_{2}\rightarrow-\phi_{2},~\phi_{3}\rightarrow\phi_{3},~\phi_{4}\rightarrow\phi_{5},~\phi_{5}\rightarrow\phi_{4},~\chi\rightarrow\chi.
\end{aligned}\label{trans2}
\end{equation}
As we see from Eqs.(\ref{trans1},\ref{trans2}), the Yukawa couplings $X_{ij}$, $Y_{\nu~ij}$ and $Y_{l~ij}$ are constrained by the $Z_{2}$ symmetry
\begin{equation}
\begin{aligned}
&Y_{l11}^{4}=Y_{l21}^{5},~Y_{l12}^{4}=Y_{l22}^{5},~Y_{l21}^{4}=Y_{l11}^{5},~Y_{l22}^{4}=Y_{l12}^{5},~Y_{l33}^{4}=Y_{l33}^{5},\\
&Y_{\nu11}^{1}=Y_{\nu22}^{1},~Y_{\nu12}^{1}=Y_{\nu21}^{1},~Y_{\nu13}^{2}=-Y_{\nu23}^{2},~Y_{\nu31}^{3}=Y_{\nu32}^{3},\\
&X_{11}=X_{22},~X_{12}=X_{21}.
\end{aligned}
\end{equation}
The resulting right-handed Majorana mass matrix $M_{R}$ and Dirac neutrino mass matrix $M_{D}$ under $Z_6 \times Z_2$-symmetry take the form
\begin{equation}
M_{R}=\left( \begin {array}{ccc} x&y&0\\ \noalign{\medskip}y&x&0
\\ \noalign{\medskip}0&0&z\end {array} \right),~~~M_{D}=\left( \begin {array}{ccc} A&B&C\\ \noalign{\medskip}B&A&-C
\\ \noalign{\medskip}D&D&E\end {array} \right).
\end{equation}
Hence, the effective Majorana mass matrix of light neutrino takes the form,
\begin{equation}
M_{\nu}=M_{D}M_{R}^{-1}M_{D}^{T}=\left(\begin{array}{ccc}
a&b&c\\
b&a&d\\
c&d&e
\end{array}\right).
\end{equation}
which is of the requested form.

We need to check that one can, to a good approximation, be in the flavor basis, as we assumed during the phenomenological study, corresponding to a diagonal charged lepton mass matrix $M_{l}$ which takes the form,
\begin{equation}
M_{l}=\left(\begin{array}{ccc}
Y_{l~11}^{4}\mathit{v}_{4}+Y^{4}_{l~21}\mathit{v}_{5}&Y_{l~12}^{4}\mathit{v}_{4}+Y^{4}_{l~22}\mathit{v}_{5}&0\\
Y_{l~21}^{4}\mathit{v}_{4}+Y^{4}_{l~11}\mathit{v}_{5}&Y_{l~22}^{4}\mathit{v}_{4}+Y^{4}_{l~12}\mathit{v}_{5}&0\\
0&0&Y_{l~33}^{4}\mathit{v}_{4}+Y^{4}_{l~33}\mathit{v}_{5}
\end{array}\right),
\end{equation}
where $\mathit{v}_{4}$ and $\mathit{v}_{5}$ are VEVs of the scalar doublets $\phi_{4}$ and $\phi_{5}$. We follow the procedure presented in \cite{Lashin_z23} and assume the hierarchy $(\mathit{v}_{4}\ll\mathit{v}_{5}\approx\mathit{v})$. Then we get
\begin{equation}
M_{l}\approx\mathit{v}\left(\begin{array}{ccc}
Y^{4}_{l21}&Y^{4}_{l22}&0\\
Y^{4}_{l11}&Y^{4}_{l12}&0\\
0&0&Y^{4}_{l33}\end{array}\right)=~\mathit{v}\left( \begin {array}{c} \mathbf{a^{T}}\\ \noalign{\medskip}\mathbf{b^{T}}\\ \noalign{\medskip}\mathbf{c^{T}}
\end {array} \right)~~~\Rightarrow~~~M_{l}M_{l}^{\dagger}=\mathit{v}^{2}\left(\begin{array}{ccc}
|\mathbf{a}|^{2}&\mathbf{a}.\mathbf{b}&0\\
\mathbf{b}.\mathbf{a}&|\mathbf{b}|^{2}&0\\
0&0&|\mathbf{c}|^{2}\end{array}\right),
\end{equation}
where the vectors $\mathbf{a}$, $\mathbf{b}$ and $\mathbf{c}$ are defined as $\mathbf{a}=(Y_{l21}^{4},Y^{4}_{l22},0)$, $\mathbf{b}=(Y_{l11}^{4},Y^{4}_{l12},0)$ and $\mathbf{c}=(0,0,Y^{4}_{l33})$. By adjusting the magnitudes of the Yukawa couplings, $M_{l}M_{l}^{\dagger}$ can be diagonalized by an infinitesimal rotation $R_\eps$ with an angle less than $10^{-2}$ determined by the ratios % $\Big|\frac{\mathbf{a}.\mathbf{b}}{|\mathbf{a}|^{2}}\Big|$ and $\Big|\frac{\mathbf{a}.\mathbf{b}}{|\mathbf{b}|^{2}}\Big|$, and where
of the magnitudes ($|\mathbf{a}| : |\mathbf{b}| : |\mathbf{c}|$)  adjusted to come in values comparable to the acute charged lepton masses hierarchies ($m_e : m_\m : m_\tau$). The question arises now as to whether or not one should update the phenomenological analysis upon
carrying out this infinitesimally small rotation $R_\eps$ of the charged leptons. Actually, since the phenomenological study
was carried out in the flavor basis, we conclude that it is valid up to small corrections of the order of
the small rotation  $R_\eps$. With this in mind, this small correction should be added to the already
anticipated one stemming from the RG loop effects upon running from the high scale, when the
symmetry was imposed, to the low scale of the experimental data.

The symmetry assignments for the other cases, $F_1$, $G_{2}$ and $G_{3}$, are summarized in the tables \ref{F1type1}, \ref{G2type1} and \ref{G3type1}.

 \begin{table}[hbtp]
\begin{center}
\scalebox{0.79}{
\begin{tabular}{|c|c|c|c|c|}
\hline
\hline
\multicolumn{5}{|c|}{The matter content}\\
\hline
    $D_{L}$  & $\nu_{R}$ & $l_{R}$ & $\phi$ & $\chi$   \\
\hline
\hline
\multicolumn{5}{|c|}{$ \mbox{Symmetry under}~Z_{6}$}\\
\hline
 $T_{D}D_{L}$ & $T_{\nu}\nu_{R}$ & $T_{l}l_{R}$ & $T_{\phi}\phi$ & $T_{\chi}\chi$
\\
 $T_{D}=\mbox{diag}\left( 1,\omega,1\right)$& $T_{\nu}=\mbox{diag}\left(1,\omega,1\right)$ & $T_{l} = \mbox{diag}\left( \omega^{3},\omega^{4},\omega^{3}\right) $& $T_{\phi} =\mbox{diag}\left(1,\omega,\omega^{3},\omega^{3}\right)$ & $T_{\chi}=\mbox{diag}\left(1,\omega^{4}\right)$ \\
\hline
\hline
\multicolumn{5}{|c|}{$ \mbox{Symmetry under}~Z_{2}$}\\
\hline
$S_{D}D_{L}$ & $S_{\nu}\nu_{R}$ & $S_{l}l_{R}$ & $S_{\phi}\phi$ & $S_{\chi}\chi$
\\
 $S_{D}=\left( \begin {array}{ccc} 0&0&1\\ \noalign{\medskip}0&1&0
\\ \noalign{\medskip}1&0&0\end {array} \right)
$& $S_{\nu}=\left( \begin {array}{ccc} 0&0&1\\ \noalign{\medskip}0&1&0
\\ \noalign{\medskip}1&0&0\end {array} \right)
$ & $S_{l} =\left( \begin {array}{ccc} 1&0&0\\ \noalign{\medskip}0&1&0
\\ \noalign{\medskip}0&0&1\end {array} \right)
 $& $S_{\Phi} =  \left( \begin {array}{cccc} 1&0&0&0\\ \noalign{\medskip}0&1&0&0
\\ \noalign{\medskip}0&0&0&1\\ \noalign{\medskip}0&0&1&0\end {array}\right)$& $S_{\chi}=\left( \begin {array}{cc} -1&0\\ \noalign{\medskip}0&1\end {array}
 \right)$ \\
\hline
\hline
\multicolumn{5}{|c|}{Mass matrices}\\
\hline
$M_{R}= \left( \begin {array}{ccc} x&0&z\\ \noalign{\medskip}0&y&0
\\ \noalign{\medskip}z&0&w\end {array} \right)$&$M_{D}= \left( \begin {array}{ccc} A&D&C\\ \noalign{\medskip}0&B&0\\ \noalign{\medskip}C&D&A\end {array} \right)$&$M_{\nu}=\left(\begin{array}{ccc}
a&b&c\\
b&d&b\\
c&b&e
\end{array}\right)$&$M_{l}\overset{\mathrm{\mathit{v}_{3}\ll\mathit{v}_{4}}}{\approx}\mathit{v}_{4}\left( \begin {array}{ccc} Y^{3}_{l31}&0&Y^{3}_{l33}\\ \noalign{\medskip}0&Y^{3}_{l22}&0
\\ \noalign{\medskip}Y^{3}_{l11}&0&Y^{3}_{l13}\end {array} \right)$&\\
\hline
\hline
\end{tabular}}
\end{center}
\caption{\small The $Z_{2}\times Z_{6}$ symmetry realization for the $F_1:M_{\nu~12}=M_{\nu~23}$ pattern within type-I seesaw scenario. S and T are the symmetry transformation matrices for $Z_2$ and $Z_6$ respectively.}
\label{F1type1}
 \end{table}
%%%%%%%%%%%%%%%%%%%%%%%%%%%%%%%%%%%%%%
 \begin{table}[hbtp]
\begin{center}
\scalebox{0.84}{
\begin{tabular}{|c|c|c|c|c|}
\hline
\hline
\multicolumn{5}{|c|}{The matter content}\\
\hline
    $D_{L}$  & $\nu_{R}$ & $l_{R}$ & $\phi$ & $\chi$   \\
\hline
\hline
\multicolumn{5}{|c|}{$ \mbox{Symmetry under}~Z_{6}$}\\
\hline
 $T_{D}D_{L}$ & $T_{\nu}\nu_{R}$ & $T_{l}l_{R}$ & $T_{\phi}\phi$ & $T_{\chi}\chi$
\\
 $T_{D}=\mbox{diag}\left( \omega,1,1\right)$& $T_{\nu}=\mbox{diag}\left(\omega,1,1\right)$ & $T_{l} = \mbox{diag}\left( \omega^{4},\omega^{3},\omega^{3}\right) $& $T_{\phi} =\mbox{diag}\left(1,\omega^{5},\omega,\omega^{3},\omega^{3}\right)$ & $T_{\chi}=\omega^{4}$ \\
\hline
\hline
\multicolumn{5}{|c|}{$ \mbox{Symmetry under}~Z_{2}$}\\
\hline
$S_{D}D_{L}$ & $S_{\nu}\nu_{R}$ & $S_{l}l_{R}$ & $S_{\phi}\phi$ & $S_{\chi}\chi$
\\
 $S_{D}=\left( \begin {array}{ccc} 1&0&0\\ \noalign{\medskip}0&0&1
\\ \noalign{\medskip}0&1&0\end {array} \right)
$& $S_{\nu}=\left( \begin {array}{ccc} 1&0&0\\ \noalign{\medskip}0&0&1
\\ \noalign{\medskip}0&1&0\end {array} \right)
$ & $S_{l} =\left( \begin {array}{ccc} 1&0&0\\ \noalign{\medskip}0&1&0
\\ \noalign{\medskip}0&0&1\end {array} \right)
 $& $S_{\Phi} =  \left( \begin {array}{ccccc} 1&0&0&0&0\\ \noalign{\medskip}0&1&0&0&0
\\ \noalign{\medskip}0&0&-1&0&0\\ \noalign{\medskip}0&0&0&0&1
\\ \noalign{\medskip}0&0&0&1&0\end {array} \right)
$& $S_{\chi}=1$ \\
\hline
\hline
\multicolumn{5}{|c|}{Mass matrices}\\
\hline
$M_{R}= \left( \begin {array}{ccc} x&0&0\\ \noalign{\medskip}0&y&z
\\ \noalign{\medskip}0&z&y\end {array} \right)$&$M_{D}= \left( \begin {array}{ccc} A&D&D\\ \noalign{\medskip}E&B&C
\\ \noalign{\medskip}-E&C&B\end {array} \right)$&$M_{\nu}=\left(\begin{array}{ccc}
a&b&c\\
b&d&e\\
c&e&d
\end{array}\right)$&$M_{l}\overset{\mathrm{\mathit{v}_{4}\ll\mathit{v}_{5}}}{\approx}\mathit{v}_{5}\left( \begin {array}{ccc} Y^{4}_{l11}&0&0\\ \noalign{\medskip}0&Y^{4}_{l32}&Y^{4}_{l33}
\\ \noalign{\medskip}0&Y^{4}_{l22}&Y^{4}_{l23}\end {array} \right)$&\\
\hline
\hline
\end{tabular}}
\end{center}
\caption{\small The $Z_{2}\times Z_{6}$ symmetry realization for the $G_{2}:M_{\nu~22}=M_{\nu~33}$ pattern within type-I seesaw scenario. S and T are the symmetry transformation matrices for $Z_2$ and $Z_6$ respectively.}
\label{G2type1}
 \end{table}
 %%%%%%%%%%%%%%%%%%%%%%%%%%%%%%%%%%%%%%%%%%%
 \begin{table}[hbtp]
\begin{center}
\scalebox{0.79}{
\begin{tabular}{|c|c|c|c|c|}
\hline
\hline
\multicolumn{5}{|c|}{The matter content}\\
\hline
    $D_{L}$  & $\nu_{R}$ & $l_{R}$ & $\phi$ & $\chi$   \\
\hline
\hline
\multicolumn{5}{|c|}{$ \mbox{Symmetry under}~Z_{6}$}\\
\hline
 $T_{D}D_{L}$ & $T_{\nu}\nu_{R}$ & $T_{l}l_{R}$ & $T_{\phi}\phi$ & $T_{\chi}\chi$
\\
 $T_{D}=\mbox{diag}\left( \omega,1,1\right)$& $T_{\nu}=\mbox{diag}\left(\omega,1,1\right)$ & $T_{l} = \mbox{diag}\left( \omega^{4},\omega^{3},\omega^{3}\right) $& $T_{\phi} =\mbox{diag}\left(1,\omega,\omega^{3},\omega^{3}\right)$ & $T_{\chi}=\mbox{diag}\left(1,\omega^{4}\right)$ \\
\hline
\hline
\multicolumn{5}{|c|}{$ \mbox{Symmetry under}~Z_{2}$}\\
\hline
$S_{D}D_{L}$ & $S_{\nu}\nu_{R}$ & $S_{l}l_{R}$ & $S_{\phi}\phi$ & $S_{\chi}\chi$
\\
 $S_{D}=\left( \begin {array}{ccc} 1&0&0\\ \noalign{\medskip}0&0&1
\\ \noalign{\medskip}0&1&0\end {array} \right)
$& $S_{\nu}=\left( \begin {array}{ccc} 1&0&0\\ \noalign{\medskip}0&0&1
\\ \noalign{\medskip}0&1&0\end {array} \right)
$ & $S_{l} =\left( \begin {array}{ccc} 1&0&0\\ \noalign{\medskip}0&1&0
\\ \noalign{\medskip}0&0&1\end {array} \right)
 $& $S_{\Phi} =  \left( \begin {array}{cccc} 1&0&0&0\\ \noalign{\medskip}0&1&0&0
\\ \noalign{\medskip}0&0&0&1\\ \noalign{\medskip}0&0&1&0\end {array}
 \right)$& $S_{\chi}=\left( \begin {array}{cc} -1&0\\ \noalign{\medskip}0&1\end {array}
 \right)$ \\
\hline
\hline
\multicolumn{5}{|c|}{Mass matrices}\\
\hline
$M_{R}= \left( \begin {array}{ccc} x&0&0\\ \noalign{\medskip}0&y&z
\\ \noalign{\medskip}0&z&w\end {array} \right)$&$M_{D}= \left( \begin {array}{ccc} A&0&0\\ \noalign{\medskip}D&B&C
\\ \noalign{\medskip}D&C&B\end {array} \right)$&$M_{\nu}=\left(\begin{array}{ccc}
a&b&b\\
b&c&d\\
b&d&e
\end{array}\right)$&$M_{l}\overset{\mathrm{\mathit{v}_{3}\ll\mathit{v}_{4}}}{\approx}\mathit{v}_{4}\left( \begin {array}{ccc} Y^{3}_{l11}&0&0\\ \noalign{\medskip}0&Y^{3}_{l32}&Y^{3}_{l33}
\\ \noalign{\medskip}0&Y^{3}_{l22}&Y^{3}_{l23}\end {array} \right)$&\\
\hline
\hline
\end{tabular}}
\end{center}
\caption{\small The $Z_{2}\times Z_{6}$ symmetry realization for the $G_{3}:M_{\nu~12}=M_{\nu~13}$ pattern within type-I seesaw scenario. S and T are the symmetry transformation matrices for $Z_2$ and $Z_6$ respectively.}
\label{G3type1}
 \end{table}
 %\newpage
We could check that one can apply the method for the $\mu$-$\tau$-transformation related patterns ($E_2, F_2$).  We could not, however, find suitable assignments for the method to be applicable for the remaining five cases ($A_1$,$B_1$,$C_1$,$D_1$, $G_1$) and their $\mu$-$\tau$-transformation related patterns ($A_2, B_2, C_2, D_2$).

\subsection{type-II seesaw}
 We present another realization method for the four patterns by using a simpler Abelian flavor symmetry within type-II seesaw scenario. We extend the standard model with $SU(2)_{L}$ scalar triplets $\Delta_{k}$, k=(1,2,...,N),
\begin{equation}
\Delta\equiv[\Delta^{++}_{k},\Delta^{+}_{k},\Delta^{0}_{k}].
\end{equation}
The gauge-invariant Yukawa interaction Lagrangian for scalar triplets $\Delta_{k}$ takes the form,
\begin{equation}
-\mathcal{L}_{\Delta}=\sum_{k=1}^{N}\sum_{i,j=1}^{3}\Gamma_{ij}^{k}\big[\Delta_{k}^{0}\nu_{Li}^{T}\mathcal{C}\nu_{Lj}+\Delta_{k}^{+}(\nu_{Li}^{T}\mathcal{C}l_{Lj}+l_{Lj}^{T}\mathcal{C}\nu_{Li})+\Delta_{k}^{++}l_{li}^{T}\mathcal{C}l_{Lj}\big],\label{Lagtype11}
\end{equation}
where $\Gamma_{ij}^{k}$ are the Yukawa couplings and $\mathcal{C}$ is the charge conjugation matrix. The light neutrino mass matrix $M_{\nu}$ takes the form,
\begin{equation}
M_{\nu~ij}=\sum_{k=1}^{N}\Gamma_{ij}^{k}\langle\Delta^{0}_{k}\rangle,
\end{equation}
where $\langle\Delta^{0}_{k}\rangle$ are the VEVs of the $\Delta_{k}^{0}$ fields. The smallness of $\langle\Delta^{0}_{k}\rangle$ is attributed to the largeness of the triplet scalars mass scale \cite{article}. As in the type-I scenario, we take a pattern $E_1$ as a concrete example. The particle content of the standard model is extended with one Higgs doublet $\phi$ and three scalar triplets $\Delta_{k}(k=1,2,3)$. We impose $Z_{2}\times Z_{2}{'}$ symmetry on Eq.(\ref{Lagtype11}) in order to generate a desired form of $M_{\nu}$. The lepton fields transform under $Z_{2}$ symmetry as
\begin{equation}
\begin{aligned}
D_{L1}\rightarrow& D_{L1},~~~~D_{L2}\rightarrow D_{L2},~~~~D_{L3}\rightarrow -D_{L3},\\
l_{R1}\rightarrow&-l_{R1},~~~~l_{R2}\rightarrow-l_{R2},~~~l_{R3}\rightarrow l_{R3},
\end{aligned}
\end{equation}
while the scalar multiplets transform as
\begin{equation}
\begin{aligned}
\Delta_{1}\rightarrow& \Delta_{1},~~~~\Delta_{2}\rightarrow -\Delta_{2},~~\Delta_{3}\rightarrow -\Delta_{3},\\
\phi_{1}\rightarrow&-\phi_{1},~~\phi_{2}\rightarrow-\phi_{2}.\label{Z2}
\end{aligned}
\end{equation}
The bilinears $\nu_{Li}\nu_{Lj}$ and $\bar{D}_{Li}l_{Rj}$ transform under $Z_{2}$ as
\begin{equation}
\nu_{Li}\nu_{Lj}\cong\left( \begin {array}{ccc} 1&1&-1\\ \noalign{\medskip}1&1&-1
\\ \noalign{\medskip}-1&-1&1\end {array} \right),~~~\bar{D}_{Li}l_{Rj}\cong \left( \begin {array}{ccc} -1&-1&1\\ \noalign{\medskip}-1&-1&1
\\ \noalign{\medskip}1&1&-1\end {array} \right).
\end{equation}
The matrix elements (1,1), (1,2), (2,1), (2,2) and (3,3) are generated in $M_{\nu}$ when the scalar triplet $\Delta^{0}_{1}$ acquiring a small vacuum expectation value while $\Delta^{0}_{2}$ and $\Delta^{0}_{3}$ enforce the non-vanishing matrix elements (1,3), (2,3), (3,1) and (3,2). The nonzero elements (1,1), (1,2), (2,1), (2,2) and (3,3) exist in $M_{l}$ through the scalar doublets $\phi_{1}$ and $\phi_{2}$. Under the action of $Z_{2}^{'}$ symmetry, the lepton fields transform as
\begin{equation}
\begin{aligned}
D_{L1}\rightarrow& D_{L2},~~~~D_{L2}\rightarrow D_{L1},~~~~D_{L3}\rightarrow D_{L3},\\
l_{R1}\rightarrow&l_{R1},~~~~~~~l_{R2}\rightarrow l_{R2},~~~~~~~l_{R3}\rightarrow l_{R3},
\end{aligned}
\end{equation}
whereas $\Delta_{k}(k=1,2,3)$ and $\phi_{i}(i=1,2)$ transform as
\begin{equation}
\begin{aligned}
\Delta_{1}\rightarrow& \Delta_{1},~~~~\Delta_{2}\rightarrow\Delta_{2},~~\Delta_{3}\rightarrow -\Delta_{3},\\
\phi_{1}\rightarrow&\phi_{2},~~~~~\phi_{2}\rightarrow\phi_{1}.\label{Z2'}
\end{aligned}
\end{equation}
Thus, the $Z'_{2}$ invariance implies the constraints:
\begin{equation}
\begin{aligned}
&\Gamma^{1}_{11}=\Gamma^{1}_{22},~~\Gamma^{1}_{12}=\Gamma^{1}_{21},~~\Gamma^{2}_{13}=\Gamma^{2}_{23},~~\Gamma^{2}_{31}=\Gamma^{2}_{32},~~\Gamma^{3}_{13}=-\Gamma^{3}_{23},~~\Gamma^{3}_{31}=-\Gamma^{3}_{32},\\
&Y_{l11}^{1}=Y_{l21}^{2},~~Y_{l12}^{1}=Y_{l22}^{2},~~Y_{l21}^{1}=Y_{l11}^{2},~~Y_{l22}^{1}=Y_{l12}^{2},~~Y_{l33}^{1}=Y_{l33}^{2}.
\end{aligned}
\end{equation}
The resulting light neutrino mass matrix takes, under $Z_2 \times Z'_2$, the form:
\begin{equation}
M_{\nu}= \left( \begin {array}{ccc} \Gamma^{1}_{11}\langle\Delta_{1}^{0}\rangle&\Gamma^{1}_{12}\langle\Delta_{1}^{0}\rangle&\Gamma^{2}_{13}\langle\Delta_{2}^{0}\rangle+\Gamma^{3}_{13}\langle\Delta_{3}^{0}\rangle\\ \noalign{\medskip}\Gamma^{1}_{12}\langle\Delta_{1}^{0}\rangle&\Gamma^{1}_{11}\langle\Delta_{1}^{0}\rangle&\Gamma^{2}_{13}\langle\Delta_{2}^{0}\rangle-\Gamma^{3}_{13}\langle\Delta_{3}^{0}\rangle\
\\ \noalign{\medskip}\Gamma^{2}_{13}\langle\Delta_{2}^{0}\rangle+\Gamma^{3}_{13}\langle\Delta_{3}^{0}\rangle&\Gamma^{2}_{13}\langle\Delta_{2}^{0}\rangle-\Gamma^{3}_{13}\langle\Delta_{3}^{0}\rangle&\Gamma^{1}_{33}\langle\Delta_{1}^{0}\rangle\end {array} \right)
\end{equation}
which is of the requested form.

The charged lepton mass matrix takes a form,
\begin{equation}
M_{l}=\left(\begin{array}{ccc}
Y_{l11}^{1}\mathit{v}_{1}+Y^{1}_{l21}\mathit{v}_{2}&Y_{l12}^{1}\mathit{v}_{1}+Y^{1}_{l22}\mathit{v}_{2}&0\\
Y_{l21}^{1}\mathit{v}_{1}+Y^{1}_{l11}\mathit{v}_{2}&Y_{l22}^{1}\mathit{v}_{1}+Y^{1}_{l12}\mathit{v}_{2}&0\\
0&0&Y_{l33}^{1}\mathit{v}_{1}+Y^{1}_{l33}\mathit{v}_{2}
\end{array}\right).
\end{equation}
As in the type-I scenario, $M_{l}$ can be diagonalized through an infinitesimal rotation with an angle less than $10^{-2}$ by considering the vacuum condition $\mathit{v}_{1}\ll\mathit{v}_{2}$. The $Z_{2}\times Z_{2}^{'}$ symmetry realization of the remaining three patterns $F_1$, $G_{1}$ and $G_{2}$ are presented in the tables \ref{F1type2}, \ref{G2type2} and \ref{G3type2}.

\begin{table}[hbtp]
\begin{center}
\scalebox{1}{
\begin{tabular}{|c|c|c|c|}
\hline
\hline
\multicolumn{4}{|c|}{The matter content}\\
\hline
    $D_{L}$  & $l_{R}$ & $\phi$ & $\Delta$   \\
\hline
\hline
\multicolumn{4}{|c|}{$ \mbox{Symmetry under}~Z_{2}$}\\
\hline
 $T_{D}D_{L}$ & $T_{l}l_{R}$ & $T_{\phi}\phi$ & $T_{\Delta}\Delta$
\\
 $T_{D}=\mbox{diag}\left( 1,-1,1\right)$& $T_{l} = \mbox{diag}\left( -1,1,-1\right) $& $T_{\phi} =\mbox{diag}\left(-1,-1\right)$ & $T_{\Delta}=\mbox{diag}\left(1,1,-1\right)$ \\
\hline
\hline
\multicolumn{4}{|c|}{$ \mbox{Symmetry under}~Z_{2}^{'}$}\\
\hline
$S_{D}D_{L}$ & $S_{l}l_{R}$ & $S_{\phi}\phi$ & $S_{\Delta}\Delta$
\\
 $S_{D}=\left( \begin {array}{ccc} 0&0&1\\ \noalign{\medskip}0&1&0
\\ \noalign{\medskip}1&0&0\end {array} \right)
$& $S_{l} =\left( \begin {array}{ccc} 1&0&0\\ \noalign{\medskip}0&1&0
\\ \noalign{\medskip}0&0&1\end {array} \right)
 $& $S_{\phi} =  \left( \begin {array}{cc} 0&1\\ \noalign{\medskip}1&0\end {array}
 \right)
$& $S_{\Delta}= \left( \begin {array}{ccc} 1&0&0\\ \noalign{\medskip}0&-1&0
\\ \noalign{\medskip}0&0&1\end {array} \right)$
 \\
\hline
\hline
\multicolumn{4}{|c|}{Mass matrices}\\
\hline
\multicolumn{4}{|c|}{
$M_{\nu}= \left( \begin {array}{ccc} \Gamma^{1}_{11}\langle\Delta_{1}^{0}\rangle+\Gamma^{2}_{11}\langle\Delta_{2}^{0}\rangle&\Gamma^{3}_{12}\langle\Delta_{3}^{0}\rangle&\Gamma^{1}_{13}\langle\Delta_{1}^{0}\rangle\\ \noalign{\medskip}\Gamma^{3}_{12}\langle\Delta_{3}^{0}\rangle&\Gamma^{1}_{22}\langle\Delta_{1}^{0}\rangle&\Gamma^{3}_{12}\langle\Delta_{3}^{0}\rangle\\\noalign{\medskip}\Gamma^{1}_{13}\langle\Delta_{1}^{0}\rangle&\Gamma^{3}_{12}\langle\Delta_{3}^{0}\rangle&\Gamma^{1}_{11}\langle\Delta_{1}^{0}\rangle-\Gamma^{2}_{11}\langle\Delta_{2}^{0}\rangle\end {array} \right),~~ M_{l}\overset{\mathrm{\mathit{v}_{1}\ll\mathit{v}_{2}}}{\approx}\mathit{v}_{2}\left( \begin {array}{ccc} Y^{1}_{l31}&0&Y^{1}_{l33}\\ \noalign{\medskip}0&Y^{1}_{l22}&0
\\ \noalign{\medskip}Y^{1}_{l11}&0&Y^{1}_{l13}\end {array} \right)$}\\
\hline
\hline
\end{tabular}}
\end{center}
\caption{\small The $Z_{2}\times Z_{2}^{'}$ symmetry realization for the $F_1:M_{\nu~12}=M_{\nu~23}$ pattern within type-II seesaw scenario. S and T are the symmetry transformation matrices for $Z_{2}^{'}$ and $Z_2$ respectively.}
\label{F1type2}
\end{table}
%%%%%%%%%%%%%%%%%%%%%%%%%%%%%%%%%%%%%%%
 \begin{table}[hbtp]
\begin{center}
\scalebox{0.90}{
\begin{tabular}{|c|c|c|c|}
\hline
\hline
\multicolumn{4}{|c|}{The matter content}\\
\hline
    $D_{L}$  & $l_{R}$ & $\phi$ & $\Delta$   \\
\hline
\hline
\multicolumn{4}{|c|}{$ \mbox{Symmetry under}~Z_{2}$}\\
\hline
 $T_{D}D_{L}$ & $T_{l}l_{R}$ & $T_{\phi}\phi$ & $T_{\Delta}\Delta$
\\
 $T_{D}=\mbox{diag}\left( -1,1,1\right)$& $T_{l} = \mbox{diag}\left( 1,-1,-1\right) $& $T_{\phi} =\mbox{diag}\left(-1,-1\right)$ & $T_{\Delta}=\mbox{diag}\left(1,-1,-1\right)$ \\
\hline
\hline
\multicolumn{4}{|c|}{$ \mbox{Symmetry under}~Z_{2}^{'}$}\\
\hline
$S_{D}D_{L}$ & $S_{l}l_{R}$ & $S_{\phi}\phi$ & $S_{\Delta}\Delta$
\\
 $S_{D}=\left( \begin {array}{ccc} 1&0&0\\ \noalign{\medskip}0&0&1
\\ \noalign{\medskip}0&1&0\end {array} \right)
$& $S_{l} =\left( \begin {array}{ccc} 1&0&0\\ \noalign{\medskip}0&1&0
\\ \noalign{\medskip}0&0&1\end {array} \right)
 $& $S_{\phi} =  \left( \begin {array}{cc} 0&1\\ \noalign{\medskip}1&0\end {array}
 \right)
$& $S_{\Delta}= \left( \begin {array}{ccc} 1&0&0\\ \noalign{\medskip}0&1&0
\\ \noalign{\medskip}0&0&-1\end {array} \right)$
 \\
\hline
\hline
\multicolumn{4}{|c|}{Mass matrices}\\
\hline
\multicolumn{4}{|c|}{
$M_{\nu}= \left( \begin {array}{ccc} \Gamma^{1}_{11}\langle\Delta_{1}^{0}\rangle&\Gamma^{2}_{12}\langle\Delta_{2}^{0}\rangle+\Gamma^{3}_{12}\langle\Delta_{3}^{0}\rangle&\Gamma^{2}_{12}\langle\Delta_{2}^{0}\rangle-\Gamma^{3}_{12}\langle\Delta_{3}^{0}\rangle\\ \noalign{\medskip}\Gamma^{2}_{12}\langle\Delta_{2}^{0}\rangle+\Gamma^{3}_{12}\langle\Delta_{3}^{0}\rangle&\Gamma^{1}_{22}\langle\Delta_{1}^{0}\rangle&\Gamma^{1}_{23}\langle\Delta_{1}^{0}\rangle\\ \noalign{\medskip}\Gamma^{2}_{12}\langle\Delta_{2}^{0}\rangle-\Gamma^{3}_{12}\langle\Delta_{3}^{0}\rangle&\Gamma^{1}_{23}\langle\Delta_{1}^{0}\rangle&\Gamma^{1}_{22}\langle\Delta_{1}^{0}\rangle\end {array} \right),~~ M_{l}\overset{\mathrm{\mathit{v}_{1}\ll\mathit{v}_{2}}}{\approx}\mathit{v}_{2}\left( \begin {array}{ccc} Y^{1}_{l11}&0&0\\ \noalign{\medskip}0&Y^{1}_{l32}&Y^{1}_{l33}
\\ \noalign{\medskip}0&Y^{1}_{l22}&Y^{1}_{l23}\end {array} \right)$}\\
\hline
\hline
\end{tabular}}
\end{center}
\caption{\small The $Z_{2}\times Z_{2}^{'}$ symmetry realization for the $G_{2}:M_{\nu~22}=M_{\nu~33}$ pattern within type-II seesaw scenario. S and T are the symmetry transformation matrices for $Z_{2}^{'}$ and $Z_2$ respectively.}
\label{G2type2}
 \end{table}
 %%%%%%%%%%%%%%%%%%%%%%%%%%%%%%%%%%%%%%%%
 \begin{table}[hbtp]
\begin{center}
\scalebox{1}{
\begin{tabular}{|c|c|c|c|}
\hline
\hline
\multicolumn{4}{|c|}{The matter content}\\
\hline
    $D_{L}$  & $l_{R}$ & $\phi$ & $\Delta$   \\
\hline
\hline
\multicolumn{4}{|c|}{$ \mbox{Symmetry under}~Z_{2}$}\\
\hline
 $T_{D}D_{L}$ & $T_{l}l_{R}$ & $T_{\phi}\phi$ & $T_{\Delta}\Delta$
\\
 $T_{D}=\mbox{diag}\left( -1,1,1\right)$& $T_{l} = \mbox{diag}\left( 1,-1,-1\right) $& $T_{\phi} =\mbox{diag}\left(-1,-1\right)$ & $T_{\Delta}=\mbox{diag}\left(1,1,-1\right)$ \\
\hline
\hline
\multicolumn{4}{|c|}{$ \mbox{Symmetry under}~Z_{2}^{'}$}\\
\hline
$S_{D}D_{L}$ & $S_{l}l_{R}$ & $S_{\phi}\phi$ & $S_{\Delta}\Delta$
\\
 $S_{D}=\left( \begin {array}{ccc} 1&0&0\\ \noalign{\medskip}0&0&1
\\ \noalign{\medskip}0&1&0\end {array} \right)
$& $S_{l} =\left( \begin {array}{ccc} 1&0&0\\ \noalign{\medskip}0&1&0
\\ \noalign{\medskip}0&0&1\end {array} \right)
 $& $S_{\phi} =  \left( \begin {array}{cc} 0&1\\ \noalign{\medskip}1&0\end {array}
 \right)
$& $S_{\Delta}= \left( \begin {array}{ccc} 1&0&0\\ \noalign{\medskip}0&-1&0
\\ \noalign{\medskip}0&0&1\end {array} \right)$
 \\
\hline
\hline
\multicolumn{4}{|c|}{Mass matrices}\\
\hline
\multicolumn{4}{|c|}{
$M_{\nu}= \left( \begin {array}{ccc} \Gamma^{1}_{11}\langle\Delta_{1}^{0}\rangle&\Gamma^{3}_{12}\langle\Delta_{3}^{0}\rangle&\Gamma^{3}_{12}\langle\Delta_{3}^{0}\rangle\\ \noalign{\medskip}\Gamma^{3}_{12}\langle\Delta_{3}^{0}\rangle&\Gamma^{1}_{22}\langle\Delta_{1}^{0}\rangle+\Gamma_{22}^{2}\langle\Delta_{2}^{0}\rangle&\Gamma^{1}_{23}\langle\Delta_{1}^{0}\rangle\\ \noalign{\medskip}\Gamma^{3}_{12}\langle\Delta_{3}^{0}\rangle&\Gamma^{1}_{23}\langle\Delta_{1}^{0}\rangle&\Gamma^{1}_{22}\langle\Delta_{1}^{0}\rangle-\Gamma_{22}^{2}\langle\Delta_{2}^{0}\rangle\end {array} \right),~~ M_{l}\overset{\mathrm{\mathit{v}_{1}\ll\mathit{v}_{2}}}{\approx}\mathit{v}_{2}\left( \begin {array}{ccc} Y^{1}_{l11}&0&0\\ \noalign{\medskip}0&Y^{1}_{l32}&Y^{1}_{l33}
\\ \noalign{\medskip}0&Y^{1}_{l22}&Y^{1}_{l23}\end {array} \right)$}\\
\hline
\hline
\end{tabular}}
\end{center}
\caption{\small The $Z_{2}\times Z_{2}^{'}$ symmetry realization for the $G_{3}:M_{\nu~12}=M_{\nu~13}$ pattern within type-II seesaw scenario. S and T are the symmetry transformation matrices for $Z_{2}^{'}$ and $Z_2$ respectively.}
\label{G3type2}
 \end{table}
 \newpage
 As in the type-I seesaw scenario, this method is not applicable for the other viable patterns ($A_1, B_1, C_1, D_1, G_1$) and their related $\mu-\tau$ transformation patterns. We note that the particle content and flavor symmetry group in the type-II seesaw construction is simpler than the  corresponding one in the type-I scheme.

 \subsection{Type (I+II) seesaw scenario}
 In this scenario, we follow the procedure of softly breaking a continuous symmetry adopted in \cite{Grimus_2012} and \cite{Grimus_2006} in order  to enforce two equal entries in $M_{\nu}$. Moreover, going along the scheme presented in \cite{GuptaDev_2011}, applied for zero textures, we use a mixed type-I and type-II seesaw scenarios. The underlying symmetry is $Z_2\times Z_2^{'}\times U(1)^3$, where $U(1)^3$ is a global separate lepton number symmetry. We assume that $U(1)^3$  symmetry is preserved by 4-D Lagrangian terms, whereas it is softly broken by 3-dim terms in order to avoid the existence of Goldstone bosons. We extend SM with two Higgs doublets ($\phi_2$, $\phi_3$), one heavy scalar singlet $\chi_{12}$ and two Higgs triplets ($\Delta_1$, $\Delta_2$). Under the action of $Z_2$, the leptons transform as
 \begin{equation}
\begin{aligned}
D_{L1}\rightarrow& D_{L1},~~~~D_{L2}\rightarrow D_{L2},~~~~D_{L3}\rightarrow -D_{L3},\\
\nu_{R1}\rightarrow&-\nu_{R1},~~\nu_{R2}\rightarrow-\nu_{R2},~~~\nu_{R3}\rightarrow\nu_{R3},\\
l_{R1}\rightarrow&l_{R1},~~~~~~~l_{R2}\rightarrow l_{R2},~~~~~~~l_{R3}\rightarrow l_{R3},
\end{aligned}\label{transLS1}
\end{equation}
while the scalar multiplets transform as
 \begin{equation}
\begin{aligned}
\phi_{1}\rightarrow& -\phi_{1},~~~~\phi_{2}\rightarrow \phi_{2},~~~~\phi_{3}\rightarrow \phi_{3},~~\chi_{12}\rightarrow\chi_{12}\\
\Delta_{1}\rightarrow&-\Delta_{1},~~~\Delta_{2}\rightarrow-\Delta_{2}
\end{aligned}\label{transLS2}
\end{equation}
We set $\chi_{12}$ to have  lepton numbers ($L_e=-1$, $\L_\mu=-1$), whereas $\Delta_1$ and $\Delta_2$ have ($L_e=-1$, $\L_\tau=-1$) and ($L_\mu=-1$, $\L_\tau=-1$) respectively. However, we set the scalar doublets $\phi_1$, $\phi_2$ and $\phi_3$ to be $U(1)^3$ singlets. Therefore, we get under $Z_2$:
\begin{align}
\nu_{Ri}^{T}\nu_{Rj}\cong&\left( \begin {array}{ccc} 1&1&-1\\ \noalign{\medskip}1&1&-1
\\ \noalign{\medskip}-1&-1&1\end {array} \right),~\bar{D}_{Li}\nu_{Rj}\cong\left( \begin {array}{ccc} -1&-1&1\\ \noalign{\medskip}-1&-1&1
\\ \noalign{\medskip}1&1&-1\end {array} \right),~\nu_{Li}^{T}\nu_{Lj}\cong\left( \begin {array}{ccc} 1&1&-1\\ \noalign{\medskip}1&1&-1
\\ \noalign{\medskip}-1&-1&1\end {array} \right),\nonumber\\
\bar{D}_{Li}l_{j}\cong&\left( \begin {array}{ccc} 1&1&1\\ \noalign{\medskip}1&1&1
\\ \noalign{\medskip}-1&-1&-1\end {array} \right).
\end{align}
The $Z_{2}\times U(1)^3$ invariant Lagrangian is
\begin{align}
-\mathcal{L}\supset&\nu_{R1}^{T}\mathcal{C}^{-1}\nu_{R1}+\nu_{R2}^{T}\mathcal{C}^{-1}\nu_{R2}+\nu_{R3}^{T}\mathcal{C}^{-1}\nu_{R3}+\nu_{R1}^{T}\mathcal{C}^{-1}\nu_{R2}+\nu_{R2}^{T}\mathcal{C}^{-1}\nu_{R1}\nonumber\\
+&Y_{12}^{\chi}\nu_{R1}^{T}\mathcal{C}^{-1}\nu_{R2}\chi+Y_{21}^{\chi}\nu_{R2}^{T}\mathcal{C}^{-1}\nu_{R1}\chi+\big[Y_{11}^{\nu 1}\bar{D}_{L1}\nu_{R1}+Y_{22}^{\nu 1}\bar{D}_{L2}\nu_{R2}+Y_{33}^{\nu 1}\bar{D}_{L3}\nu_{R3}\big]\tilde{\phi}_{1}\nonumber\\
+&\big[Y_{11}^{l2}\bar{D}_{L1}l_{1}+Y_{22}^{l2}\bar{D}_{L2}l_{2}\big]\phi_{2}+\big[Y_{11}^{l3}\bar{D}_{L1}l_{1}+Y_{22}^{l3}\bar{D}_{L2}l_{2}\big]\phi_{3}+Y_{33}^{l1}\bar{D}_{L3}l_{3}\phi_{1}\nonumber\\
+&\big[Y_{13}^{\Delta1}\nu_{L1}^{T}\mathcal{C}^{-1}\nu_{L3}+Y_{31}^{\Delta1}\nu_{L3}^{T}\mathcal{C}^{-1}\nu_{L1}\big]\Delta_{1}+\big[Y_{23}^{\Delta2}\nu_{L2}^{T}\mathcal{C}^{-1}\nu_{L3}+Y_{32}^{\Delta2}\nu_{L3}^{T}\mathcal{C}^{-1}\nu_{L2}\big]\Delta_{2}+h.c.
\end{align}
Under the action $Z_2^{'}$ symmetry, we assume the lepton fields to transform as
\begin{equation}
\begin{aligned}
D_{L1}\rightarrow& D_{L2},~~~~D_{L2}\rightarrow D_{L1},~~~~D_{L3}\rightarrow D_{L3},\\
\nu_{R1}\rightarrow&\nu_{R2},~~~~~~\nu_{R2}\rightarrow\nu_{R1},~~~~~\nu_{R3}\rightarrow\nu_{R3},\\
l_{R1}\rightarrow&l_{R2},~~~~~~~l_{R2}\rightarrow l_{R1},~~~~~~~l_{R3}\rightarrow l_{R3},
\end{aligned}\label{transLS3}
\end{equation}
while the scalar multiplets transform as
 \begin{equation}
\begin{aligned}
\phi_{1}\rightarrow& \phi_{1},~~~~\phi_{2}\rightarrow \phi_{2},~~~~\phi_{3}\rightarrow -\phi_{3},~~\chi_{12}\rightarrow\chi_{12}\\
\Delta_{1}\rightarrow&\Delta_{2},~~~\Delta_{2}\rightarrow\Delta_{1}
\end{aligned}\label{transLS4}
\end{equation}
Thus, the $Z_{2}^{'}$ implies the constraints
\begin{align}
Y_{12}^{\chi}=&Y_{21}^{\chi},~Y_{11}^{\nu 1}=Y_{22}^{\nu1},~Y_{11}^{l2}=Y_{22}^{l2},~Y_{11}^{l3}=-Y_{22}^{l3},~Y_{13}^{\Delta1}=Y_{23}^{\Delta2},~Y_{31}^{\Delta1}=Y_{32}^{\Delta2}.
\end{align}
When $\phi_1$ and $\chi$ take vevs, then $M_{D}$ and  $M_R$ are given by
\begin{equation}
M_{D}=\left( \begin {array}{ccc} A&0&0\\ \noalign{\medskip}0&A&0
\\ \noalign{\medskip}0&0&B\end {array} \right),~M_{R}=\left( \begin {array}{ccc} x&y&0\\ \noalign{\medskip}y&x&0
\\ \noalign{\medskip}0&0&z\end {array} \right)
\end{equation}
The neutrino mass matrix which comes from the type-I seesaw is given by
\begin{equation}
M_{I}=-M_{D}M_{R}^{-1}M_{D}^{T}=\left( \begin {array}{ccc} -{\frac {{A}^{2}x}{{x}^{2}+{y}^{2}}}&{
\frac {{A}^{2}y}{{x}^{2}+{y}^{2}}}&0\\ \noalign{\medskip}{\frac {{A}^
{2}y}{{x}^{2}+{y}^{2}}}&-{\frac {{A}^{2}x}{{x}^{2}+{y}^{2}}}&0
\\ \noalign{\medskip}0&0&-{\frac {{B}^{2}}{z}}\end {array} \right),
\end{equation}
where $A=Y_{11}^{\nu 1}\mathit{v}_{1}^{\phi},~B=Y_{33}^{\nu 1}\mathit{v}_{1}^{\phi},~y=Y_{12}^{\chi}\mathit{v}^{\chi}$. We see that $M_{\nu~11}=M_{\nu~22}$ in $M_I$.
In order to enforce non-zero degenerate elements (1,3) and (2,3) in $M_{\nu}$, we have to go to type-II seesaw. When $\Delta_{1}$ and $\Delta_{2}$ take VEVs, the type-II neutrino mass matrix is given by
\begin{equation}
M_{II}=\left( \begin {array}{ccc} 0&0&X\\ \noalign{\medskip}0&0&Y
\\ \noalign{\medskip}X&Y&0\end {array} \right)
\end{equation}
where $X=Y_{13}^{H1}\mathit{v}_{1}^{\Delta}$ and $Y=Y_{13}^{H1}\mathit{v}_{2}^{\Delta}$. The effective neutrino mass matrix $M_{\nu}$ after the mixed type (I+II) seesaw mechanism becomes
\begin{equation}
M_{\nu}=M_{I}+M_{II}=\left(\begin{array}{ccc}
a&b&c\\
b&a&d\\
c&d&e
\end{array}\right).
\end{equation}
As to the charged lepton mass matrix, it is by construction diagonal, meaning we are in the flavor basis:
\begin{equation}
M_{l}=\left(\begin{array}{ccc}
Y^{l_2}_{11} v^\phi_2+Y^{l_3}_{11} v^\phi_3 &0&0\\
0&Y^{l_2}_{11} v^\phi_2-Y^{l_3}_{11} v^\phi_3&0\\
0&0&Y^{l_1}_{33} v^\phi_1
\end{array}\right).
\end{equation}
where $v^\phi_k$ is the VEV of $\phi_k$, and one can arrange so that the diagonal elements come in hierarchical ratios corresponding to the charged lepton masses hierarchy.

The symmetry realization for the other cases $F_1$,  $G_{2}$ and $G_{3}$ are summarized in the tables \ref{I+IIF1},  \ref{I+IIG2} and \ref{I+IIG3}.
%%%%%%%%%%%%%%%%%%%%%%%%%%%%%%%%%%%%
%%%%%%%%%%%%%%%%%%%%%%%%%%%%%%%%%%%
\newgeometry{left=-0.0cm,right=-1cm}
\begin{landscape}
 \begin{table}[hbtp]
\scalebox{0.92}{
{\small
\begin{tabular}{|c|c|c|c|c|c|c|c|c|c|c|c|c|c|c|}
\hline
\hline
\multicolumn{14}{|c|}{matter content $F_{1}$ pattern}\\
\hline
    $D_{L1}$  &$D_{L2}$ & $D_{L3}$& $l_{1}$ &$l_{2}$ & $l_{3}$& $\nu_{R1}$ & $\nu_{R2}$ & $\nu_{R3}$ & $\Delta_{1}$ & $\Delta_{2}$& $\phi_{1} $ &$\phi_{2}$ &$\phi_{3}$    \\
\hline
\hline
\multicolumn{14}{|c|}{$ \mbox{Becoming under}\; Z_{2}\; \mbox{symmetry}$}\\
\hline
  $-D_{L1}$ & $-D_{L2}$ & $-D_{L3}$ &$-l_{1}$ & $l_{2}$ & $-l_{3}$ & $\nu_{R1}$ & $\nu_{R2}$ & $\nu_{R3}$ & $\Delta_{1}$ & $\Delta_{2} $& $-\phi_{1} $ &$\phi_{2}$ &$\phi_{3}$    \\
\hline
 \hline
 %%%
 %%%
\multicolumn{14}{|c|}{$ \mbox{Becoming under}\; Z_{2}^{'}\; \mbox{symmetry}$}\\
\hline
$ D_{L3}$  &$ D_{L2}$ & $D_{L1}$& $ l_{3}$ &$l_{2}$ & $l_{1}$& $\nu_{R3}$ & $\nu_{R2}$ & $\nu_{R1}$ & $ \Delta_{2}$ & $ \Delta_{1} $& $\phi_{1} $ &$\phi_{2}$ &$-\phi_{3}$    \\
\hline
\hline
\multicolumn{14}{|c|}{$ \mbox{the field charge}\; {\bf q}\; \mbox{ under } U(1)^{3} \; \mbox{ symmetry}$}\\
\hline
$(1,0,0)$ & $(0,1,0)$ & $(0,0,1)$ & $(1,0,0)$ & $(0,1,0)$ & $(0,0,1)$ & $(1,0,0)$ & $(0,1,0)$ & $(0,0,1)$ & $(-2,0,0)$ & $(0,0,-2)$& ${\bf 0}$ & ${\bf 0}$ & ${\bf 0}$ \\
\hline
\hline
\end{tabular}
}}
 \caption{\small  The $Z_{2}\times Z_{2}^{'} \times U(1)^{3}$ symmetry realization for $F_1:M_{\nu~12}=M_{\nu~23}$ pattern within type I+II seesaw scenario .}\label{I+IIF1}
 \end{table}
 %\end{landscape}
 %%%%%%%%%%%%%%%%%%%%%%%%%%%%%%%%%%%%%%%%%%%%%%%%%
 %%%%%%%%%%%%%%%%%%%%%%%%%%%%%%%%%%%%%%%%%%%%%%%%%%
 %\begin{landscape}
 \begin{table}[hbtp]
\scalebox{0.91}{
{\small
\begin{tabular}{|c|c|c|c|c|c|c|c|c|c|c|c|c|c|c|}
\hline
\hline
\multicolumn{15}{|c|}{matter content $G_{2}$ pattern}\\
\hline
    $D_{L1}$  &$D_{L2}$ & $D_{L3}$& $l_{1}$ &$l_{2}$ & $l_{3}$& $\nu_{R1}$ & $\nu_{R2}$ & $\nu_{R3}$ & $\Delta_{1}$ & $\Delta_{2} $ & $\chi_{23}$ & $\phi_{1} $ &$\phi_{2}$ &$\phi_{3}$    \\
\hline
\hline
\multicolumn{15}{|c|}{$ \mbox{Becoming under}\; Z_{2}\; \mbox{symmetry}$}\\
\hline
  $-D_{L1}$ & $D_{L2}$ & $D_{L3}$ &$l_{1}$ & $l_{2}$ & $l_{3}$ & $\nu_{R1}$ & $-\nu_{R2}$ & $-\nu_{R3}$ & $-\Delta_{1}$ & $-\Delta_{2} $ & $\chi_{23}$ & $-\phi_{1} $ &$\phi_{2}$ &$\phi_{3}$    \\
\hline
 \hline
 %%%
 %%%
\multicolumn{15}{|c|}{$ \mbox{Becoming under}\; Z_{2}^{'}\; \mbox{symmetry}$}\\
\hline
$ D_{L1}$  &$D_{L3}$ & $D_{L2}$& $l_{1}$ &$l_{3}$ & $l_{2}$& $\nu_{R1}$ & $\nu_{R3}$ & $\nu_{R2}$ & $ \Delta_{2}$ & $\Delta_{1} $ & $\chi_{23}$ & $\phi_{1} $ &$\phi_{2}$ &$-\phi_{3}$    \\
\hline
\hline
\multicolumn{15}{|c|}{$ \mbox{the field charge}\; {\bf q}\; \mbox{ under } U(1)^{3} \; \mbox{ symmetry}$}\\
\hline
$(1,0,0)$ & $(0,1,0)$ & $(0,0,1)$ & $(1,0,0)$ & $(0,1,0)$ & $(0,0,1)$ & $(1,0,0)$ & $(0,1,0)$ & $(0,0,1)$ & $(-1,-1,0)$ & $(-1,0,-1)$ & $(0,-1,-1)$ & ${\bf 0}$ & ${\bf 0}$ & ${\bf 0}$ \\
\hline
\hline
\end{tabular}
}}
 \caption{\small  The $Z_{2}\times Z_{2}^{'} \times U(1)^{3}$ symmetry realization for $G_{2}:M_{\nu~22}=M_{\nu~33}$ pattern within type I+II seesaw scenario .}\label{I+IIG2}
 \end{table}
 %\end{landscape}
 %%%%%%%%%%%%%%%%%%%%%%%%%%%%%%%%%%%%%%
 %%%%%%%%%%%%%%%%%%%%%%%%%%%%%%%%%%%%%%
 %\begin{landscape}
 \begin{table}[hbtp]
\scalebox{0.85}{
{\small
\begin{tabular}{|c|c|c|c|c|c|c|c|c|c|c|c|c|c|c|}
\hline
\hline
\multicolumn{14}{|c|}{matter content $G_{3}$ pattern}\\
\hline
    $D_{L1}$  &$D_{L2}$ & $D_{L3}$& $l_{1}$ &$l_{2}$ & $l_{3}$& $\nu_{R1}$ & $\nu_{R2}$ & $\nu_{R3}$ & $\Delta_{1}$ & $\Delta_{2}$& $\phi_{1} $ &$\phi_{2}$ &$\phi_{3}$    \\
\hline
\hline
\multicolumn{14}{|c|}{$ \mbox{Becoming under}\; Z_{2}\; \mbox{symmetry}$}\\
\hline
  $-D_{L1}$ & $-D_{L2}$ & $-D_{L3}$ &$l_{1}$ & $-l_{2}$ & $-l_{3}$ & $\nu_{R1}$ & $\nu_{R2}$ & $\nu_{R3}$ & $\Delta_{1}$ & $\Delta_{2} $& $-\phi_{1} $ &$\phi_{2}$ &$\phi_{3}$    \\
\hline
 \hline
 %%%
 %%%
\multicolumn{14}{|c|}{$ \mbox{Becoming under}\; Z_{2}^{'}\; \mbox{symmetry}$}\\
\hline
$ D_{L1}$  &$ D_{L3}$ & $D_{L2}$& $ l_{1}$ &$l_{3}$ & $l_{2}$& $\nu_{R1}$ & $\nu_{R3}$ & $\nu_{R2}$ & $ \Delta_{2}$ & $ \Delta_{1} $& $\phi_{1} $ &$\phi_{2}$ &$-\phi_{3}$    \\
\hline
\hline
\multicolumn{14}{|c|}{$ \mbox{the field charge}\; {\bf q}\; \mbox{ under } U(1)^{3} \; \mbox{ symmetry}$}\\
\hline
$(1,0,0)$ & $(0,1,0)$ & $(0,0,1)$ & $(1,0,0)$ & $(0,1,0)$ & $(0,0,1)$ & $(1,0,0)$ & $(0,1,0)$ & $(0,0,1)$ & $(0,-2,0)$ & $(0,0,-2)$& ${\bf 0}$ & ${\bf 0}$ & ${\bf 0}$ \\
\hline
\hline
\end{tabular}
}}
 \caption{\small  The $Z_{2}\times Z_{2}^{'} \times U(1)^{3}$ symmetry realization for $G_{3}:M_{\nu~12}=M_{\nu~13}$ pattern within type I+II seesaw scenario .}\label{I+IIG3}
 \end{table}
  \end{landscape}
 \restoregeometry

\subsection{Realization `indirect' method related to Zero-textures}
\subsubsection{General strategy}
We saw in section 4 how one could classify all the one-equality textures into 4 classes. Relating these textures to one-zero textures amounts to finding a unitary matrix $U$ linking the one-equality texture matrix $M$ with a one-zero texture matrix $M^0$ via the `adjoint' action \footnote{In the terminology of footnote \ref{rep}, we seek a unitary matrix $U$ such that the `adjoint' representation $\mathcal{D}$ evaluated at it $\mathcal{D}_U$ is a bijection transforming $M$ into $M^0$, whereas $\mathcal{D}_{U^\dagger}$ denotes the inverse bijection transforming $M^0$ into $M$.}, as follows.
\bea U^T. M^0. U &=& M\eea
As an example, we give the unitary matrix that relates the one zero texture ($M^0_{12}=0$) to the $E_{1}$ texture satisfying $(M_{11}=M_{22})$  as follows.
\begin{equation}
 U=\left( \begin {array}{ccc} \frac{1}{\sqrt{2}}&-\frac{1}{\sqrt {2}}&0
\\ \noalign{\medskip}\frac{i}{\sqrt{2}}&\frac{i}{\sqrt {2}}&0
\\ \noalign{\medskip}0&0&1\end {array} \right)~~\Rightarrow~~U^T\left( \begin {array}{ccc} A&0&C\\ \noalign{\medskip}0&D&E
\\ \noalign{\medskip}C&E&F\end {array} \right)U=\!\!\left( \begin {array}{ccc} \frac{1}{2}(A-D)&-\frac{1}{2}(A+D)&\frac{1}{\sqrt{2}}(C+iE)\\ \noalign{\medskip}-\frac{1}{2}(A+D)&\frac{1}{2}(A-D)&-\frac{1}{\sqrt{2}}(C-iE)
\\ \noalign{\medskip}\frac{1}{\sqrt{2}}(C+iE)&-\frac{1}{\sqrt{2}}(C-iE)&F\end {array}\!\! \right) \!\!\! \in E_1.
 \end{equation}
It is easy to show that only the two classes ($Y_1, Y_2$) can be thus related to zero matrix, whereas the other two classes ($Y_3, Y_4$) can not be related by the `adjoint' action to zero textures. We show in Table \ref{unitary_matrices_relating_to_zero_textures}, for each of the remaining 5 cases of ($Y_1 \cup Y_2$) a possible unitary matrix which can act on a specified one-zero texture in order to get a matrix of the desired form.

\begin{table}[hbtp]
\begin{center}
\scalebox{1}{
\begin{tabular}{|c|c|c|c|c|}
\hline
\hline
\multicolumn{5}{|c|}{ One-equality texture}\\
\hline
    $E_2$  & $G_2$ & $F_1$ & $F_2$ & $G_3$    \\
\hline
\hline
\multicolumn{5}{|c|}{One-zero texture $M^0$}\\
\hline
 $M^0_{13}=0$ & $M^0_{23}=0$ & $M^0_{23}=0$ & $M^0_{13}=0$ & $M^0_{13}=0$\\
\hline
\hline
\multicolumn{5}{|c|}{Unitary matrix  $U$}\\
\hline
$\left( \begin {array}{ccc} \frac{1}{\sqrt{2}}&0&-\frac{1}{\sqrt {2}}
\\  \noalign{\medskip} 0&1&0
\\  \noalign{\medskip} \frac{1}{\sqrt{2}}&0&\frac{1}{\sqrt {2}}\end {array} \right)$
& $\left( \begin {array}{ccc} 1&0&0
\\ \noalign{\medskip}0&\frac{1}{\sqrt{2}}&\frac{1}{\sqrt{2}}
\\ \noalign{\medskip}0&\frac{-1}{\sqrt{2}}&\frac{1}{\sqrt {2}}\end {array} \right)$
 &$\left( \begin {array}{ccc} \frac{1}{\sqrt{2}}&0&\frac{1}{\sqrt {2}}
\\ \noalign{\medskip}0&1&0
\\ \noalign{\medskip}-\frac{1}{\sqrt{2}}&0&\frac{1}{\sqrt {2}}\end {array} \right)$
& $\left( \begin {array}{ccc} \frac{1}{\sqrt{2}}&-\frac{1}{\sqrt {2}} &0
\\ \noalign{\medskip}\frac{1}{\sqrt{2}}&\frac{1}{\sqrt {2}} &0
\\ \noalign{\medskip}0&0&1\end {array} \right)$
& $\left( \begin {array}{ccc} 1&0&0
\\ \noalign{\medskip}0& \frac{1}{\sqrt{2}}&\frac{1}{\sqrt {2}}
\\ \noalign{\medskip}0& -\frac{1}{\sqrt{2}}&\frac{1}{\sqrt {2}} \end {array} \right)$\\
\hline
\hline
\end{tabular}}
\end{center}
\caption{\small The unitary matrix $U$ relating a one-zero texture $M^0$ to one-equality texture $M$ via $(U^T. M^0. U = M)$ }
\label{unitary_matrices_relating_to_zero_textures}
\end{table}

Once we find the linking unitary matrix, then any realization model of one-zero texture can be transformed automatically, but not trivially, into a realization model for a one-equality texture, by `adjoint' action on the corresponding fields. If a field $f^0$ transforms according to $T^0_f$ in the realization of $M^0$, then a corresponding field $f$ transforms according to the rule
 \bea \label{adjoint_rule} T_f &=& U_{ex}^\dagger T^0_f U_{ex} \eea
 where $U_{ex}$ is a possible `extension' of the unitary matrix $U$ applied onto the space of field $f$.  We refer the reader, for details, to  \cite{Lashin_2017,ismael}, while we briefly apply the method for the texture $E_1$ within seesaw type-II scenario.

\subsubsection{Indirect realization of $E_{1}$ pattern with $Z_{5}$ symmetry}
The symmetry realization for the one zero texture $M^0_{12}=0$ is given in the Table \ref{t1}, with $\Omega=e^{2\pi i/5}$, where the matter content includes four  $SU(2)_{L}$ Higgs triplets $H_{i}$, three SM-like Higgs doublets $\phi_b$,  while $D_{L1}$, $D_{L2}$ and $D_{L3}$ refer to the left-handed leptons in the three generations, and $l_{R}$ for the corresponding right-handed charged leptons.
\begin{table}[hbtp]
\begin{center}
\begin{tabular}{|c|c|c|c|c|c|c|c|c|c|c|}
\hline
\hline
\multicolumn{11}{|c|}{matter content}\\
\hline
    $H_{1}$  &$H_{2}$ & $H_{3}$& $H_{4}$ &$D_{L1}$ & $D_{L2}$& $D_{L3}$&$l_{R}$&$\phi_1$&$\phi_2$&$\phi_3$ \\
\hline
\hline
\multicolumn{11}{|c|}{$ \mbox{transforming, by multiplication, under}\; Z_{5}\; \mbox{symmetry}$}\\
\hline
  1 & $\Omega^{3}$ & $\Omega^{2}$ &$\Omega$ & 1 & $\Omega$ & $\Omega^{2}$&1&1 & $\Omega$ & $\Omega^{2}$ \\
\hline
 \hline
 \end{tabular}
 \end{center}
 \caption{\small  The $Z_{5}$ symmetry assignments of the matter fields in the one zero texture $M_{12}=0$.  $H_{a},a=1,\ldots,4$ are $SU(2)_{L}$ Higgs triplets, $D_{Li}=(\n_{Li},l_{Li})^T,i=1,2,3$ refer to the left-handed leptons in the three generations, $l_{R}$ are the right-handed charged leptons in the three generations. $\phi_b,b=1,2,3$ are SM-like Higgs doublets, and $\Omega=e^{2\pi i/5}$.}
 \label{t1}
 \end{table}

 The `rotated' symmetry, which imposes the equality condition corresponding to $E_1$ pattern, is given, thus, by the following non-simple fields transformations:
 \begin{align}
 \Omega_{f}=&U^{\dagger}\Omega_{f}^{0}U,\\
  \Omega_{\phi}=&U^{\dagger}\Omega_{\phi}^{0}U,\\
 \Omega_{H}=&U_{ex}^{\dagger}\Omega_{H}^{0}U,
 \end{align}
 where $\Omega_{f}^{0}$, $\Omega_{\phi}^{0}$ and $\Omega_{H}^{0}$ are the `unrotated' symmetry, imposing $M^0_{12}=0$, diagonal transformation matrices for the fermions, SM Higgs doublets and $SU(2)_{L}$ scalar triplets respectively. We take $U_{ex}=\mbox{diag}(U,~1_{1\times1})$ as an extension of $U$ to match the finite-dimensional space of the fields $H$. Thus, we obtain the non-diagonal, possibly, transformation matrices of fields for the `rotated' symmetry leading to $E_1$ texture.
 \begin{align}
 \Omega_{D_L}=&\Omega_{\phi}=\left( \begin {array}{ccc} \frac{1}{2}(1+\Omega)&-\frac{1}{2}(1-\Omega)&0
\\ \noalign{\medskip}-\frac{1}{2}(1-\Omega)&\frac{1}{2}(1+\Omega)&0
\\ \noalign{\medskip}0&0&{\Omega}^{2}\end {array} \right)\label{DL},\\
\Omega_{H}=&\left( \begin {array}{cccc} \frac{1}{2}(1+\Omega^{3})&-\frac{1}{2}(1-\Omega^{3})&0&0
\\ \noalign{\medskip}-\frac{1}{2}(1-\Omega^{3})&\frac{1}{2}(1+\Omega^{3})&0&0
\\ \noalign{\medskip}0&0&{\Omega}^{2}&0
\\ \noalign{\medskip}0&0&0&\Omega\end {array} \right)\label{H},\\
\Omega_{l_R}=&I. \label{la}
 \end{align}
 Invariance  under $Z_5$ symmetry of the following Lagrangian:
  \bea \label{type2Lagrangian}
\mathcal{L} &\supseteq& \mathcal{L}_{H,L} + \mathcal{L}_{l} \nonumber \\  &=& \sum_{i,j=1}^3\sum_{a=1}^4\,Y_ {ij}^{\n a}\,\left[ H_a^0 \n_{Li}^T\, \mathcal{C}^{-1}\, \n_{Lj} + H_a^+ \left(\n_{Li}^T\, \mathcal{C}^{-1}\, l_{Lj}
+ l_{Lj}^T \,\mathcal{C}^{-1}\, \n_{Li}\right) + H_a^{++} l_{Li}^T \,\mathcal{C}^{-1}\, l_{Lj}\right]
\nonumber \\
&&+\sum_{i,j=1}^3\sum_{b=1}^3\,Y_ {ij}^{l b}\,   \overline{D}_{Li}\; \phi_b \; l_{Rj}   ,
\eea
   implies a constraint on the Yukawa coupling as follows:
 \begin{equation}
 Y^{\n b}=\Omega_{Hab}\Omega^T_{\nu_{L}}Y^{\n a}\Omega_{\nu_{L}},\label{const}
 \end{equation}
 where $(a,b=1...4)$.
  Now using Eqs. (\ref{DL},\ref{H}), we can solve Eq. \ref{const} to get the form of the neutrino mass matrix, when the scalar triplets acquire VEVs:
 \begin{equation}
M_{\nu}=\left( \begin {array}{ccc} Y_{11}^{\n 1}\mathit{v}_{1}^{H}+Y_{11}^{\n 2}\mathit{v}_{2}^{H}&Y_{11}^{\n 2}\mathit{v}_{1}^{H}+Y_{11}^{\n 1}\mathit{v}_{2}^{H}&Y_{13}^{\n 1}(\mathit{v}_{1}^{H}+\mathit{v}_{2}^H)+Y_{13}^{\n 3}\mathit{v}_{3}
\\ \noalign{\medskip}-~-~-~-&Y_{11}^{\n 1}\mathit{v}_{1}^{H}+Y_{11}^{\n 2}\mathit{v}_{2}^{H}&-Y_{13}^{\n 1}(\mathit{v}_{1}^{H}+\mathit{v}_{2}^H)+Y_{13}^{\n 3}\mathit{v}_{3}
\\ \noalign{\medskip}-~-~-~-&-~-~-~-&Y_{33}^{\n 4}\mathit{v}_{4}\end {array}
 \right),
\end{equation}
which is of the desired form.

What remains to be checked is whether one can, to a good approximation, argue that the charged lepton mass matrix is diagonal. Actually, in order o preserve the $Z_5$ symmetry, the Yukawa couplings for the charged lepton must take the form
\begin{equation}
Y^{l b}=\Omega_{\phi ab}\Omega^{\dagger}_{D_{L}}Y^{l a}\Omega_{l_{R}},\label{const2}
\end{equation}
where $(a,b=1...3)$. By solving Eq. \ref{const2} with Eqs. (\ref{DL},\ref{H},\ref{la}), we obtain
\begin{equation}
M_{l}\approx\mathit{v}\left( \begin {array}{ccc} Y_{11}^{l2}&Y_{12}^{l2}&Y_{13}^{l2}
\\ \noalign{\medskip}Y_{21}^{l2}&Y_{22}^{l2}&Y_{23}^{l2}
\\ \noalign{\medskip}Y_{31}^{l2}+Y_{31}^{l3}&Y_{32}^{l3}&Y_{33}^{l3}\end {array}
 \right)=\mathit{v}\left( \begin {array}{c} \mathbf{{a}^{T}}\\ \noalign{\medskip}\mathbf{{b}^{T}}
\\ \noalign{\medskip}\mathbf{{c}^{T}}\end {array} \right)\Rightarrow M_{l}M_{l}^{\dagger}\approx\mathit{v}^{2}\left( \begin {array}{ccc}|\textbf{a}|^{2}&\textbf{a}.\textbf{b}&\textbf{a}.\textbf{c}\\ \noalign{\medskip}\textbf{b}.\textbf{a}&|\textbf{b}|^{2}&\textbf{b}.\textbf{c}
\\ \noalign{\medskip}\textbf{c}.\textbf{a}&\textbf{c}.\textbf{b}&|\textbf{c}|^{2}\end {array} \right).
\end{equation}
where we assumed  $\mathit{v}_{1}^{\phi}\ll\mathit{v}_{2}^{\phi}\approx\mathit{v}_{3}^{\phi}\approx\mathit{v}$. We see now that if we take the  natural assumption on the norms of the vectors
\begin{align}
||\textbf{a}||/||\textbf{c}||=&m_{e}/m_{\tau}=3\times 10^{-4},\\
||\textbf{b}||/||\textbf{c}||=&m_{e}/m_{\tau}=6\times 10^{-2},
\end{align} then
$M_{l}M_{l}^{\dagger}$ can be diagonalized by an infinitesimal rotation, which means we are to a good approximation in the flavor basis.

 \section{Summary and conclusion}
 In this work, we have studied a specific texture characterized by one equality between two independent elements in the neutrino mass matrix. We find that all the 15 possible one-Equality textures can accommodate the experimental data in the case of normal ordering ($m_{1}<$  $m_{2}<$  $m_{3}$), whereas only fourteen one-Equality patterns are viable in the inverted hierarchy type ($m_{3}<$ $m_{1}<$ $m_{2}$). The pattern $C_{2}$, however, is viable for the latter type only at the $3\sigma$-level and with a limited acceptable region. We give the analytical formulae for the A's coefficients Eq.(\ref{coeff}) corresponding to each independent texture, and present approximative analytical expressions for the mass ratios and other neutrino physical parameters provided they are not too cumbersome to be presented. The analytical expressions for the remaining six cases can be deduced from the $\mu-\tau$ permutation transformation. The singular (non-invertible) neutrino mass matrices with one-equality texture  are expected for four patterns in normal type and ten patterns in inverted type. The allowed ranges of the neutrino physical parameters at all $\sigma$-levels for each hierarchy type of all possible patterns are listed in the Tables  (\ref{numerical1},\ref{numerical2}). In Fig.\ref{A1fig}-\ref{G3fig}, we introduce 9 or 12 correlation plots, at the $3\sigma$-level, for all possible textures with either hierarchy type, whereas the correlations ($m_{ee}$-LNM) are put in Figures (\ref{LNM vs mee (N)} and \ref{LNM vs mee (I)}) for normal and inverted type respectively.

 Finally, we present symmetry realizations for four viable patterns, from nine pattern not related by $\mu$-$\tau$-symmetry,  in different seesaw scenarios. In the framework of the type-I scenario, we use a $Z_{2}\times Z_{6}$ discrete flavor symmetry to enforce one equality in the neutrino mass matrix, while $Z_{2}\times Z_{2}^{'}$ symmetry is used in the type-II scheme. The realization methods proposed in the different seesaw scenarios are applicable for only six texture structures out of fifteen. We find the particle content and flavor symmetry group in the type-II seesaw construction is simpler than the corresponding one in the type-I scheme. Thus, and for simplicity, we also adopt type-II scenario in explaining `indirect' methods of realizations by linking the one-equality texture to one-zero textures.

 In the realization methods, we have not yet touched the question of the scalar potential and finding its general form under the imposed symmetry. Nor did we deal with the radiative corrections effect on the phenomenology and whether or not it can spoil the form of the texture while running from the ``ultraviolet'' scale where the seesaw scale imposes the texture form to the low scale where phenomenology was analyzed. Some studies \cite{ismael} argue that, unlike zero textures, Equality textures do not keep their form under RG running, and so this question is worthy of a separate study. Likewise, introducing many scalars may lead to rich phenomenology at colliders, and asking for just one SM-like Higgs at low scale requires a situation where fine tuning of the many parameters in the scalar potential, so that to ensure new scalars are out of reach at current experiments, is heavily called upon. For completeness, we treat in appendix the question of how to find the most general form of the scalar potential, within type II seesaw scenario, in one specific texture, noting that a corresponding rich colliders' phenomenology needs a complete analysis of such a setup, which goes beyond the current study.

\section*{{\large \bf Acknowledgements}}
 E.I. Lashin and N. Chamoun acknowledge support from ICTP-Associate program. N.C. acknowledges support of the Alexander von Humboldt
Foundation and is grateful for the kind hospitality
of the Bethe Center for Theoretical Physics at Bonn University. E.L.'s work was partially supported by the STDF project 37272.

 \appendix{
 \section{Scalar potential for $E_{1}$ texture within $Z_2 \times Z_2^{'}$ symmetry}
 In this appendix, we write down  the $ Z_ {2} \times Z_ {2} ^ {'} $ invariant terms for the type-II scalar potential. For the sake of illustration, we take $ E_1 $ texture as a concrete example. We extend the SM with three $ SU (2) _L $ Higgs triplets $ (\Delta_1, \Delta_2, \Delta_3) $ besides one Higgs doublet. The $Z_2$ and $Z_2^{'}$ symmetry transformation for the Higgs multiplets are given by Eqs. (\ref{Z2} and \ref{Z2'}). The most renormalizable gauge invariant type-II scalar potential takes the following form \cite{Grimus_2000, Chaudhuri_2014}:
 \begin{align}
 V(\phi,\Delta)=&V_{\phi}+V_{\Delta}+V_{\phi\Delta},\nonumber\\
 V_{\phi}=&a_{ij}\phi_{i}^{\dagger}\phi_{j}+c_{ijkl}\phi^{\dagger}_{i}\phi_{j}\phi_{k}^{\dagger}\phi_{l},\nonumber\\
 V_{\Delta}=&b_{ij}\mbox{Tr}(\Delta_{i}^{\dagger}\Delta_{j})+d_{ijkl}\mbox{Tr}(\Delta_{i}^{\dagger}\Delta_{j})\mbox{Tr}(\Delta_{k}^{\dagger}\Delta_{l})+f_{ijkl}\mbox{Tr}(\Delta_{i}^{\dagger}\Delta_{j}^{\dagger})\mbox{Tr}(\Delta_{k}\Delta_{l}),\nonumber\\
 V_{\phi\Delta}=&\bigg(\frac{e_{ijkl}-h_{ijkl}}{2}\bigg)\phi_{i}^{\dagger}\phi_{j}\mbox{Tr}(\Delta_{k}^{\dagger}\Delta_{l})+h_{ijkl}\phi_{i}^{\dagger}\Delta_{j}^{\dagger}\Delta_{k}\phi_{l}+\big(t_{ijk}\phi_{i}^{\dagger}\Delta_{j}\tilde{\phi}_{k}+h.c.\big).
 \end{align}

 The lower indices of $\phi$'s run from 1 to 2, whereas the lower indices of $\Delta$'s run from 1 to 3. The scalar triplet $\Delta$ is written as
 \begin{equation}
\Delta=\vec{\Phi}.\vec{\tau}=\left( \begin {array}{cc} H^{+}&\sqrt{2}H^{++}\\ \noalign{\medskip}
\sqrt{2}H^{0}&-H^{+}\end {array} \right),~\mbox{where}~\Phi_{i}=\left( \begin {array}{c} \frac{1}{\sqrt{2}}\big(H^{0}+H^{++}\big)\\ \noalign{\medskip} -\frac{i}{\sqrt{2}}\big(H^{0}-H^{++}\big)
\\ \noalign{\medskip}H^{+}\end {array} \right),
\end{equation}
and $\tau_{i}~(i=1,2,3)$ are the Pauli matrices. However, the scalar doublet $\phi$ is given by
\begin{equation}
\phi=\left( \begin {array}{c} \phi^{+}\\ \noalign{\medskip}\phi^{0}
\end {array} \right).
\end{equation}
The VEVs of the scalar multiplets are
\begin{equation}
\langle\phi_{i}\rangle_{0}=\frac{1}{\sqrt{2}}\left( \begin {array}{c} 0\\ \noalign{\medskip}|\mathit{v}_{i}|e^{i\alpha_{i}}
\end {array} \right),~~\langle\Delta_{i}\rangle_{0}=\left( \begin {array}{cc} 0&0\\ \noalign{\medskip}|\omega_{i}|e^{i\beta_{i}}&0\end {array} \right).
\end{equation}
The $Z_{2}\times Z_{2}^{'}$ invariant $V_{\phi}$ terms are given by
\begin{align}
V_{\phi}=&a_{11}\big(\phi^{\dagger}_{1}\phi_{1}+\phi^{\dagger}_{2}\phi_{2}\big)+a_{12}\big(\phi^{\dagger}_{1}\phi_{2}+\phi^{\dagger}_{2}\phi_{1}\big)+c_{1111}\big[(\phi^{\dagger}_{1}\phi_{1})^{2}+(\phi^{\dagger}_{2}\phi_{2})^{2}\big]+c_{1122}\phi_{1}^{\dagger}\phi_{1}\phi_{2}^{\dagger}\phi_{2}+c_{1221}\phi_{1}^{\dagger}\phi_{2}\phi_{2}^{\dagger}\phi_{1}\nonumber\\+&c_{1212}\big[(\phi_{1}^{\dagger}\phi_{2})^{2}+(\phi_{2}^{\dagger}\phi_{1})^{2}\big]+c_{1112}\big[\phi_{1}^{\dagger}\phi_{1}\phi_{1}^{\dagger}\phi_{2}+\phi_{2}^{\dagger}\phi_{1}\phi_{2}^{\dagger}\phi_{2}+\phi_{2}^{\dagger}\phi_{1}\phi_{1}^{\dagger}\phi_{1}+\phi_{1}^{\dagger}\phi_{2}\phi_{2}^{\dagger}\phi_{2}\big].
\end{align}
We see that a's and c's parameters are real. Thus, we get
\begin{align}
\bra{0}V_{\phi}\ket{0}=&\frac{a_{11}}{2}\big(|\mathit{v}_{1}|^2+|\mathit{v}_{2}|^2\big)+a_{12}~\mbox{Re}\big(\mathit{v}_{1}^{*}\mathit{v}_{2}\big)+\frac{c_{1111}}{4}\big(|\mathit{v}_{1}|^4+|\mathit{v}_{2}|^4\big)
+\frac{c_{1122}}{4}|\mathit{v}_{1}\mathit{v}_{2}|^2+\frac{c_{1221}}{4}|\mathit{v}_{1}\mathit{v}_{2}|^2\nonumber\\
+&\frac{c_{1212}}{2}~\mbox{Re}\big(\mathit{v}_{1}^{*2}\mathit{v}_{2}^{2}\big)+\frac{c_{1112}}{2}\big(|\mathit{v}_{1}|^2+|\mathit{v}_{2}|^2\big)~\mbox{Re}\big(\mathit{v}_{1}^{*}\mathit{v}_{2}\big).
\end{align}
The $Z_2\times Z_{2}^{'}$ invariant $V_{\Delta}$ is
\begin{align}
V_{\Delta}=&b_{11}\mbox{Tr}(\Delta^{\dagger}_{1}\Delta_{1})+b_{22}\mbox{Tr}(\Delta^{\dagger}_{2}\Delta_{2})+b_{33}\mbox{Tr}(\Delta^{\dagger}_{3}\Delta_{3})+d_{1111}\big[\mbox{Tr}(\Delta^{\dagger}_{1}\Delta_{1})\big]^2+d_{1122}\mbox{Tr}(\Delta_{1}^{\dagger}\Delta_{1})\mbox{Tr}(\Delta_{2}^{\dagger}\Delta_{2})\nonumber\\
+&d_{2222}\big[\mbox{Tr}(\Delta^{\dagger}_{2}\Delta_{2})\big]^2+d_{3333}\big[\mbox{Tr}(\Delta^{\dagger}_{3}\Delta_{3})\big]^2+d_{1133}\mbox{Tr}(\Delta_{1}^{\dagger}\Delta_{1})\mbox{Tr}(\Delta_{3}^{\dagger}\Delta_{3})+d_{2233}\mbox{Tr}(\Delta_{2}^{\dagger}\Delta_{2})\mbox{Tr}(\Delta_{3}^{\dagger}\Delta_{3})\nonumber\\+&d_{1221}\mbox{Tr}(\Delta_{1}^{\dagger}\Delta_{2})\mbox{Tr}(\Delta_{2}^{\dagger}\Delta_{1})+d_{1331}\mbox{Tr}(\Delta_{1}^{\dagger}\Delta_{3})\mbox{Tr}(\Delta_{3}^{\dagger}\Delta_{1})+d_{2332}\mbox{Tr}(\Delta_{2}^{\dagger}\Delta_{3})\mbox{Tr}(\Delta_{3}^{\dagger}\Delta_{2})\nonumber\\+&d_{1212}\big[(\mbox{Tr}(\Delta^{\dagger}_{1}\Delta_{2})^2+h.c.\big]+d_{2323}\big[(\mbox{Tr}(\Delta^{\dagger}_{2}\Delta_{3})^2+h.c.\big]+d_{1313}\big[(\mbox{Tr}(\Delta^{\dagger}_{1}\Delta_{3})^2+h.c.\big]\nonumber\\
+&f_{1111}\mbox{Tr}(\Delta^{\dagger}_{1}\Delta^{\dagger}_{1})\mbox{Tr}(\Delta_{1}\Delta_{1})+f_{2222}\mbox{Tr}(\Delta^{\dagger}_{2}\Delta^{\dagger}_{2})\mbox{Tr}(\Delta_{2}\Delta_{2})+f_{3333}\mbox{Tr}(\Delta^{\dagger}_{3}\Delta^{\dagger}_{3})\mbox{Tr}(\Delta_{3}\Delta_{3})\nonumber\\
+&f_{1212}\mbox{Tr}(\Delta^{\dagger}_{1}\Delta^{\dagger}_{2})\mbox{Tr}(\Delta_{1}\Delta_{2})+f_{1313}\mbox{Tr}(\Delta^{\dagger}_{1}\Delta^{\dagger}_{3})\mbox{Tr}(\Delta_{1}\Delta_{3})+f_{2323}\mbox{Tr}(\Delta^{\dagger}_{2}\Delta^{\dagger}_{3})\mbox{Tr}(\Delta_{2}\Delta_{3})\nonumber\\
+&f_{1122}\big[\mbox{Tr}(\Delta^{\dagger}_1\Delta^{\dagger}_1)Tr(\Delta_2\Delta_2)+h.c.\big]+f_{1133}\big[\mbox{Tr}(\Delta^{\dagger}_1\Delta^{\dagger}_1)\mbox{Tr}(\Delta_3\Delta_3)+h.c.\big]\nonumber\\
+&f_{2233}\big[\mbox{Tr}(\Delta^{\dagger}_2\Delta^{\dagger}_2)\mbox{Tr}(\Delta_3\Delta_3)+h.c.\big]
\end{align}
where $f$ and $d$ parameters are all real. Then, we obtain
\begin{align}
\bra{0}V_{\Delta}\ket{0}=&b_{11}|\mathit{\omega}_{1}|^2+b_{22}|\mathit{\omega}_{2}|^2+b_{33}|\mathit{\omega}_{3}|^2+d_{1111}|\mathit{\omega}_{1}|^4+d_{2222}|\mathit{\omega}_{2}|^4+d_{3333}|\mathit{\omega}_{3}|^4
+(d_{1122}+d_{1221})|\mathit{\omega}_{1}^{2}\mathit{\omega}_{2}^{2}|\nonumber\\+&(d_{1133}+d_{1331})|\mathit{\omega}_{1}^{2}\mathit{\omega}_{3}^{2}|+(d_{2233}+d_{2332})|\mathit{\omega}_{2}^{2}\mathit{\omega}_{3}^{2}|+2d_{1212}~\mbox{Re}\big(\omega_{1}^{*2}\omega_{2}^{2}\big)+2d_{2323}~\mbox{Re}\big(\omega_{2}^{*2}\omega_{3}^{2}\big)\nonumber\\
+&2d_{1313}~\mbox{Re}\big(\omega_{1}^{*2}\omega_{3}^{2}\big).
\end{align}
The $Z_2\times Z_{2}^{'}$ invariant interaction potential $V_{\phi\Delta}$ is given by
\begin{align}
V_{\phi\Delta}=&\bigg(\frac{e-h}{2}\bigg)_{1111}\bigg[\phi_{1}^{\dagger}\phi_{1}\mbox{Tr}(\Delta_{1}^{\dagger}\Delta_1)+\phi_{2}^{\dagger}\phi_{2}\mbox{Tr}(\Delta_{1}^{\dagger}\Delta_1)\bigg]+\bigg(\frac{e-h}{2}\bigg)_{2222}\bigg[\phi_{2}^{\dagger}\phi_{2}\mbox{Tr}(\Delta_{2}^{\dagger}\Delta_2)+\phi_{1}^{\dagger}\phi_{1}\mbox{Tr}(\Delta_{2}^{\dagger}\Delta_2)\bigg]\nonumber\\
+&\bigg(\frac{e-h}{2}\bigg)_{1133}\bigg[\phi_{1}^{\dagger}\phi_{1}\mbox{Tr}(\Delta_{3}^{\dagger}\Delta_3)+\phi_{2}^{\dagger}\phi_{2}\mbox{Tr}(\Delta_{3}^{\dagger}\Delta_3)\bigg]+\bigg(\frac{e-h}{2}\bigg)_{1222}\bigg[\phi_{1}^{\dagger}\phi_{2}\mbox{Tr}(\Delta_{2}^{\dagger}\Delta_2)+\phi_{2}^{\dagger}\phi_{1}\mbox{Tr}(\Delta_{2}^{\dagger}\Delta_2)\bigg]\nonumber\\
+&\bigg(\frac{e-h}{2}\bigg)_{1233}\bigg[\phi_{1}^{\dagger}\phi_{2}\mbox{Tr}(\Delta_{3}^{\dagger}\Delta_3)+\phi_{2}^{\dagger}\phi_{1}\mbox{Tr}(\Delta_{3}^{\dagger}\Delta_3)\bigg]+\bigg(\frac{e-h}{2}\bigg)_{1211}\bigg[\phi_{1}^{\dagger}\phi_{2}\mbox{Tr}(\Delta_{1}^{\dagger}\Delta_1)+\phi_{2}^{\dagger}\phi_{1}\mbox{Tr}(\Delta_{1}^{\dagger}\Delta_1)\bigg]\nonumber\\
+&h_{1111}\bigg[\phi_1^{\dagger}\Delta_{1}^{\dagger}\Delta_{1}\phi_{1}+\phi_2^{\dagger}\Delta_{1}^{\dagger}\Delta_{1}\phi_{2}\bigg]+h_{2222}\bigg[\phi_2^{\dagger}\Delta_{2}^{\dagger}\Delta_{2}\phi_{2}+\phi_1^{\dagger}\Delta_{2}^{\dagger}\Delta_{2}\phi_{1}\bigg]+h_{1331}\bigg[\phi_1^{\dagger}\Delta_{3}^{\dagger}\Delta_{3}\phi_{1}+\phi_2^{\dagger}\Delta_{3}^{\dagger}\Delta_{3}\phi_{2}\bigg]\nonumber\\
+&h_{1222}\bigg[\phi_1^{\dagger}\Delta_{2}^{\dagger}\Delta_{2}\phi_{2}+\phi_2^{\dagger}\Delta_{2}^{\dagger}\Delta_{2}\phi_{1}\bigg]+h_{1332}\bigg[\phi_1^{\dagger}\Delta_{3}^{\dagger}\Delta_{3}\phi_{2}+\phi_2^{\dagger}\Delta_{3}^{\dagger}\Delta_{3}\phi_{1}\bigg]+h_{1112}\bigg[\phi_1^{\dagger}\Delta_{1}^{\dagger}\Delta_{1}\phi_{2}+\phi_2^{\dagger}\Delta_{1}^{\dagger}\Delta_{1}\phi_{1}\bigg]\nonumber\\
+&\bigg[\bigg(\frac{e-h}{2}\bigg)_{2223}\bigg(\phi_{2}^{\dagger}\phi_{2}\mbox{Tr}(\Delta_{2}^{\dagger}\Delta_3)-\phi_{1}^{\dagger}\phi_{1}\mbox{Tr}(\Delta_{2}^{\dagger}\Delta_3)\bigg)+\bigg(\frac{e-h}{2}\bigg)_{1223}\bigg(\phi_{1}^{\dagger}\phi_{2}\mbox{Tr}(\Delta_{2}^{\dagger}\Delta_3)-\phi_{2}^{\dagger}\phi_{1}\mbox{Tr}(\Delta_{2}^{\dagger}\Delta_3)\bigg)\nonumber\\+&h_{2232}\bigg(\phi_2^{\dagger}\Delta_{2}^{\dagger}\Delta_{3}\phi_{2}-\phi_1^{\dagger}\Delta_{2}^{\dagger}\Delta_{3}\phi_{1}\bigg)+h_{1232}\bigg(\phi_1^{\dagger}\Delta_{2}^{\dagger}\Delta_{3}\phi_{2}-\phi_2^{\dagger}\Delta_{2}^{\dagger}\Delta_{3}\phi_{1}\bigg)+t_{111}\big(\phi_1^{\dagger}\Delta_1\tilde{\phi}_1+\phi_2^{\dagger}\Delta_1\tilde{\phi}_2\big)\nonumber\\+&t_{112}\big(\phi_1^{\dagger}\Delta_1\tilde{\phi}_2+\phi_2^{\dagger}\Delta_1\tilde{\phi}_1\big)+h.c.\bigg].
\end{align}
We used $\big(\frac{e_{ijkl}-h_{ijkl}}{2}\big)=\big(\frac{e-h}{2}\big)_{ijkl}$. We see that all the $e-$ and $h-$ parameters are real except $e_{2223}$, $e_{1223}$, $h_{2223}$, $h_{1223}$, $h_{2232}$, $h_{1232}$, $t_{111}$ and $t_{112}$ are complex. Thus, we get
\begin{align}
\bra{0}V_{\phi\Delta}\ket{0}=&\bigg[\bigg(\frac{e-h}{4}\bigg)_{1111}|\mathit{\omega_1}|^2+\bigg(\frac{e-h}{4}\bigg)_{2222}|\mathit{\omega_2}|^2+\bigg(\frac{e-h}{4}\bigg)_{1133}|\mathit{\omega_3}|^2\bigg]\big(|\mathit{v}_{1}|^2+|\mathit{v}_{2}|^2\big)\nonumber\\
+&\bigg[\bigg(\frac{e-h}{2}\bigg)_{1211}|\mathit{\omega_1}|^2+\bigg(\frac{e-h}{2}\bigg)_{1222}|\mathit{\omega_2}|^2+\bigg(\frac{e-h}{2}\bigg)_{1233}|\mathit{\omega_3}|^2\bigg]\mbox{Re}\big(\mathit{v}_{1}^{*}\mathit{v}_{2}\big)\nonumber\\
+&\mbox{Re}\bigg[\big(e-h\big)_{2223}\omega_{2}^{*}\omega_{3}\bigg]\frac{|\mathit{v}_{2}|^2-|\mathit{v}_{1}|^2}{2}+\frac{1}{2}\mbox{Re}\bigg[\big(e-h\big)_{1223}\omega_{2}^{*}\omega_{3}\mbox{Im}\big(\mathit{v}_{1}^{*}\mathit{v}_{2}\big)\bigg]\nonumber\\
+&|t_{111}\mathit{v}_{1}^{2}\omega_{1}|\cos(2\alpha_1-\beta_1-\gamma_1)+|t_{111}\mathit{v}_{2}^{2}\omega_{1}|\cos(2\alpha_2-\beta_1-\gamma_1)\nonumber\\
+&2|t_{112}\omega_1\mathit{v}_1\mathit{v}_2|\cos(\alpha_1+\alpha_2-\beta_1-\gamma_2)
\end{align}
where $t_{111}=|t_{111}|e^{i\gamma_1},~t_{112}=|t_{112}|e^{i\gamma_2}$.

In all, using the above expressions, one can now construct the $3\times 3$ mass matrix for the doubly charged scalars ($H^{++}_k, k\in\{1,2,3\}$), and the $5 \times 5$ mass matrix for the singly charged scalars ($H^+_k, \phi^+_j, j\in\{1,2\}$), whereas for the neutral scalars ($H^0_k,\phi^0_j$) with real and imaginary parts, we have a $10 \times 10$ mass matrix. For the latter,  and due to $SU(2)$ gauge invariance we should have at least three vanishing eigen masses corresponding to the three Goldstone bosons.

There are many ways to simplify the calculations, such as assuming all the VEVs are real so that no CP violation originating from them, or extending the gauge symmetry by an extra $U(1)$, with respect to which the neutrinos are neutral so that $M_\nu$ structure remains invariant, in order to kill the cubic interaction terms. Furthermore, one can also assume ad hoc certain vanishing parameters so that no coupling between scalars and pseudo scalars.  Such a full study of the scalar potential is interesting and necessary for colliders' phenomenology, however it is beyond the scope of this paper.
 }

\newpage
\bibliographystyle{mybst}
\bibliography{ref}
\end{document}